\documentclass[12pt]{iopart}
\usepackage{graphicx}  
\def\gsimeq{\,\,\raise0.14em\hbox{$>$}\kern-0.76em\lower0.28em\hbox  
{$\sim$}\,\,}  
\def\lsimeq{\,\,\raise0.14em\hbox{$<$}\kern-0.76em\lower0.28em\hbox  
{$\sim$}\,\,} 

\def\gsimeq{\,\,\raise0.14em\hbox{$>$}\kern-0.76em\lower0.28em\hbox
{$\sim$}\,\,}
\def\lsimeq{\,\,\raise0.14em\hbox{$<$}\kern-0.76em\lower0.28em\hbox
{$\sim$}\,\,}
\def\chem#1#2{$\rm{}^{#1}\kern-0.8pt#2$}
\def\reac#1#2#3#4#5#6{$\rm\,{}^{#1}\kern-0.8pt{#2}\,({#3}\,,{#4})\,
{}^{#5}\kern-0.8pt{#6}\,$}

\begin{document}

\title{Nuclear astrophysics}

\author{M  Arnould \dag\footnote[3]{to whom correspondence should be 
addressed.} and K  Takahashi\ddag\footnote[4]{supported by the
SFB 375, Astro-Particle Physics, of the
Deutsche Forschungsgemeinschaft}}

\address{\dag\ Institut d'Astronomie et d'Astrophysique, Universit\'e 
Libre de Bruxelles, Campus Plaine, CP 226, B-1050 Bruxelles, Belgium
\ \ \ marnould@astro.ulb.ac.be}

\address{\ddag\ Max-Planck-Institut f\"ur Astrophysik,
Karl-Schwarzschild-Str. 1, D-85740 Garching, Germany
\ \ \ kjt@mpa-garching.mpg.de}
 
\begin{abstract}
 
Nuclear astrophysics is that branch of astrophysics which helps 
understanding the Universe, or at least some of its many faces, through 
the knowledge of the microcosm of the atomic nucleus.
It attempts to find as many nuclear physics imprints as possible in the 
macrocosm, and to decipher what those messages are telling us 
about the varied constituent objects in the Universe at present and in 
the past.
In the last decades much advance has been made in nuclear astrophysics 
thanks to the sometimes spectacular progress made in the modelling of
the structure and evolution of the stars, in the quality and diversity
of the astronomical observations, as well as in the experimental and 
theoretical understanding of the atomic nucleus and of its spontaneous 
or induced transformations. 
Developments in other sub-fields of physics and chemistry have also 
contributed to that advance. 
Notwithstanding the accomplishment, many long-standing problems remain 
to be solved, and the theoretical understanding of a large variety of 
observational facts needs to be put on safer grounds. 
In addition, new questions are continuously emerging, and new facts 
endangering old ideas.

This review shows that astrophysics has been, and still is, highly  
demanding to nuclear physics in both its experimental and theoretical 
components. 
On top of the fact that large varieties of nuclei have to be dealt with,  
these nuclei are immersed in highly unusual environments which may have  
a significant impact on their static properties, the diversity of their 
transmutation modes, and on the probabilities of these modes.  
In order to have a chance of solving some of the problems nuclear 
astrophysics is facing, the astrophysicists and nuclear physicists  
are obviously bound to put their competence in common, and have  
sometimes to benefit from the help of other fields of physics, like  
particle physics, plasma physics or solid-state physics.  
Given the highly varied and complex aspects,
we pick here some specific nuclear physics topics 
which largely pervade nuclear astrophysics.
\end{abstract}

\maketitle
\vfill\eject
\noindent
{\bf Contents}
\vskip0.5truecm
\noindent
1.\ Introduction
 
\noindent
2.\ Observational foundation
 
\noindent\ \ \
2.1. The Hertzsprung-Russell diagram (HRD)
 
\noindent\ \ \
2.2. Electromagnetic spectra and abundance determinations
 
\noindent\ \ \ \indent
2.2.1. The spatio-temporal evolution of the composition of our 
Galaxy
 
\noindent\ \ \ \indent
2.2.2. The composition of other galaxies
 
\noindent\ \ \ \indent
2.2.3. Abundances in evolved or exploding stars  
 
\noindent\ \ \ \indent
2.2.4. A special electromagnetic message from the sky: 
$\gamma$-ray line astrophysics
 
\noindent\ \ \
2.3. The composition of the solar system
 
\noindent\ \ \ \indent
2.3.1. The bulk solar-system composition
 
\noindent\ \ \ \indent
2.3.2. Isotopic anomalies in the solar composition
 
\noindent\ \ \ \indent
2.3.3. The composition of cosmic rays and solar energetic 
particles
 
\noindent\ \ \
2.4. Neutrino astronomy
 
\noindent
3.\ The contributors to nuclei in the Cosmos
 
\noindent\ \ \
3.1. Some generalities about stellar structure and evolution
 
\noindent\ \ \ \indent
3.1.1. $M \gsimeq 10$ M$_{\odot}$ stars
 
\noindent\ \ \ \indent
3.1.2. $0.45 \lsimeq M \lsimeq 8$ M$_{\odot}$ stars
 
\noindent\ \ \ \indent
3.1.3. $8 \lsimeq M \lsimeq 10$ M$_{\odot}$ stars
 
\noindent\ \ \ \indent
3.1.4. Binary stars
 
\noindent\ \ \
3.2. Spallation reactions
 
\noindent\ \ \
3.3. The Big Bang contribution to nucleosynthesis
 
\noindent\ \ \
3.4. The chemical evolution of galaxies
 
\noindent
4.\ Nuclear data needs for astrophysics: Nuclear masses, decay modes
 
\noindent\ \ \
4.1. The masses of ``cold'' nuclei
 
\noindent\ \ \
4.2. Nuclei at high temperatures
 
\noindent\ \ \
4.3. Nuclei at high densities
 
\noindent\ \ \
4.4. Nuclear decays and reactions via weak interaction
 
\noindent\ \ \ \indent
4.4.1. Various $\beta$-decay modes in astrophysics
 
\noindent\ \ \ \indent
4.4.2. Neutrino reactions
 
\noindent\ \ \ \indent
4.4.3. $\beta$-decays from nuclear excited states
 
\noindent\ \ \ \indent
4.4.4. $\beta$-decays at high ionization
 
\noindent\ \ \
4.5. Nuclear decays and reactions via electromagnetic interaction
 
\noindent\ \ \
4.6. Nuclear decays via strong interaction
 
\noindent
5.\ Thermonuclear reactions in non-explosive events
 
\noindent\ \ \
5.1. Energy production in the Sun, and the solar neutrino problem
 
\noindent\ \ \
5.2. Non-explosive stellar evolution and concomitant nucleosynthesis
 
\noindent\ \ \ \indent
5.2.1. Hydrogen burning
 
\noindent\ \ \ \indent
5.2.2. Helium burning and the s-process
 
\noindent\ \ \ \indent
5.2.3. Carbon burning
 
\noindent\ \ \ \indent
5.2.4. Neon, oxygen, and silicon burning
 
\noindent
6.\ Thermonuclear reactions in explosive events
 
\noindent\ \ \
6.1. Big Bang nucleosynthesis
 
\noindent\ \ \
6.2. The hot modes of hydrogen burning
 
\noindent\ \ \ \indent
6.2.1. The hot p-p mode
 
\noindent\ \ \ \indent
6.2.2. The hot CNO and NeNa-MgAl chains
 
\noindent\ \ \ \indent
6.2.3. The rp- and $\alpha$p-processes
 
\noindent\ \ \
6.3. The He to Si explosive burnings
 
\noindent\ \ \
6.4. The $\alpha$-process and the r-process
 
\noindent\ \ \
6.5. The p-process
 
\noindent
7.\ Nuclear data acquisition for astrophysics
 
\noindent\ \ \
7.1. Nuclear binding 
 
\noindent\ \ \ \indent
7.1.1. Nuclear mass models
 
\noindent\ \ \ \indent
7.1.2. Nuclear equation of state at high temperatures/densities
 
\noindent\ \ \
7.2. Nuclear decay properties 
 
\noindent\ \ \ \indent
7.2.1. Beta-decay half-lives and strength functions
 
\noindent\ \ \ \indent
7.2.2. Bound-state $\beta^-$ decays
 
\noindent\ \ \
7.3. Charged-particle induced reactions: Experiments
 
\noindent\ \ \ \indent
7.3.1. Direct cross-section measurements
 
\noindent\ \ \ \indent
7.3.2. Indirect cross-section measurements
 
\noindent\ \ \
7.4. Neutron capture reactions: Experiments
 
\noindent\ \ \
7.5. Thermonuclear reaction rates: Models
 
\noindent\ \ \ \indent
7.5.1. Microscopic models
 
\noindent\ \ \ \indent
7.5.2. The potential and DWBA models
 
\noindent\ \ \ \indent
7.5.3. Parameter fits
 
\noindent\ \ \ \indent
7.5.4. The statistical models
 
\noindent
8.\ Selected topics
 
\noindent\ \ \
8.1. Heavy-element nucleosynthesis by the s- and r-processes 
of neutron captures
 
\noindent\ \ \ \indent
8.1.1. Defining the s-process
 
\noindent\ \ \ \indent
8.1.2. Defining the r-process
 
\noindent\ \ \ \indent
8.1.3. The s- and r-process contributions to the
solar-system composition
 
\noindent\ \ \ \indent
8.1.4. Astrophysical sites for the s- and r-processes
 
\noindent\ \ \ \indent
8.1.5. Heavy elements in low-metallicity stars
 
\noindent\ \ \
8.2. Cosmochronometry
 
\noindent\ \ \ \indent
8.2.1. Nucleo-cosmochronology: generalities
 
\noindent\ \ \ \indent
8.2.2. The trans-actinide clocks
 
\noindent\ \ \ \indent
8.2.3. The \chem{187}{Re} - \chem{187}{Os} chronometry
 
\noindent\ \ \
8.3. Type-II supernovae
 
\noindent\ \ \ \indent
8.3.1. Evolution of massive stars leading to neutrino-driven supernovae
 
\noindent\ \ \ \indent
8.3.2. Nucleosynthesis in the hot bubble: Can the r-process occur ?
 
\noindent\ \ \ \indent
8.3.3. Signatures of a large-scale mixing of nucleosynthesis products
 
\noindent
9.\ Summary

\noindent 
References
 
\vfill\eject
\section{Introduction}
%
Uniting astronomy and physics, astrophysics aims at deciphering the 
macro-structure of the Universe and of its various constituents. 
To reach this goal the physical laws investigated on earth are 
systematically applied to the vast and diverse laboratory of space, and 
a key concept pervading the whole field is its being 
interdisciplinary.
In particular, cosmology, every branch of astronomy, astronautics, 
elementary-particle-, nuclear-, atomic- and-molecular physics, and 
geo- and cosmo-chemistry have to take their worthy share to the common 
adventure. 

More often than not the cosmic objects exhibit macroscopic properties 
that bear clear fingerprints of the micro-physics of the elementary 
particles or nuclei making up the matter. 
This review deals with the very special interplay between nuclear 
physics and astrophysics, which is embodied into a field commonly 
referred to as ``nuclear astrophysics.'' 
Its main goal is to explain the huge energy output from some cosmic 
objects, and especially from stars, as well as to provide a coherent 
picture for the spatial and temporal variations of the abundances of 
the nuclides in the Universe and its various constituent objects. 
Its adjacent and derivative sub-field of astrophysics is ``particle 
astrophysics.'' 
It is not that nuclear astrophysics does not need to be well aware of 
the basic properties of the elementary particles in addition to those of 
the nuclei. 
But the very origin of these particles is outside of its realm.  
When it searches for the origin of the nuclides, for instance, nuclear 
astrophysics is quite content  with the existence of the proton, and 
leaves the responsibility of unravelling its very origin to 
``baryosynthesis'' models (\cite{Barrow92}).\footnote{Though
bulky it may be, the reference list is
by no means meant to be complete. 
Throughout the text, therefore, the readers are kindly requested to
interpret a doubly bracketed citation ([\#]...) as (see e.g.~[\#] and
the references therein...)}

The hypothesis that the energy production in the Sun and other stars 
results from the energy output of nuclear reactions was formulated, 
apparently for the first time and soon after the first measurements of 
atomic masses, by Russell \cite{Russell19}, followed shortly by Perrin
\cite{Perrin20} (see \cite{Schatzmann93} for historical 
developments). 
Following the clarification of the composition of the atomic nuclei in 
1932 with the discovery of the neutron, Gamow, von Weizs\"acker, Bethe 
and others put that idea on a quantitative basis. 
In particular, the energy source of the Sun was ascribed to the 
so-called ``p-p chain'' of reactions, the net effect of which being the 
``burning'' of four protons into a \chem{4}{He} nucleus. 
The energy release is about 6 MeV per proton (meaning about $2 \times 
10^{19}$ kg or $10^{-11}$ M$_{\odot}$ of protons burning per year 
presently in the Sun, where the mass of the Sun M$_\odot \approx 2 \times
10^{30}$ kg). 
A myriad of further works have substantiated these early ideas beyond 
doubt.

The focal role played by the nuclear reactions in the ``alchemy'' of
the Universe also started to be recognized, leading to the development 
of a major chapter of nuclear astrophysics referred to as the ``theory 
of nucleosynthesis.''
Some nucleosynthesis models developed in the late 1940s 
assumed that the nuclides were built in a primordial ``fireball'' at the 
beginning of the Universe (\cite{Alpher53}). 
In spite of some attractive features those models failed to explain 
the mounting evidence that all stars do not exhibit the same surface 
composition. 
They were also unable to explain the presence of the unstable element 
technetium (Tc) discovered by Merrill \cite{Merrill52} at the surface of 
certain giant (``S-type'') stars.
(No technetium isotope lives more than a few 
million years.)

The problems encountered by those models of primordial nucleosynthesis 
put to the forefront the idea previously expressed by Hoyle 
\cite{Hoyle46} that stars are likely to be major nucleosynthesis agents. 
By the late 1950s the stellar nucleosynthesis model, substantiated by 
some seminal works including the famed B$^2$FH \cite{BBFH57}, was 
recognized as being able to explain the origin of the vast majority of 
the naturally-occurring nuclides with mass numbers $A \geq 12$. 
One key theoretical step in the development of these ideas was the 
identification \cite{Opik51,Salpeter52} of the so-called ``$3{\alpha}$''  
nuclear transformation enabling to bridge in stars the gap of 
stable nuclides
at mass number $A = 8$. In this process the unbound $^8$Be nucleus 
created in
equilibrium with the $\alpha + \alpha$ system can capture an 
$\alpha$-particle
to produce \chem{12}{C}, which is the starting point for the synthesis of 
heavier species.  
In considering the relative abundances of $^4$He, $^{12}$C
and $^{16}$O,   Hoyle went so far as to predict the existence of a
 7.7 MeV
0$^+$   excited state of $^{12}$C as a resonant state in the $^8$Be$ + 
\alpha$ reaction that was soon discovered  experimentally 
\cite{Hoyle53} - \cite{Hoyle54}. 
Despite these early successes, the natural abundances of the light 
nuclides (D, \chem{3}{He}, \chem{4}{He}, \chem{6}{Li}, \chem{7}{Li}, 
\chem{9}{Be}, \chem{10}{B}, \chem{11}{B}) were difficult to explain 
in terms
of stellar thermonuclear processes, and their very  origin has
remained puzzling for some time.

Since those pioneering works, nuclear astrophysics has advanced at a 
remarkable pace and has achieved an impressive record of success. 
Factors having contributed to the rapid developments include the 
progress in  experimental and theoretical nuclear physics, ground-based 
or space astronomical observations, and in astrophysical modelling. 
In fact, nuclear astrophysics has constantly been challenged, and at the 
same time inspired, by new  discoveries, many of them marking epochs in 
the history of science and leading to the birth of a new sub-field of 
research. 
We name here a few of such events in the observational front that have 
had a great impact on nuclear 
astrophysics:\footnote{Omitted here, the references will be given  
later along with further discussions of individual items} 
 
\noindent (1) the discovery in 1965 of the 3K microwave background, which
provided a major support of the Big Bang model and opened a new era of
cosmology.  
In the realm of nuclear astrophysics it resurrected the ``primordial 
nucleosynthesis'' idea. 
However, in contradistinction to the earlier ideas referred to above, 
it appeared soon to be efficient for some light nuclides only; 
 
\noindent (2) the detection in the late 1960s of the neutrinos from the 
Sun, providing the first ``vision'' of the very interior of a star. 
With time it appeared more and more convincing that the observed flux 
was lower than predicted by the solar models. 
The infamous ``solar neutrino problem'' was born, implying our incomplete 
understanding of the structure and evolution of the Sun (and likely of 
stars in general), of nuclear reactions in the solar plasma, of the
neutrino properties, or possibly any combination thereof; 
 
\noindent (3) the discovery in minute fractions of the meteoritic 
material of a suite of chemical elements exhibiting isotopic compositions 
that  differ from those characterizing the bulk solar-system material. 
Quite remarkably, it was soon realized that some of these ``isotopic 
anomalies'' are due to the {\it in situ} decay of by now-extinct 
short-lived radionuclides with half-lives ($t_{1/2}$) ranging from 
about $10^5$ to $10^8$ years. 
A notable example of this type concerns \chem{26}{Al}
($t_{1/2}=7.4\times 10^5$ y).
Even the record of the {\it in situ} decay of the ``ultra-short-lived''
\chem{22}{Na} ($t_{1/2}=2.6$ y) and \chem{44}{Ti} ($t_{1/2}=60$ y)
seems to have been kept in some meteoritic material; 
 
\noindent (4) the discovery in the interstellar medium of a $\gamma$-ray
line from the de-excitation of \chem{26}{Mg} produced by the 
$\beta$-decay of \chem{26}{Al}. 
This detection has led to the vigorous development of a new type of
astronomy, i.e. the $\gamma$-ray line astronomy; 

\noindent (5) the  supernova SN1987A in the Large Magellanic Cloud has 
been a milestone for many fields of astrophysics. 
The detection of some neutrinos from  this explosion has opened the 
new chapter of the astrophysics of non-solar neutrinos. 
More has been added to the growing $\gamma$-ray line astrophysics, as  
well as to nuclear astrophysics and particularly
 to the theory of nucleosynthesis in 
explosive environments.
 
In order to take up the continuous challenge from new observational facts,
nuclear astrophysics concepts and models have to be put on a firmer and 
firmer footing. 
To achieve this goal a deeper and more precise understanding of the  
many nuclear physics processes operating in the astrophysical 
environments 
is crucial, along with improved astrophysical modelling.
Naturally, the acquisition of new nuclear physics data is indispensable 
in the process. 
This quest is, however, easier said than done, given the fact that it is  
generally very difficult, if not impossible, to simulate in the laboratory
the behaviour of a nucleus under relevant astrophysical conditions, or 
even to produce nuclei that might be involved in astrophysical processes. 
Consequently, the development of novel experimental techniques is not 
sufficient, and has to be complemented by the progress of the theoretical 
modelling of the nucleus. 
Both experimental and theoretical approaches face great difficulties of 
their own.  
It may be worth noting here that, although initially motivated by  
astrophysics, some experimental and theoretical nuclear physics efforts  
have provided on many occasions unexpected intellectual rewards in 
nuclear physics itself.

The items of relevance to nuclear astrophysics listed above by no  
means exhaust the questions this field of research has to tackle. 
They are in reality so varied that it appears impossible to review them 
all here in any decent way (this diversity is illustrated by the many
contributions to recent nuclear astrophysics conferences; 
see e.g.~\cite{NICII}  - \cite{Ringberg98}).
We will consequently limit ourselves to a
presentation of a selected set  of questions involving what we consider to
represent the earnest efforts  made in recent years in order to 
decipher the
nuclear fingerprints in the Universe with the help of the improved
 knowledge
of nuclear data of  astrophysical interest. In many instances,
some astrophysics basics and motivations will also be   provided in
 order to
help the non-expert readers, who may very profitably  refer to the
 textbooks
\cite{Clayton68} - \cite{Pagel97} for many more  details.  

Section~2 reinforces with generalities the above discussion on 
observational facts having important bearings upon nuclear astrophysics.
Section~3 presents an overview of the identified main contributing agents 
to nuclei in the Cosmos. 
In Sects.~4 to 7 we address the following broad questions:
(1) what sorts of nuclear data are necessary for which astrophysical
scenarios, (2) what are the changes in nuclear properties inferred when
going from the laboratory to astrophysical sites, and (3) how can one 
acquire such data by laboratory experiments, or supplementarily by
theoretical models ?
For the sake of clarity we attempt to answer those questions successively,
although considerable overlaps are inevitable.  
 
For that, we first examine in Sect.~4 
the static and decay properties
of nuclei, and furthermore make the reader aware of the fact that
astrophysics forces us to tackle some of those very basic nuclear
physics questions in quite an unconventional way.
With the basic nuclear properties provided, the most crucial knowledge 
in quest concerns the rates of various nuclear reactions in energy
domains of astrophysical interest, which is a focus of continuing,
dedicated laboratory studies.
 
The nuclear reactions to be dealt with are of ``thermonuclear'' or
``spallative'' nature.
The latter ones act in low-temperature, low-density media through the 
interaction of non-thermally accelerated particles with the interstellar
medium (``ISM'' in the following),
or with the material (gas or grains) at stellar surfaces and in 
circumstellar shells (Sect.~3.2).
The former ones have developed at the cosmological level (Big Bang),
and continue to take place in stars.
They are charged-particle induced reactions, neutron-capture reactions, 
and photodisintegrations.
We do not deal in this review with the nuclear physics aspects of 
spallation reactions, and concentrate instead on thermonuclear reactions.
Sections~5 and 6 are devoted to some general considerations about such
transmutations, which are divided into two categories:
one concerned 
with non-explosive  events, and the other with explosive phenomena. 
This is done in view of the   different experimental and theoretical 
problems raised by the determination of the rates of the relevant 
reactions, as well as of the different and complementary  roles they 
play in astrophysics.
 
In Sect.~7 we summarize
 the experimental and theoretical techniques for the  
acquisition of the relevant nuclear data. 
Section~8 presents a few topics that illustrate quite vividly the  
beauties, as well as the complexities, of nuclear astrophysics and of 
its relation to various sub-fields of astrophysics.
A brief summary and outlook is given in Sect.~9.   
%

\section{Observational foundation}
%
The observational foundation of nuclear astrophysics, and more 
specifically of the theory of nucleosynthesis, rests largely upon the 
determination of elemental and isotopic abundances in the broadest 
possible variety 
of cosmic objects, as well as upon the study of as complete a
set  as possible of observables that help characterizing the objects. 
This knowledge  relies almost entirely on the detailed study of the light 
originating from a large diversity of emitting locations: our Galaxy 
(non-exploding or exploding stars of various types, the ISM), external 
galaxies, and perhaps even the early Universe. 
Recent progress in optical astronomy, paralleled by the advent of a 
variety of ``new'' 
(in particular: infrared, UV, X- and $\gamma$-ray) astronomies,
 has led to the unprecedented vision we now have of the sky 
at all wave-lengths, ranging from radio-frequencies to $\gamma$-ray 
energies (up to the TeV and PeV domain).  
Very often these dramatic advances are directly related to those of the 
space technologies (\cite{Lena88}). 
The characteristics of the emitted light are often associated with the 
temperature in the environment (``heat radiation''), the most exemplary 
situation of this type being the black-body radiation. 
Non-thermal sources have also been identified, which emit generally 
high-energy photons (e.g.~X- and $\gamma$-rays).  
Bremsstrahlung, synchrotron radiation and nuclear radioactivity are good 
examples of this sort.

The studies of the electromagnetic radiation are complemented with the 
careful analysis of the minute amount of matter of the Universe 
accessible to humankind. 
This matter is comprised for its very largest part in various types of 
solar-system solids. 
The rest is in the form of (extra-)galactic cosmic rays. 
The observation of solar and non-solar neutrinos has also been a major 
step for nuclear astrophysics as well as for many other fields of 
astrophysics.

It is impossible to review here the myriad of observational data that are
relevant to nuclear astrophysics. 
In what follows we illustrate by way of selected examples the 
nuclear-astrophysics importance 
of the deciphering of some electromagnetic 
messages from the sky. 

\subsection{The Hertzsprung-Russell diagram (HRD)}
%
The HRD represents a given sample of stars in a (colour,~surface 
brightness) plane, or any equivalent plane, where the colour is replaced 
by a ``colour index,'' ``spectral type'' or ``effective temperature,'' 
and the surface brightness (``luminosity'') by some ``magnitude'' scale 
(see \cite{Boehm89} for the definitions of these various basic 
quantities, and for the relations among them).
The adoption of the magnitude and colour scales leads to a 
``colour-magnitude'' (C-M) representation of the HRD, examples of which 
are provided in Figs.~1 and 2.  

\begin{figure}
\center{\includegraphics[width=1.00\hsize,height=0.5\hsize]{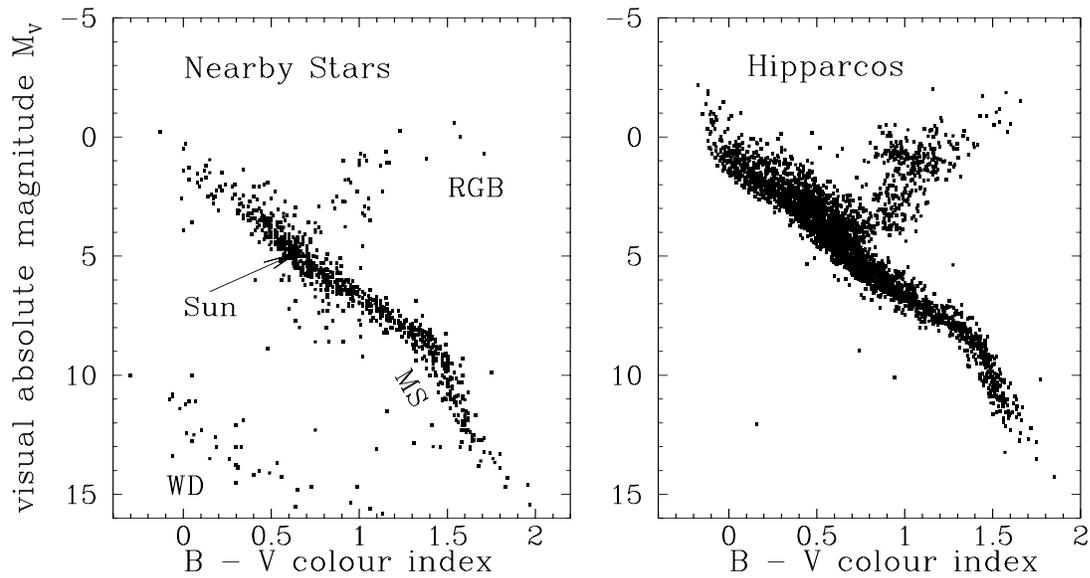}}
\caption{Examples of colour-magnitude diagrams. {\it Left panel:} for a 
sample of {\it ca.} 1000 nearby  stars \cite{Glies69} with 
distances less than about 20 pc [1 pc 
(parsec or parallax second) $= 3.086 
\times 10^{13}$ km].      
{\it Right panel:} for a sample of {\it ca.} 5000 stars (most  
located in the  60 - 70 pc range) as observed by the HIPPARCOS
satellite \cite{Perryman97}. Each point represents a sample
star.  The labels MS, RGB and WD refer to ``Main Sequence,'' ``Red Giant
Branch'' and 
``White Dwarf'' stars. 
The magnitudes displayed in ordinate are defined as decreasing by five 
units for a one hundred-fold increase of the luminosity (emitted power 
through the whole stellar surface), so that the lower the magnitude, 
the brighter the star at the origin.
Thus the intrinsically
brighter (dimmer) stars are located in the upper (lower) part 
of the displayed diagram. 
The observed luminosity (at the top of the Earth atmosphere), or
the ``apparent magnitude,'' decreases quadratically with the distance 
from the star.   
Given this distance, the ``absolute magnitude'' is defined to be what one 
would observe if the star were re-located at 10 pc.
The visual absolute magnitude $M_V$ and its counterpart (visual apparent 
magnitude) $V$ are associated with the stellar brightness in the yellow 
(V) band of the widely used UBV trichromatic system 
(\cite{Boehm89}). 
The abscissa displays the most commonly used colour index. 
It expresses the difference between the apparent magnitudes in the blue 
(B) and visual (V) bands. 
The lower the index, the bluer the star. 
The bluer (redder) stars are thus located more to the left (right) side 
in the diagram. 
For confrontations with model predictions it is operationally convenient 
to convert magnitudes into absolute luminosity
values, and colour indices into 
``effective temperatures,'' $T_{\rm{eff}}$, defined as the temperatures 
of black bodies that would radiate the same flux of energy as the 
considered stars  
(\cite{Boehm89} for details) 
}
\end{figure}
%
\begin{figure}
\center{\includegraphics[width=1.00\hsize,height=0.5\hsize]{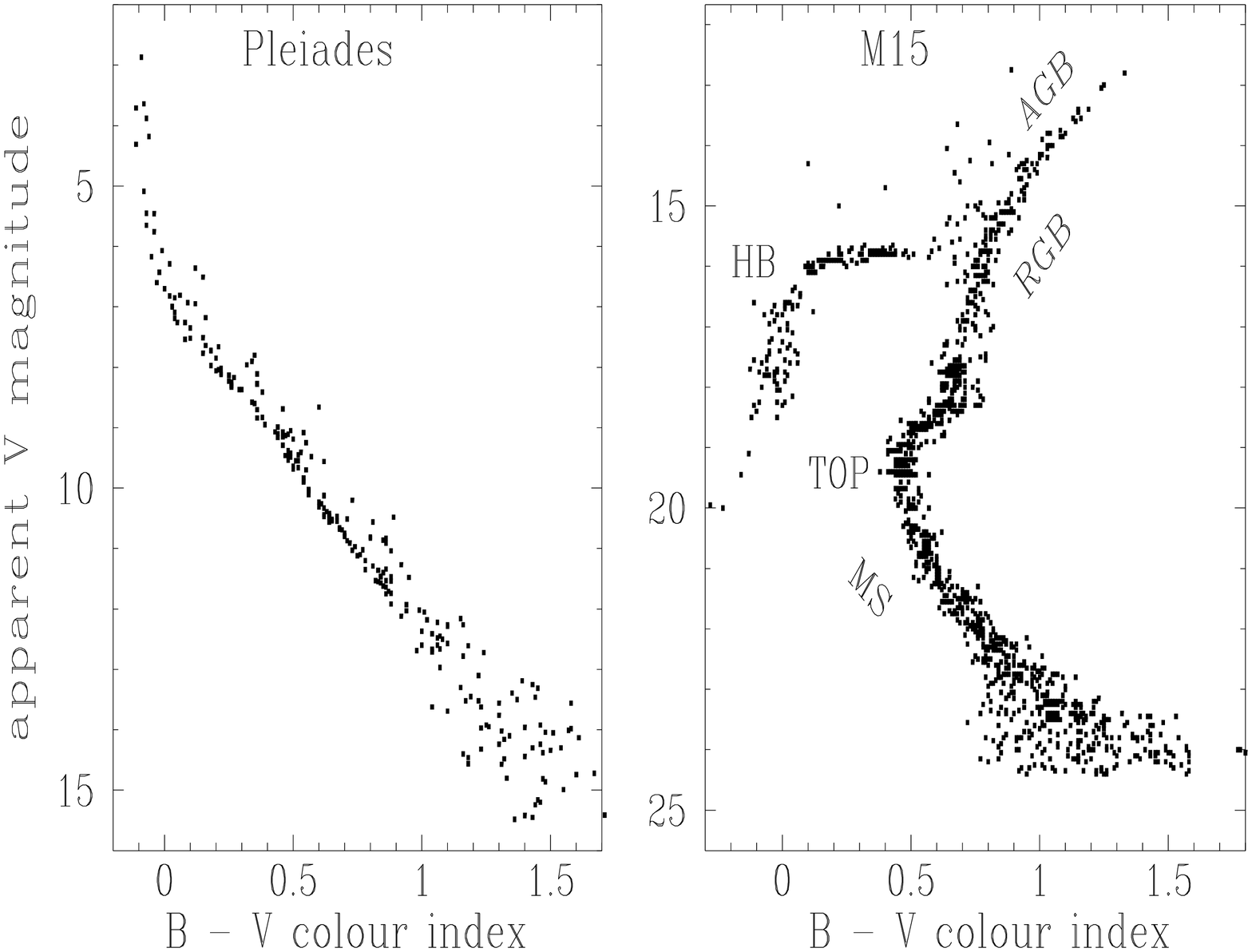}}
\caption{Same as Fig.~1, but for the Pleiades, a young galactic 
(``open'') 
cluster ({\it left panel}), and for the old ``globular cluster'' 
M15 ({\it right panel}). 
The labels ``HB'' and ``AGB'' refer to ``Horizontal Branch'' and 
``Asymptotic Giant Branch'' stars. 
The label ``TOP'' identifies the top of the MS branch. 
Note that all member stars of a globular or open cluster are at
essentially 
equal distance from the observer, so that the transformation from 
apparent to absolute magnitudes in the HRD for a given cluster would 
just correspond to an equal shift of all the points by an amount dictated 
by the (squared) distance of that cluster.
With the use of the observed distances,
the ordinates
are adjusted here
 in such a way that the $M_V$ magnitudes for both clusters
have the same  scale height.
The data points for the Pleiades are from \cite{Johnson58}.
(Omitted are half a dozen 
of the brightest stars, whose specific locations  
in the diagram require special attention because of their rapid 
rotation \cite{Hazlehurst70}.)
The data for M15 are from \cite{Durrell93}
} 
\end{figure}
%
The HRD is generally considered as the ``Rosetta stone'' of stellar 
evolution for the key role it has played in the development of the theory 
of stellar structure and evolution. 
The most remarkable feature of the HRD is undoubtedly the existence of 
concentrations of stars from a given sample along correlation 
{\it lines}. 
Various such lines can be identified. 
The most spectacular one is certainly the ``Main Sequence (MS)'' running 
diagonally in the HRD from its red low-luminosity sector to its blue 
high-luminosity one. 
Other concentrations are also visible, like the ``Horizontal Branch 
(HB),'' the ``Red Giant Branch (RGB)'' or ``Asymptotic Giant Branch 
(AGB).'' 
It is also very important to recognize that the relative populations of 
different correlation lines, or even of different portions of a given 
line, depend upon the selected star sample. 
For example, the RGB and AGB branches are almost totally vanishing, 
while the blue portion of the MS is well developed, in the HRD for 
samples of rather young stars located in the galactic disc (like the
nearby stars shown in Fig.~1), or stars belonging to so-called open 
clusters (left panel of Fig.~2). 
In contrast, the blue end of the MS almost vanishes, while the RGB or 
AGB branches are strikingly apparent, in the HRD for globular clusters
populating the galactic halo and made of relatively old stars (right 
panel of Fig.~2). 

The theory of stellar structure and evolution is now able to explain 
these specific characteristics of the topology of the HRD, often in a
quite quantitative way (\cite{Chiosi92}). 
It is well beyond the scope of this review to discuss these questions at 
length. 
Let us just stress two basic findings: (1) {\it formally} it can be 
demonstrated that the very existence of correlation {\it lines} (as 
opposed to broad correlation {\it surfaces}) is the direct signature of 
stellar structures in which the energy output can be counter-balanced 
by a 
nuclear energy production, this situation being generally referred to in 
astrophysics as a state of ``thermal equilibrium'' (\cite{Cox68} 
for a discussion of this 
statement).\footnote{The situation concerning the white dwarf (WD) 
clustering involves a different physics which does 
not call for a nuclear energy source (\cite{Kawaler97})
}
%
So, the observed HRD correlations bring nuclear physics and astrophysics 
closely together; and (2) {\it numerically} it appears that the MS, RGB, 
HB and AGB concentrations correspond in fact to stages of central H 
burning, shell H burning, central He burning, and double shell H-He 
burning. 
For a given star, the H-burning phase is by far the longest. 
On the other hand, calculations show that the duration of each burning 
stage decreases with increasing stellar mass. 
In between these phases the stars are out of thermal equilibrium and 
are predicted to evolve quickly, and more so with increasing stellar
 mass.  
These dissimilarities in evolutionary time-scales between thermal 
equilibrium and non-equilibrium and among stars of different masses 
account very well for the differences in the populations of the diverse 
correlation lines of the HRD for a given sample of stars, as well as for
the changes in the population of a given correlation line between HRDs 
for samples of, for instance, old and young stars.
(\cite{Schwarzschild58} for details.)  In particular, the 
location of the
top of the MS of globular clusters  (Fig.~2, right panel) is used to
 evaluate
the age of these clusters  (\cite{VandenBerg96,Chaboyer98}; 
Sect.~8.2)

\subsection{Electromagnetic spectra and abundance determinations}
%
Very roughly speaking, the light from the sky appears to demonstrate some
uniformity of composition of the objects in the Universe, which is most 
strikingly exemplified by the fact that H and He are by far more abundant 
than the heavier species in the whole observable Universe. 
However, they also point to a great diversity of elemental and/or 
isotopic 
abundances that superimposes on that uniformity at all scales ranging 
from stars to galaxies and galaxy clusters.
They imply diverse classes of objects, as well as a diversity of the 
objects belonging to a given class. 
It is impossible here to do justice to the richness of the information
gained by now in this field, and we just limit ourselves to some guiding 
considerations.

\subsubsection{The spatio-temporal evolution of the composition of our 
Galaxy}\ \ \
%
This evolution can be traced by deriving the surface abundances for a 
suitably selected and large ensemble of stars. 
These stars are chosen on grounds of two requirements: (1) they have to 
span  substantial ranges of ages and locations in our Galaxy. 
So, old halo or disc stars, old stars in globular clusters, as well 
as young disc stars have been under active scrutiny; and (2) their 
surfaces are likely not contaminated with nuclear-processed matter from 
their interiors, and have well preserved the composition of the Galaxy 
at the place and time of their birth. 
This is clearly an essential requirement in order for these stars to 
witness the large-scale time and space variations of the nuclear content 
of the Galaxy. 
Such information from stars is very usefully complemented with the
analysis of the composition of the present-day ISM at various galactic 
locations.
\begin{figure}
\center{\includegraphics[width=0.80\hsize,height=0.4\hsize]{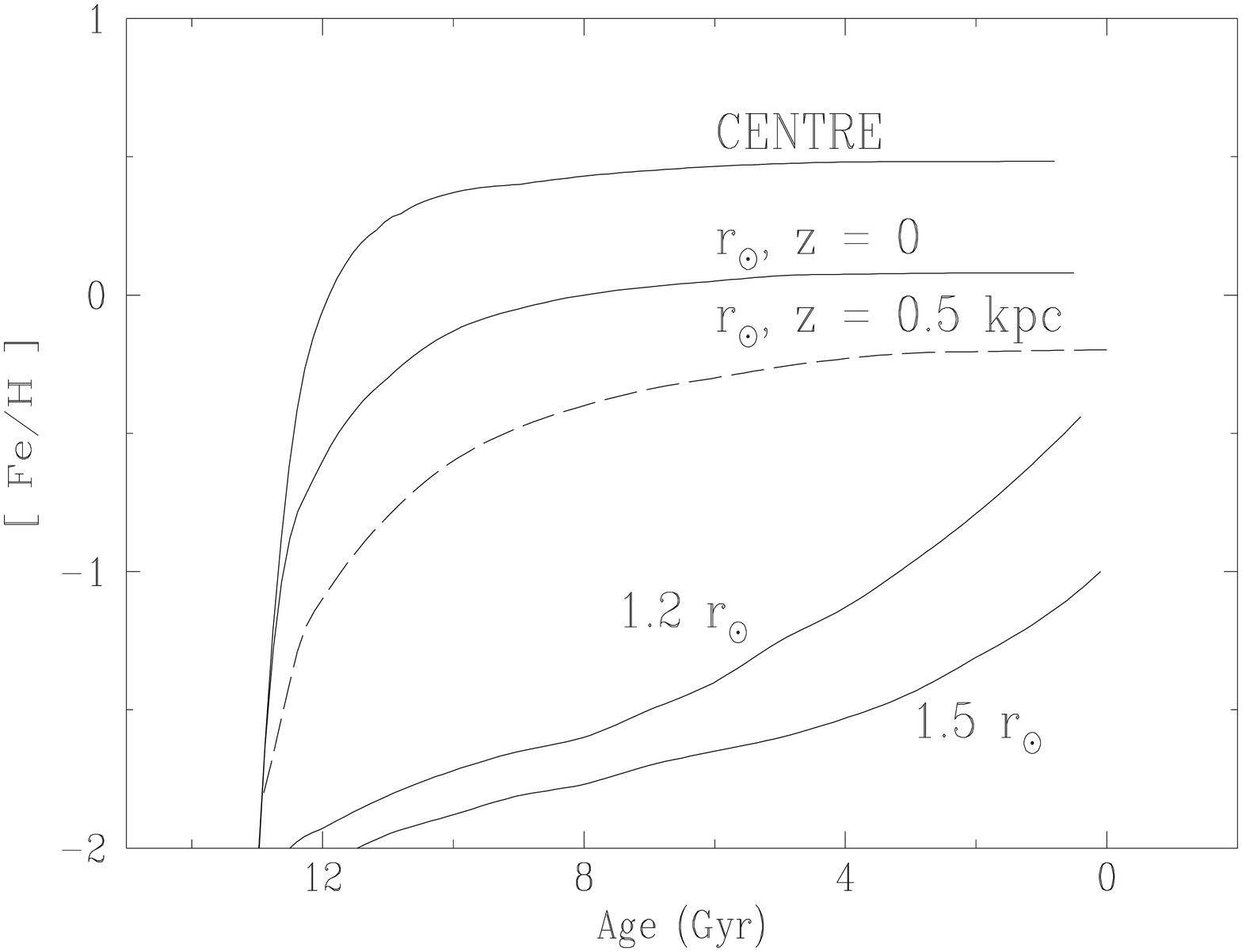}}
\caption{A ``schematic suggestion'' \cite{Sandage88} of the 
age-metallicity relationships for five locations in the Galaxy: the 
galactic centre (CENTRE); at the distance of the Sun ($r_\odot$)
measured from the galactic centre in midplane ($z = 0$) of the galactic 
disc; at $r_\odot$, but at a distance $z = 500$ pc from the galactic 
midplane; and 1.2 and 1.5~$r_\odot$ at $z = 0$. 
The metallicity in ordinate is expressed in the usual stellar 
spectroscopic notation [Fe/H] $\equiv$ log$_{10}$(Fe/H) - 
log$_{10}$(Fe/H)$_{\odot}$, where the chemical symbols represent 
abundances by number, and the subscript $\odot$ refers to the Sun, with 
log$_{10}$(Fe/H)$_{\odot} \approx -4.5$ \cite{Anders89}
}
\end{figure}

>From the large body of abundance observations, some very general trends
emerge, which can just be sketched here: 

\noindent (1) At a global scale the  
``metallicity''\footnote{The metallicity is generally measured by the 
sum of the abundances of all nuclides with mass numbers $A \geq 12$. 
An oft-used metallicity indicator is Fe. 
For certain classes of stars, and in particular old stars, this choice 
is probably not the most suitable one, and O may replace Fe as
 the reference 
element
}
%
 increases with time at a pace that itself varies with time and place in 
the Galaxy, as sketched in Fig.~3. 
This leads to metallicity differences among the various galactic 
subsystems: halo, (thick and thin) disc, bulge, centre (for a discussion 
of the structure of our Galaxy, see Chap.~2 of \cite{Gilmore89}). 
Also metallicity gradients can exist in these subsystems, along
with  possible more local variations; 

\noindent (2) Trends
as well as scatters with respect to the  metallicity index 
are identified in
the abundances of a large variety of elements.
These data concern elements ranging from Li, Be and B of cosmological 
importance (\cite{Reeves94}) to the rare-earth  elements 
(\cite{McWilliam97}).
Figure~4 illustrates the metallicity dependence of the so-called
``$\alpha$-elements,'' 
each of which has the most abundant isotope (in the 
bulk solar-system at least; see Sect.~2.3) with a mass number equal to a 
multiple of four ($A = 4n \leq 56$). 
These variations again demonstrate that different elements or groups of 
elements accumulate at different rates at different locations of the 
Galaxy. 
This is certainly the signature of different nucleosynthesis processes 
acting with unequal spatial and temporal efficiencies.

\begin{figure}
\center{\includegraphics[width=1.00\hsize,height=0.3\hsize]{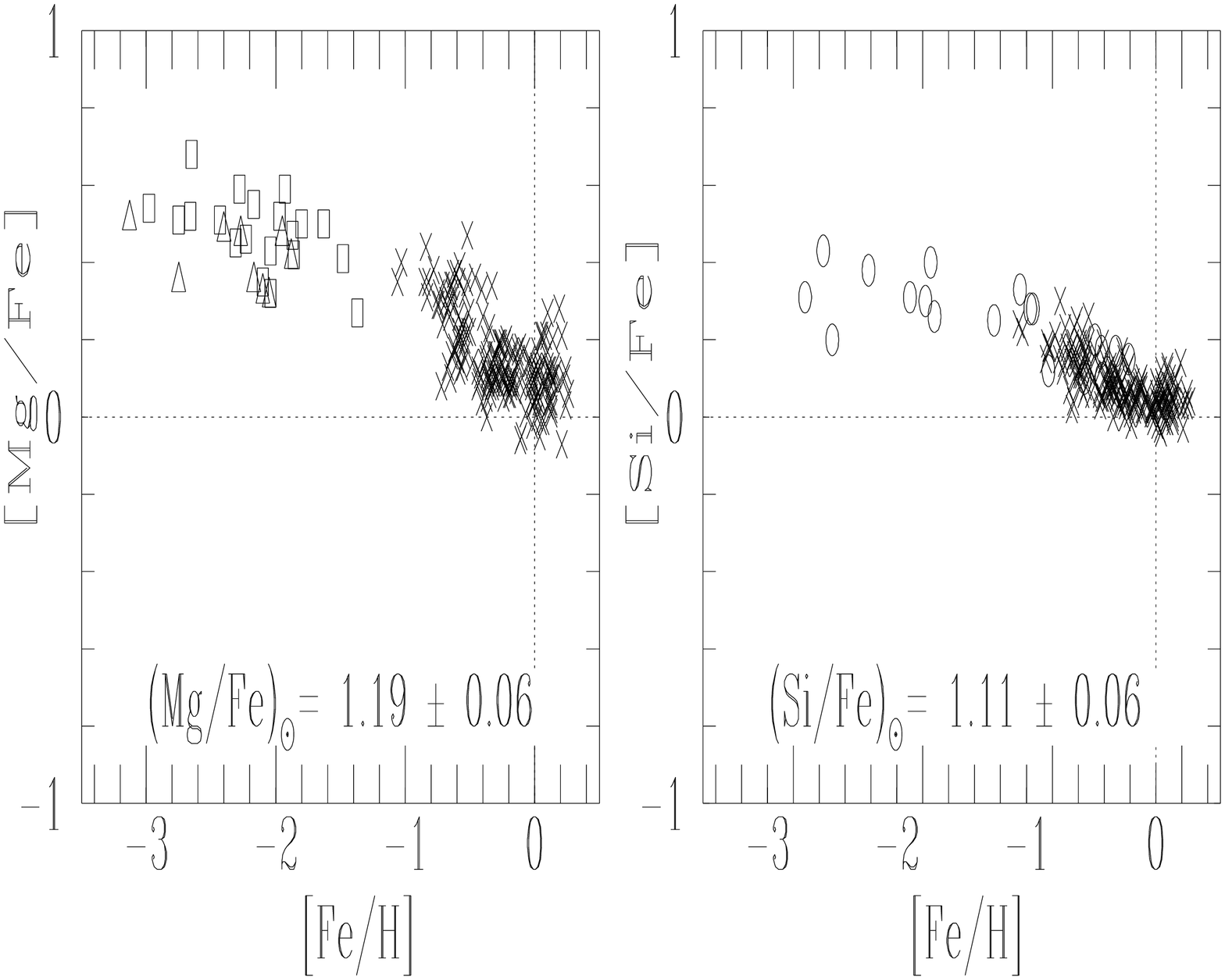}}
\vskip10pt
\center{\includegraphics[width=1.00\hsize,height=0.3\hsize]{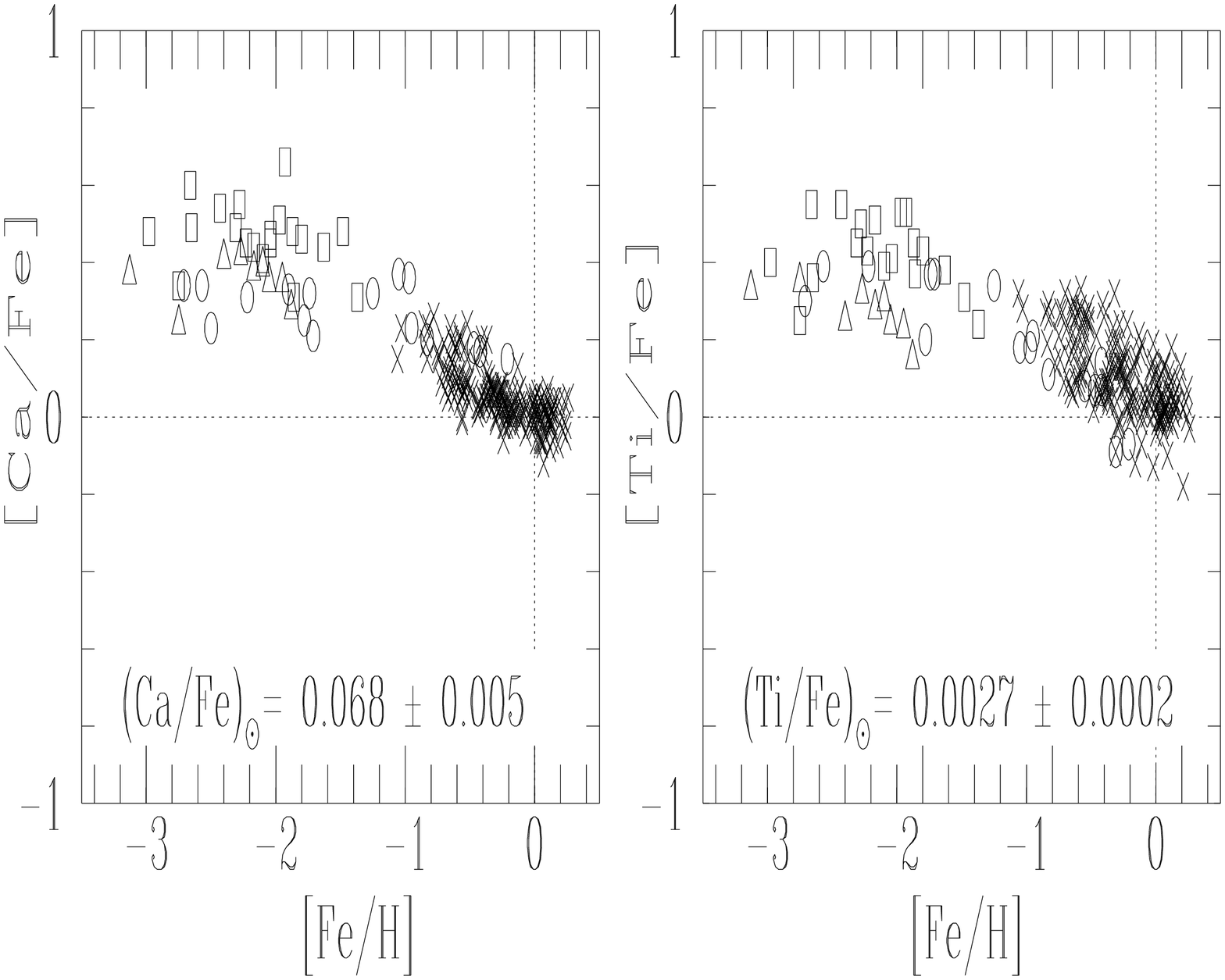}}
\caption{Metallicity dependence of various ``$\alpha$-elements,'' showing 
that the Galaxy has been enriched in the displayed elements more 
rapidly than in Fe.
Roughly speaking, stars with [Fe/H] \lsimeq - 1 are in the halo, and 
more metal-rich stars are in the disc.
The data are taken from: \cite{Magain89} ({\it squares}); 
\cite{Gratton91} ({\it circles}); 
\cite{Edvardsson93} ({\it crosses}); and \cite{Nissen94} ({\it triangles})
}
\end{figure}
%
A clear indication also emerges that mixing processes among the various 
galactic subsystems, or even at smaller scales within a given subsystem, 
may have had a limited efficiency. 
Among the demonstrated composition inhomogeneities within a given 
subsystem, let us note a very interesting difference in abundance 
patterns in certain elements between two of the oldest star populations 
in the halo, i.e., the field halo stars and those which are members of 
globular clusters (\cite{Smith97}); 
 
\noindent (3) The abundance determinations obtained from the analysis of
photospheric spectra are very usefully complemented with data derived 
from the study of the gas and grain components of galactic interstellar 
clouds and circumstellar envelopes. 
These observations are done at radio-wavelengths (\cite{Kahane95}) 
as well as in the UV domain (\cite{Savage96}). 
They provide in particular a very interesting piece of information on the 
existence or absence in the galactic disc of radial gradients of isotopic 
ratios for some major elements, including C, N and O, and thus essential 
constraints to modern models for the evolution of the chemical content of 
the Galaxy. 

\subsubsection{The composition of other galaxies}\ \ \  
%
In addition to the very many abundance data concerning the various 
constituents of our Galaxy, information on other galaxies is now 
accumulating very rapidly.
The analysis of such data forces the conclusion that abundances and their
spatial trends may vary quite significantly from galaxy to galaxy, 
even within
a given galaxy class. Substantial local variations are  
also observed in many instances.  One of the most significant recent 
advances
in the study of galactic  abundances concerns  high-redshift galaxies 
[referred
to as ``Damped  Lyman
$\alpha$ systems (DLAs)''], which provide information on their 
 composition at
an early phase in their evolution: the highest the  redshift, the shortest
the evolution history of their constituent elements 
from their  pristine stage. 
An ``age-metallicity relation'' constructed from a sample of DLAs 
spanning a range of redshifts indicates that a DLA of a given age has 
a lower metallicity than do the disc stars of our Galaxy at the same age
(Fig.~5). 
Data start accumulating on the DLA abundances for a variety of elements 
ranging up to Ni \cite{Lu96}.
These observations will certainly provide a wealth of information,
which the 
theory of nucleosynthesis and of the chemical evolution of galaxies will 
have to cope with. 
%
\begin{figure}
\centerline{\includegraphics[width=0.95\hsize,height=0.6\hsize]{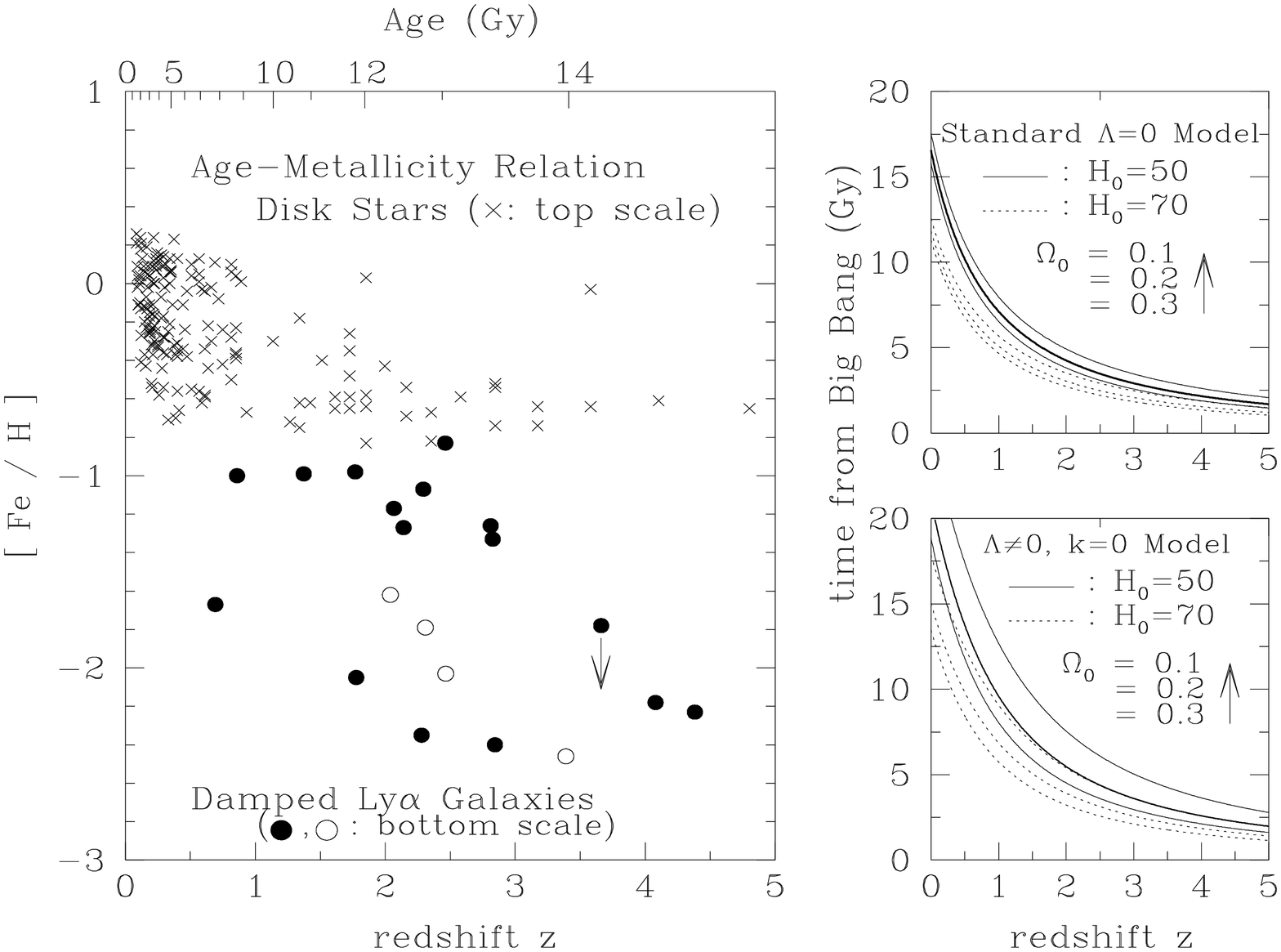}}
\vskip-0.7truecm
\caption{{\it Left panel}: Age-metallicity relations for disc stars of 
our Galaxy \cite{Edvardsson93} ({\it crosses} referring to the top axis
in Gy $\equiv 10^9$ y), 
and for DLA galaxies at various redshifts \cite{Lu96} ({\it circles} 
referring to the bottom axis, with  
open ones corresponding to cases where [Cr/H] replaces [Fe/H]). 
{\it Upper right panel}: age -- redshift relations in the standard 
cosmological model, which assumes a Universe with zero cosmological
parameter $\Lambda$. {\it Lower right panel}: the same but 
in a flat ($k = 0$) Universe with non-zero $\Lambda$. 
The parameters are the present-day Hubble constant $H_0$ (expressed in 
km/s/Mpc), and density $\Omega_0$ relative to the ``critical density'' 
necessary for ``closing'' the Universe. 
The oft-used ``deceleration parameter'' relates to $\Omega_0$ through 
$q_0 = \Omega_0/2$ in the standard model, and through 
$q_0 = (3/2)\Omega_0 
- 1$ in the other model considered here. 
As in \cite{Lu96}, the left panel is constructed with the choice of 
$H_0 = 50$ and $\Omega_0 = 0.2$ (the thick line in the upper right 
panel). See  \cite{Tayler86} - \cite{Arnould90} for introductory 
reviews of the cosmological models; also see Sect.~3.3
}
\end{figure}
 
\subsubsection{Abundances in evolved or exploding stars}\ \ \  
%
On top of the spatio-temporal abundance variations that exist at all 
scales in galaxies or galaxy clusters, significant abundance differences 
are also well documented at the stellar scale, where individual 
{\it evolved} stars exhibit major abundance differences. 
In contrast to the un-evolved stars referred to in the previous 
subsections, the evolved stars are considered to have been able to 
contaminate their surfaces at one point or another during their evolution 
with material processed in their very interiors. 
Various classes of ``chemically peculiar'' evolved stars have been 
identified. 
A non-exhaustive list of such objects includes the Barium stars or the 
various sub-classes of stars belonging to the RGB or AGB branches 
of the HRD (\cite{Smith89}). 
Let us also note that star-to-star variations, as well as
correlations or anti-correlations, in the abundances of C, N, O, Na, 
Mg and Al
are almost ubiquitous within red-giant globular clusters, with a
clear additional diversity from one cluster to the other. 
These observations
may imply the combination of initial abundance differences 
whose precise
origin remains to be identified, and of intrinsic variable surface
contaminations which are not predicted by standard stellar evolutionary
 models
(\cite{Denissenkov98,Kraft98}).
 
In addition, exploding objects, like novae (\cite{Gehrz98}) and 
supernovae (\cite{Filip97}), also exhibit peculiar abundance 
patterns.  
It has to be noticed that a rich variety of chemical peculiarities are  
also observed in stars which would be suspected to belong to the class 
of un-evolved stars in the sense that they are not supposed to be able to 
transport nuclear-processed material to their surfaces. 
That is in particular the case for many classes of stars along the MS in 
the HRD. 
It is by now acknowledged that these composition peculiarities have 
nothing to do with nuclear physics, but are explicable in terms of 
diffusion-type processes operating in the stellar atmospheres 
(\cite{Vauclair82}). 
In this respect, nuclear astrophysics has to define its limits very
carefully, and perhaps with modesty !

\subsubsection{A special electromagnetic message from the sky: 
$\gamma$-ray line astrophysics}\ \ \  
%
At the beginning of the 1980s it was discovered that the electromagnetic 
message from the sky was even richer and more diverse than previously 
thought. 
In fact the ISM was seen to emit a $\gamma$-ray line resulting from the 
de-excitation of the 1.8 MeV level of \chem{26}{Mg} fed by the nuclear
$\beta$-decay of \chem{26}{Al} (\cite{Prantzos96}).
This discovery has been followed by the observation that the famed 
supernova SN1987A was emitting $\gamma$-ray lines originating from the 
\chem{56}{Ni}$\rightarrow$\chem{56}{Co}$\rightarrow$\chem{56}{Fe} decay 
chain and from the \chem{57}{Co}$\rightarrow$\chem{57}{Fe} decay.
The decay of \chem{44}{Ti} in the young Cas(siopeia)-A supernova remnant 
has also been observed (\cite{Diehl98,Diehl98a}).
These observations provide an essential source of information, as well as 
constraints, on the operation of nuclear reactions in 
astrophysical sites. 
In particular the observed emission of $\gamma$-ray lines from supernovae 
was immediately recognized as the clearest demonstration of the 
operation of explosive nucleosynthesis processes (Sect.~6).
 
\subsection{The composition of the solar system}
%
The understanding of the composition of the solar system has always held 
a very special place in nuclear astrophysics. 
This relates directly to the fact that it provides a body of abundance 
data whose quantity, quality and coherence remain unmatched, despite the 
spectacular progress made in astronomical abundance observations. 
This concerns especially isotopic compositions, which are the prime 
fingerprints of astrophysical nuclear processes.

\subsubsection{The bulk solar-system composition}\ \ \  
%
A milestone in the solar-system studies of astrophysical relevance was 
the realization that, in spite of large differences between the 
elemental compositions of constituent members, it was possible to derive 
a meaningful set of abundances likely representative for the composition 
with which the solar system formed some 4.6 Gy ago. 
Such an elemental abundance distribution is displayed in Fig.~6. 
It is largely based on abundance analyses in a special class of rare
meteorites, the CI1 carbonaceous chondrites, which are considered as the 
least-altered samples of primitive solar matter presently available 
(\cite{Anders89,Grevesse93}). 
Solar spectroscopic data, which now come in quite good agreement with the 
CI1 data for a large variety of elements, have to be used for 
the volatile 
elements H, He, C, N, O and Ne, whereas interpolations guided by 
theoretical considerations are still required in some cases (Ar, Kr, 
Xe, Hg).

\begin{figure}
\center{\includegraphics[width=1.00\hsize,height=0.4\hsize]{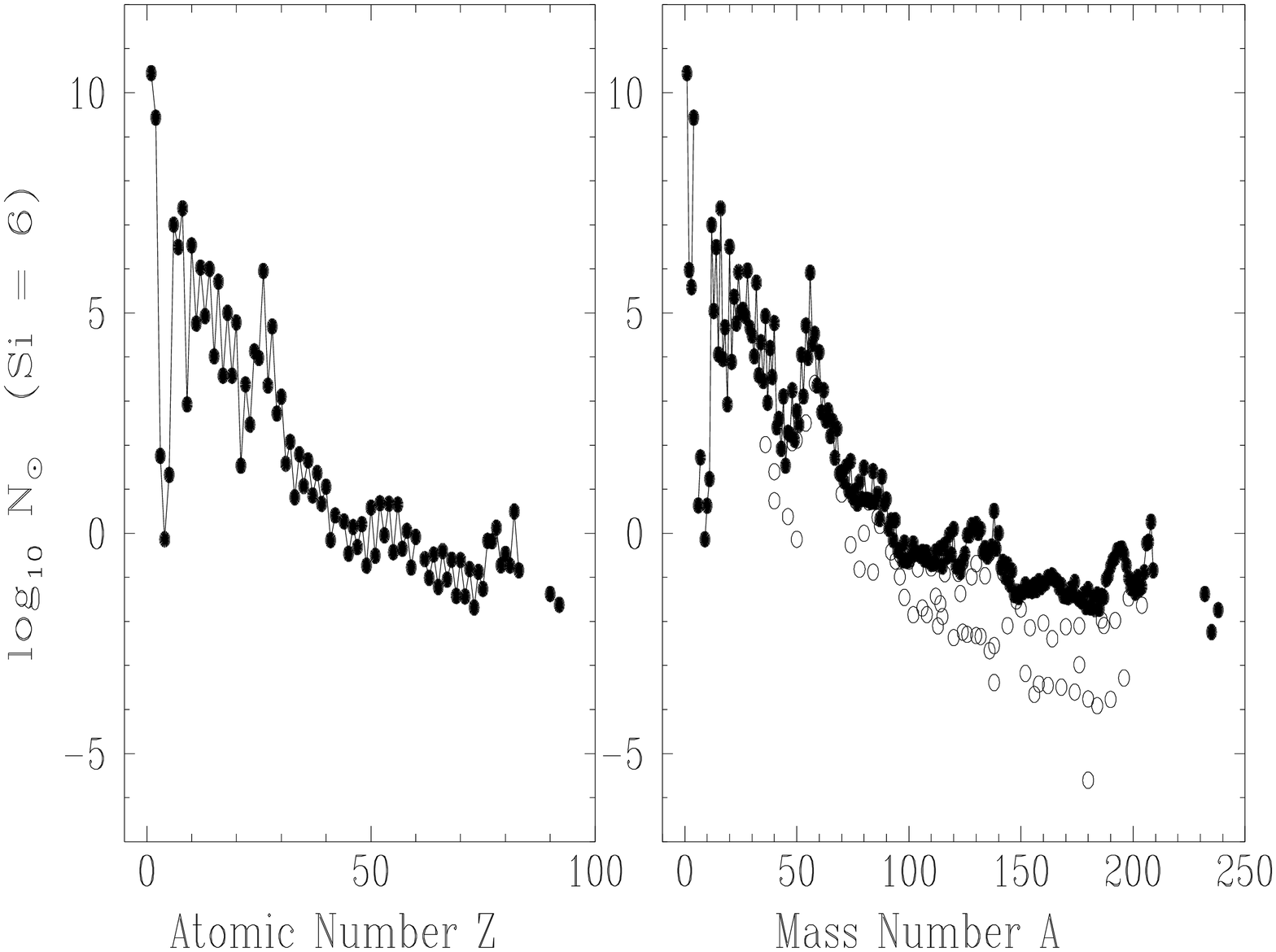}}
\caption{Bulk elemental ({\it left panel}) and nuclear 
({\it right panel}) compositions of the galactic material from which the 
solar system formed about 4.6 Gy ago \cite{Anders89}. 
The abundances are normalized to $10^6$ Si atoms. 
The elemental abundances are largely based on the analysis of meteorites 
of the CI1 carbonaceous chondrite type. 
The nuclear composition is derived from those elemental abundances
with the use of the terrestrial isotopic composition of the elements, 
except for H and the noble gases.
For a given $A$, the most abundant isobar is shown by a {\it dot}, and 
less abundant ones, if any, by {\it open circles}
}
\end{figure}
%
Starting from the composition displayed in Fig.~6, it is possible to 
account for the differences between the {\it elemental} compositions of 
the various solar-system solid constituents in terms of a large variety  
of secondary physico-chemical and geological processes. 
In contrast these secondary processes seem to have played only a minor 
role as far as the isotopic composition is concerned, save some specific 
cases.  
This fact manifests itself through the extremely high homogeneity of the 
bulk isotopic composition of most elements within the solar system. 
For this, terrestrial materials have been classically adopted as the 
primary standard for the isotopic composition characteristic of the 
primitive solar nebula.  
However, the choice of the most representative isotopic composition of 
H and the noble gases raises certain specific problems.  

Even without going into details, some characteristics of the bulk
solar-system composition (Fig.~6) are worth noticing. 
In particular, H and He are by far the most abundant species, while Li, 
Be, and B are extremely under-abundant with respect to the neighbouring 
light nuclides. 
On the other hand, some abundance peaks are superimposed on a curve which 
is decreasing with increasing mass number $A$. 
Apart from the most important one centred around \chem{56}{Fe} 
(``iron peak''), peaks are found at the locations of 
the ``$\alpha$-elements.'' 
In addition a broad peak is observed in the $A \approx 80-90$ region, 
whereas double peaks show up at $A = 130\sim 138$ and $195\sim 208$. 
 
It has been realized very early that these abundance data provide a clear
demonstration that a close correlation exists between solar-system
abundances and nuclear properties. 
As a simple example, a nuclide is observed to be more abundant than the 
neighbouring ones if it is more stable in the sense of nuclear physics. 
An intense nuclear-astrophysics activity has been devoted to unravel the 
details of the ``alchemy'' that has led to that nuclear imprint.
In a word, very light ($A < 12$) nuclides are produced by Big Bang
nucleosynthesis (Sect.~3.3) and/or by ``spallation'' reactions
(Sect.~3.2), whereas the heavier ones result from the nuclear ``cooking''
inside stars. 
As will be detailed in Sect.~3.1, various charged-particle induced 
reactions are responsible for the synthesis of the vast majority of the
nuclides up to the ``Fe peak.'' In contrast, neutron capture chains, 
referred to as the ``s(low)-'' and ``r(apid)-'' processes (see Sect.~8.1 
for their definitions), are called for in order to synthesize the heavier 
species. 
An additional mechanism, referred to as the ``p-process,'' is dominated 
by photodisintegrations of pre-existing nuclides (Sect.~6.5). 
 
\subsubsection{Isotopic anomalies in the solar composition}\ \ \   
%
The solar-system composition has raised further astrophysical interest
and excitement with the discovery that a minute fraction of the 
solar-system material has an isotopic composition which differs from 
that of the bulk. 
Such ``isotopic anomalies'' are observed in quite a large suite of 
elements ranging from C to Nd (including the rare gases), and are now 
known to be carried by high-temperature inclusions of primitive 
meteorites (\cite{Harper93}), as well as by various types of grains 
(diamond, graphite, SiC, corundum, refractory carbides of Ti, Mo, Zr and 
Fe, Si$_3$N$_4$) found in meteorites (\cite{Bernatowicz97} for many 
review papers). 
The inclusions are formed from solar-system material out of equilibrium 
with the rest of the solar nebula. 
The grains are considered to be of circumstellar origin and have 
survived the process of incorporation into the solar system.

These anomalies contradict the canonical model of a homogeneous and 
gaseous protosolar nebula, and provide new clues to many astrophysical 
problems, like the physics and chemistry of interstellar dust grains, the 
formation and growth of grains in the vicinity of objects with active 
nucleosynthesis, the circumstances under which stars (and in particular 
solar-system-type structures) can form, as well as the early history of 
the Sun (in the so-called ``T-Tauri'' phase) and of the solar-system  
solid bodies. 
Last but not least, they raise the question of their nucleosynthesis 
origin and offer the exciting perspective of complementing the 
spectroscopic data for chemically peculiar stars in the confrontation 
between abundance observations and nucleosynthesis models for a very 
limited number of stellar sources, even possibly a single one.
This situation is in marked contrast with the one encountered when trying 
to understand the bulk solar-system composition, which results from the 
mixture of a large variety of nucleosynthesis events, and consequently 
requires the modelling of the chemical evolution of the Galaxy
(Sect.~3.4).

Some further astrophysical excitement has followed the 
realization that some of the anomalies result from the decays within the
solar system itself of radioactive nuclides with half-lives in excess of 
about $10^5$ y.
This is in particular the case for \chem{26}{Al} (\cite{Podosek97}).
The possible presence of such short-lived radioactive nuclides in live  
form within the early solar-system has far reaching consequences for the 
understanding of its (pre)history and for the nuclear astrophysical 
modelling of the stellar origins of these radionuclides.
 
\subsubsection{The composition of cosmic rays and solar energetic 
particles}\ \ \
%
Much advance has been made in the knowledge of the elemental or isotopic 
compositions of the solar energetic particles (\cite{MeyerJP93}) and 
galactic cosmic rays (\cite{MeyerJP97,Ellison97}). 
The accumulated data have had some significant impact on our 
understanding of the possible sources of cosmic rays and of their 
composition. 
Concomitantly they have triggered some nuclear astrophysics activities.
  
\subsection{Neutrino astronomy}
%
For years, neutrino astrophysics has built upon experiments designed to
detect neutrinos emitted by the Sun, leading to
the ``solar neutrino problem'' (Sect.~5.1).
This type of astrophysics has entered a new era with the detection by 
the IMB and Kamiokande II collaborations of a neutrino burst from the 
supernova SN1987A \cite{Bionta87,Hirata87}.
This remarkable observation seems to validate standard models of neutrino 
emission from supernovae, but the limited number of neutrino events 
observed apparently makes it difficult to validate or invalidate at a 
{\it quantitative} level the various choices of input physics 
in supernova 
modelling (\cite{Arnett96,Arnett89}). 
In any case, that observation has without any doubt opened the door 
to many new neutrino observatories. 
These developments have received a strong support from the speculation
that high-energy neutrinos may originate from a large variety of cosmic
objects, including active galactic nuclei or gamma-ray bursts, as well 
as from cosmological structures, like ``strings'' (\cite{Gaisser95}).

Clearly, a better understanding of the Universe and its constituents 
through the various astrophysical or cosmochemical approaches sketched 
above requires progress to be made in a large variety of observational or 
experimental devices. 
In very many instances it also goes through experimental and theoretical 
improvements in astrophysics and nuclear physics, as will be described  
later.

\section{The contributors to nuclei in the Cosmos}
%
The very nature of the phenomena responsible for the transformation of the
composition of stellar surfaces and of the galaxies at various scales has
been the subject of much work. 
We summarize here the main characteristics of some of the major 
nucleosynthesis agents. 
%
\begin{figure} 
\center{\includegraphics[width=0.9\textwidth,height=0.7\textwidth]{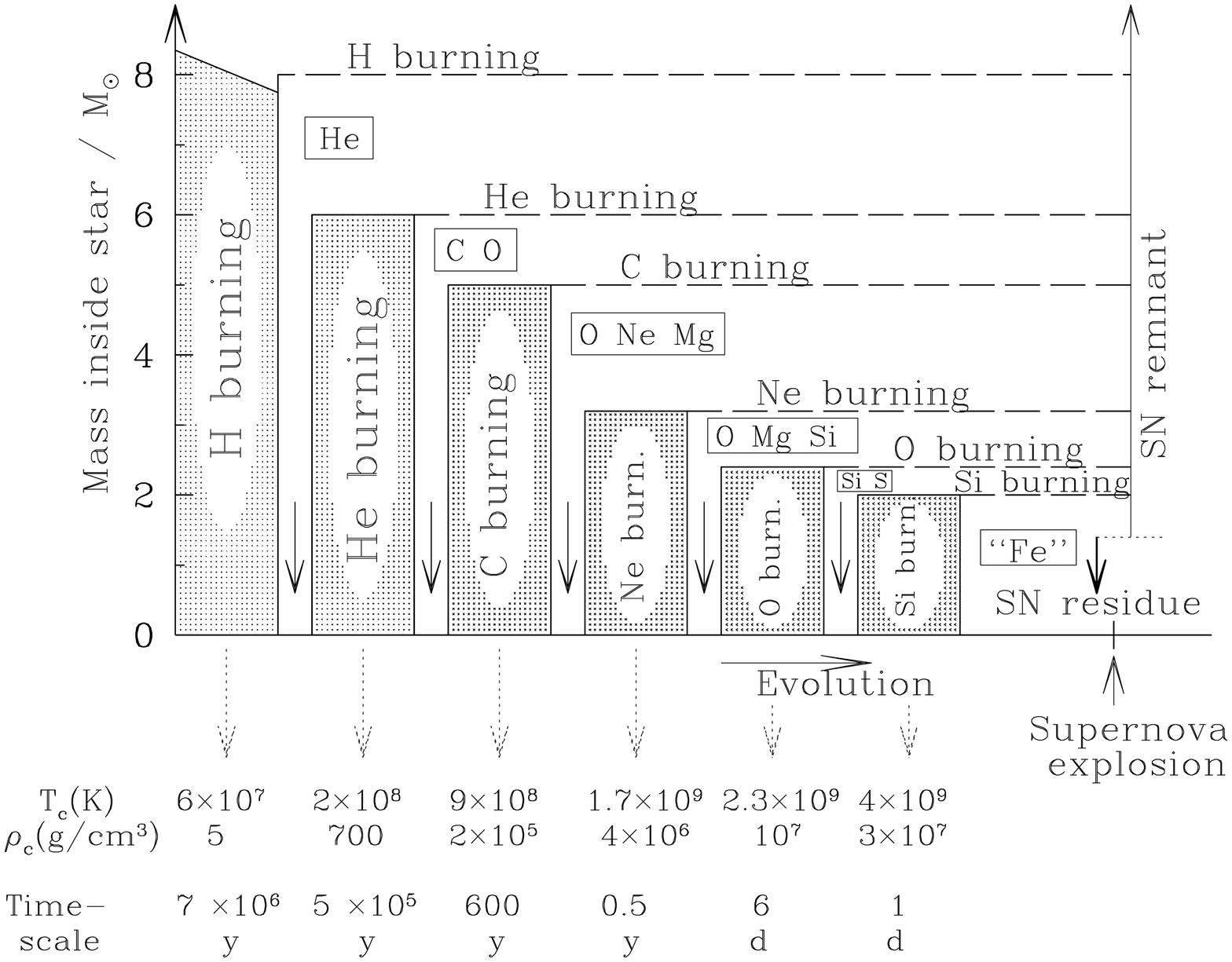}}
\caption{Schematic representation of the evolution of the internal
structure of a spherically-symmetric massive ($M \approx 25$ 
M$_{\odot}$) star.
The shaded zones correspond to nuclear burning stages. 
A given burning phase starts in the central regions (the central 
temperatures $T_c$ and densities $\rho_c$ are 
indicated at the bottom of the figure), 
and then migrates into thin peripheral burning shells. 
In between the central nuclear burning phases are episodes of 
gravitational contraction (downward arrows). 
The chemical symbols represent the most abundant nuclear species left 
after each nuclear-burning mode (``Fe'' symbolizes the iron-peak
nuclei with $50 \lsimeq A \lsimeq 60$). 
If the star explodes as a (Type-II) supernova, the most central 
parts may leave a ``residue,'' while the rest of the stellar material
is ejected into the ISM, where it is observed as a supernova
 ``remnant''
}
\end{figure}

\subsection{Some generalities about stellar structure and evolution}
%
In short, and as pictured very schematically in Fig.~7, the evolution of 
the central regions of a star is made of successive ``controlled'' 
thermonuclear burning stages and of phases of gravitational contraction.
The latter phases are responsible for a temperature increase, while the
former ones produce nuclear energy through charged-particle induced 
reactions. 
Of course, composition changes also result from these very same reactions,
as well as, at some stages at least, from neutron-induced reactions, 
which in contrast do not play any significant role in the stellar energy 
budget. 
The nuclear reactions involved in the different nuclear-burning 
stages are 
discussed in greater detail in Sect.~5.   
Let us simply emphasize here that such a sequence develops in time with  
nuclear fuels of increasing charge number $Z$ and at temperatures 
increasing from several tens of $10^6$ K to about $4 \times 10^9$ K. 
Concomitantly the duration of the successive nuclear-burning phases 
decreases in a dramatic way. 
This situation results from the combination of (i) a decreasing energy 
production when going from H burning to the later burning stages and 
(ii) an increasing neutrino production, and the consequent energy losses,
with temperatures exceeding about $5 \times 10^8$ K 
(see \cite{Arnett96} Chap.~10). 
Figure~7 also depicts schematically that a nuclear burning phase, once
completed in the central regions, migrates into a thin peripheral shell. 
As a consequence the deep stellar regions look like an onion with  
various ``skins'' of different compositions. 

It is quite important to notice that all stars do not necessarily 
experience all the burning phases displayed in Fig.~7 
(see \cite{Arnett96} Chap.~6.6 for details): while massive ($M \gsimeq
10$ M$_{\odot}$) stars go through all those burning episodes, low-mass
($M \lsimeq 8$ M$_{\odot}$) stars stop their nuclear history already 
after completion of central He burning. 
Stars in the $8 \lsimeq M \lsimeq 10$ M$_{\odot}$ range represent 
complicated intermediate cases. 
It has also to be stressed that the true stellar structure is certainly 
much more complicated than sketched in Fig.~7 (cf.~Figs.~10.5-6 
of \cite{Arnett96}) even when effects like deviations from spherical 
symmetry (induced by rotation 
or certain mechanisms of transport of matter)
are neglected. 
This spherically symmetric picture of a star may break down, especially 
during the advanced stages of the evolution of massive stars, and would 
lead to a dramatic growing of the complication of the stellar structure 
and evolution. 
This increased complexity is demonstrated by recent multi-dimensional 
simulations of the structure of massive stars 
(\cite{Arnett97,Bazan98}).
The consideration of rotation of course brings additional difficulties 
(\cite{Pinson97}). 
Finally, steady mass loss from a star may also affect its evolution in 
various ways (\cite{Chiosi86,Dupree86}).
 
\subsubsection{$M \gsimeq 10$ M$_{\odot}$ stars}\ \ \  
%
The evolution of these stars and the concomitant nucleosynthesis have
been the subject of much computation, at least in spherically 
symmetric cases (\cite{Arnett96,Weaver93,Hashimoto95}).
After having experienced all the burning phases depicted in Fig.~7, these 
stars are seen to develop an Fe core that is lacking further nuclear 
fuels, any transformation of the strongly-bound Fe nuclei being 
endothermic.  
In fact this core becomes dynamically unstable and implodes as a result 
of free-electron captures and Fe photodisintegration, the former 
transformation playing an especially important role at the low end of the 
relevant stellar-mass range.
Through a very complex chain of physical events the implosion can, in 
certain cases at least, turn into a catastrophic supernova explosion 
referred to as a Type-II 
%
supernova\footnote{Type-II supernovae are defined as those showing 
H-lines in their spectra.  
It is likely that most, if not all, of the exploding massive stars still 
have some H-envelope left, and thus exhibit such a feature. 
In contrast, Type-I supernovae lack H in their ejecta.
Specific spectral features have led to the identification of different 
Type-I subclasses, among which are the Types-Ia, -Ib and -Ic. 
See \cite{Chevalier97,Nomoto97} for the classification scheme
of supernovae
} 
%
(see \cite{Arnett96} Chaps.~12 and 13 in particular, for a detailed 
discussion of the implosion-explosion mechanism and for the model 
predictions of the supernova observables). 
In this sequence of physical events, neutrino production and diffusion 
through the supernova core material are now known to play a key role.
These processes crucially help the generation and powering of a shock wave
propagating outward through most of the supernova layers.
This shock wave compresses the various traversed layers, heats them up
before pushing them outward until their ejection into the ISM.
This expansion is of course accompanied by a cooling of the material. 
This heating and cooling process of the layers hit by the supernova shock 
wave allows some nuclear transformations to take place during a quite 
brief time, modifying more or less significantly the pre-explosion 
composition of the concerned layers. 
The study of the composition of the ejected material that makes up the 
supernova remnant is one of the main chapters of the theory 
of ``explosive nucleosynthesis.''  

In this supernova picture not the whole of the stellar mass is returned 
to the ISM.  
The innermost parts are bound into a ``residue,'' which may be
a neutron star (observable as a pulsar if it is magnetized and rapidly 
rotating) or even a black hole (see  \cite{Srinivasan97,Novikov97} for a 
detailed discussion of the physics of these objects).  
If a general consensus has been reached on the above description of
massive star explosions, many very complicated astrophysics and 
nuclear physics questions have yet to be answered. 
 
The Type-II supernova explosion sketched above is not the only way for 
massive stars to return material to the ISM. 
Hot stars are indeed observed to suffer steady (non-explosive) stellar 
winds which may carry non-explosively processed material, e.g., H- and 
He-burning products in the case of the so-called Wolf-Rayet stars 
(\cite{Arnould97}). 
These stars eventually explode as supernovae with Type-Ib or -Ic spectral 
features, and thus add their share to the contamination of the ISM with 
explosive burning products \cite{Woosley95}.

Quite clearly, massive stars, through their pre-supernova and supernova 
evolution, are essential agents to the evolution of nuclides in galaxies. 
It is also considered nowadays that some isotopically anomalous grains 
identified in meteorites originate from supernovae. 
However, many detailed aspects of this contamination remain more or less 
uncertain. 
Some of these problems are of nuclear physics origin, while others are of 
purely astrophysics nature. 
In this respect it has to be stressed that the modelling of a 
spherically-symmetric
supernova event already raises so many severe difficulties that 
very limited numbers of multi-dimensional simulations have been attempted 
to date (\cite{Janka96}).

\subsubsection{$0.45 \lsimeq M \lsimeq 8$ M$_{\odot}$ stars}\ \ \  
As noted above the nuclear history of these stars is essentially limited 
to H and He burnings that take place in the central regions before 
affecting peripheral shells.  
This nuclear activity is considered to be responsible for the many 
chemical peculiarities observed at the surface of a variety of RGB or 
AGB stars. 
Indeed these stars are expected to have their surfaces more or less 
severely contaminated with H- and He-burning ashes as well as with the 
products of a neutron-capture process known as the ``s-process'' 
(Sects.~5.2.2 and 8.1). 
This contamination is a direct consequence of episodes of extended mixing 
between nuclear-processed deep layers and the stellar surface 
which are referred to as ``dredge-up'' phases (\cite{Lattanzio97}).
The level of this contamination is also influenced by the considerable 
mass losses those stars experience during their AGB phase. 
This wind is also responsible for the formation of planetary nebulae, the 
cores of which evolve to white dwarfs essentially made of C and 
O, as well as for the enrichment of the galaxies with nuclear-processed 
material.  
Last but not least, some anomalous meteoritic grains are also suspected
to form in AGB circumstellar wind-ejected shells. 

The modelling of the stars in the considered mass range remains uncertain 
in various respects.
Uncertainties in predicted surface composition, pre- or post-AGB 
evolution and galactic contamination relate in particular to the 
efficiency of the dredge-up episodes, or to the extent of the stellar 
winds during the AGB phase. 
The mass loss rates also influence the upper limit (taken here 8 
M$_{\odot}$) for the initial masses of stars which leave C-O white 
dwarfs as residues (\cite{Bloecker95}).
 
\subsubsection{$8 \lsimeq M \lsimeq 10$ M$_{\odot}$ stars}\ \ \  
%
The evolution of these stars is expected to exhibit special and complex 
features that appear to depend very sensitively on the precise stellar 
mass in the considered range. 
The non-explosive history of some of these stars has been followed in 
detail recently \cite{Iben97}. 
The central and peripheral H- and He-burning episodes are complemented 
with a C-burning stage that starts off-centre before migrating to the 
more central regions. 
At the end of the C-burning phase an O+Ne-rich core is produced, 
while the 
surface compositions are expected to be modified by the dredge-up to the 
surface of the products of the nuclear burnings. 
The post C-burning evolution might be dominated by an increasing 
degeneracy of the free electrons, the Fermi energy of which could exceed 
the threshold for electron captures (Sect.~4.4.1) by dominant core 
constituents such as \chem{20}{Ne}. 
These captures would result in a deficit of electron pressure, and 
consequently cause core collapse \cite{Nomoto87}. 
The details of this collapse are somewhat different from those of the 
implosion induced by iron photodisintegration in the cores of more 
massive stars, even if the end product may also be a Type-II supernova.
Save the possibility of an O-deflagration leading to its complete
disruption, such a star appears to form a neutron star
of less than about 1.3  M$_{\odot}$ (\cite{Nomoto87}).
 
Extensive mass loss prior to an explosion might also have several 
interesting implications such as the possibility of production of an 
O+Ne white dwarf. 
This scenario would avoid a catastrophic supernova fate, at least if this 
white dwarf cannot
increase its mass through the accretion of some matter 
from a companion in a binary system in particular (see below).      
The initial stellar-mass range for the ``electron-capture triggered''
supernovae may also be subject to changes depending on the (uncertain)
treatment of 
convection (\cite{Portinari98}).

\subsubsection{Binary stars}\ \ \ 
%
Roughly two-third of all stars in our Galaxy belong to multiple- 
(mainly binary-) systems, and many different types of observed 
astronomical events are now interpreted in terms of phenomena occurring 
in such systems. 
The effects of a binary companion on the evolution of a star are 
complicated and far from being fully explored. 
This concerns both purely structural problems and highly specific
nuclear-physics questions (see \cite{Nussbaumer94} for extended 
reviews).
 We just highlight here
some characteristics of direct relevance to nuclear astrophysics.

One of the most important characteristics of the evolution of binary 
systems is the existence of episodes during which matter is transferred
from one component to the other. 
Such mass exchanges offer a rich variety of distinct astrophysical and 
nuclear scenarios. 
Various instabilities of nuclear origin may indeed develop in the 
transferred material at the surface of the accreting component. 
In particular, explosive H-burning at the surface of an accreting 
white dwarf or neutron star may be responsible for the occurrence of 
certain types of novae or of X-ray bursts, respectively. 
At least in the nova case, some nuclear-processed material is known to be 
recycled into the ISM.  
In addition to such ``surface'' effects, accreting WDs may also 
experience in their interiors explosive He- or C-combustions  
of the deflagration or detonation types, different burnings and different 
regimes possibly developing in a single object and in a sequential way.
The net result may be the explosive disruption of the accreting WDs 
as Type-Ia supernovae (\cite{Nomoto97}).
These are important contributors to the evolution of the composition of
galaxies. 
They have also been viewed as possible providers of certain anomalous 
presolar grains. 
Under certain circumstances the white dwarf may collapse into a neutron
star.
 
\setcounter{footnote}{0}
\subsection{Spallation reactions}
%
Spallation 
reactions\footnote{The terminology ``spallation reaction'' is used here  
in its historical context, independently of its exact meaning in nuclear 
physics, as a synonym for nuclear reactions induced by the interaction of 
primary particles with relative energies in excess of some tens of 
MeV/nucleon} 
%
are essential agents in the shaping of the elemental and isotopic  
composition of the relativistic galactic cosmic rays (GCRs), 
and play a central role in ``propagation models'' attempting to determine 
the GCR abundance variations as the result of the interaction of the GCRs 
with the interstellar matter when they travel from the GCR source and the 
Earth (\cite{Vernois96}). 
An important subset of the predictions of astrophysical interest from the 
propagation models concerns the production by spallation of the galactic 
Li, Be and B content. 
More specifically, it has been shown that the GCRs interacting with 
interstellar matter prior to the isolation of the solar-system material 
from the general galactic pool are responsible for the bulk solar-system 
\chem{6}{Li}, \chem{9}{Be}, \chem{10}{B}, most of the \chem{11}{B}, and  
at least some  of the \chem{7}{Li} \cite{Reeves94}. 

Various non-exploding  stars (including the Sun) or exploding ones are 
also known to accelerate particles into the approximate 10 to 
200 MeV/nucleon range.
These ``stellar energetic particles'' can interact with the material 
(gas or grains) at the stellar surfaces, in circumstellar shells, or in 
the local ISM. 
The outcome of these nuclear interactions has recently  been the subject  
of a renewed interest following the determination of the Li, Be and B 
content of very old stars \cite{Ramaty96} - \cite{Parizot97}.
Some of these studies have also revisited the question of the light 
isotopes in the solar system with the conclusion that their solar 
energetic-particle  production could cure the predicted insufficient
\chem{11}{B} production in the GCRs, while making still other 
\chem{7}{Li} 
sources mandatory. 
  
Spallation reactions also play an important role in other astrophysical 
questions, like the production by GCRs or solar energetic 
particles of stable and radioactive nuclides  in extraterrestrial 
solar-system matter such as planetary surfaces, meteorites or cosmic dust 
\cite{Michel98}.

The study of the exact role of spallation reactions in astrophysics
raises very many specific questions concerning, in particular, the 
determination of the spectra of particles accelerated by various cosmic 
objects, or the experimental and theoretical studies of nuclear reactions 
taking place above the Coulomb barrier (\cite{Vernois96,Michel98}).
      
\subsection{The Big Bang contribution to nucleosynthesis}
%
The standard hot Big Bang (SBB) model provides a very successful and 
economical description of the evolution of the (observable) Universe from
temperatures as high as $T \approx 10^{12}$ K ($t\approx10^{-4}$ s after 
the ``bang'') until the present epoch ($t\approx 10-20$ Gy). 
This model has many far-reaching implications not only 
in cosmology and particle 
physics, but also in
high-energy nuclear physics (like the relativistic heavy-ion 
physics in relation with the quark-hadron phase transition), low-energy 
nuclear physics (like the physics of thermonuclear reaction rates in 
relation to the primordial nucleosynthesis episode), and in  astrophysics 
(through the problems of the formation and initial composition of the 
galaxies). Many of these exciting questions, as well as the details of the
thermodynamics of  the SBB are dealt with at
length in \cite{Sarkar96}. 
Here we just limit ourselves to a brief account of some basic 
aspects of the Big Bang, and in particular of the main features of the
nucleosynthesis epoch of most direct relevance to nuclear astrophysics.

The observational evidence testifying to the  validity  of the SBB model 
is threefold:
 
\noindent (1) the universal expansion discovered by Hubble in 1929: all
galaxies, except those of the Local Group (a gravitationally bound group
of about 20 galaxies to which our Milky Way Galaxy belongs), are receding 
from us (and from each other) with velocities proportional to their 
distances. 
The factor of proportionality is the Hubble parameter $H(t)$, the current 
value of which being the Hubble constant $H_0$.
The precise value of $H_0$ has yet to be known in the approximate
50-100 km/s/Mpc range, although many recent determinations point to
$H_0 \approx 70$ km/s/Mpc (\cite{Hogan96});
 
\noindent (2) the cosmic (microwave) background radiation, discovered by 
Penzias and Wilson in 1965 
\cite{Penzias65},\footnote{As for the theoretical
predictions preceding the discovery, see \cite{Alpher88}, a 
bitter-sweet memoir
} 
%
which is a unique laboratory for studying the 
initial conditions that gave rise to the observed Universe. 
It has a spectrum that fits with astonishing precision a black-body of 
temperature $T_0 \approx 2.7$ K, and its angular uniformity 
($\Delta T/T<10^{-4}$) combined with its presumed homogeneity 
(``Cosmological Principle'') strongly argues for a hot and homogeneous 
early Universe, where matter and radiation were in equilibrium 
(\cite{White94}). 
Matter and radiation are predicted to have ``de-coupled'' a few 10$^5$ 
years after the ``bang,'' when the temperature had decreased to about 
3000 K. 
After this epoch, radiation cooled during the universal expansion to its 
present value of about 2.7 K more rapidly than matter did; 
 
\noindent (3) most, if not all, of the universal content of D and 
\chem{4}{He} are considered to originate from the hot early Universe. 
This is likely true also for the \chem{7}{Li} observed in old un-evolved 
stars which are probably the good sites to have maintained the signature 
of the \chem{7}{Li} content of the forming galaxies. 
The situation of \chem{3}{He} appears to be less clear-cut, some being 
possibly of direct Big Bang origin, some coming from the stellar 
transformation of the primordial D, and some coming from H burning in 
stars. 
The nucleosynthesis epoch of the Big Bang is expected to have developed  
when the temperatures decreased to values in the approximate 
$10^9 \gsimeq T \gsimeq 10^8$ K range (corresponding to times  
$10^2 \lsimeq t \lsimeq 10^3$ s).

The confrontation between the SBB nucleosynthesis expectations and 
observations necessitates the consideration of a chain of theoretical and
observational links, each of which bringing its share of uncertainties and
difficulties.  
On the observational side, the determination of the ``primordial'' 
galactic abundances is far from being a trivial matter
(\cite{Reeves94,Sarkar96}).  It necessitates not only direct abundance
determinations by observations,  but also an evaluation of the role of the
oldest observable stars or of  spallation reactions in modifying the early
galactic content of nuclides  like D, \chem{3}{He} or \chem{7}{Li}. 
The controversy over discrepant determinations (from the analyses of the 
spectra of distant quasars) of the D abundance in intergalactic clouds 
supposedly made of primordial material is an additional dramatic 
illustration of the problems one has to face in this field 
(\cite{Sarkar96}, \cite{Burles96} - \cite{Levshakov98}). 
On the theoretical side, the SBB nucleosynthesis offers a variety of 
pleasing features. 
In particular, for the present temperature of the cosmic background 
radiation is quite accurately known, the calculated abundances depend
only upon the ratio $\eta$ of the total number of (bound and free) 
nucleons to the number of photons (which remains constant following the 
e$^+$e$^-$ annihilation episode in the Big Bang thermodynamics). 
This is true at least in the Standard Model (SM) of particle physics,  
which involves three kinds of weakly-interacting massless neutrinos. 
The quantity $\eta$ cannot be reliably determined on observational 
grounds, and is considered as a (the only) free parameter in the SBB+SM 
framework. 

Due account being taken of the uncertainties of nuclear nature affecting 
the predicted abundances for a given $\eta$, the SBB+SM model is often 
regarded as being able to account for the presumed initial galactic 
content of D, \chem{4}{He} and \chem{7}{Li} for values of $\eta$ in the 
approximate $2 \times 10^{-10} \lsimeq \eta \lsimeq 9 \times 10^{-10}$, 
the upper limit being imposed in particular by the low primordial 
D abundance claimed by some authors 
(see \cite{Reeves94,Sarkar96,Olive96} 
for a more thorough discussion; see also \cite{Levshakov98}). 
This converging agreement for the primordial abundances, and the 
concomitant constraints on $\eta$ are generally considered as a real 
triumph of the SBB model. 
These constraints have also far-reaching implications concerning the
nature of the dark matter which is observed to dominate the dynamics of 
individual galaxies, as well as groups and clusters of galaxies 
(\cite{Ashman92}). 
In particular, little room would be left for baryonic matter (in 
any form) for high values of the Hubble constant and high primordial D 
abundances. 
This conclusion would be invalidated if the initial galactic D content 
were indeed low (\cite{Sarkar96}).

This success notwithstanding, many variants of the SBB + SM model have
been developed and studied to a varied extent, some of them stemming 
from the claim  of a ``SBB crisis'' caused by a discrepancy between 
predicted and observed abundances \cite{Hata95}. 
Some of these variants call for gross departures from the standard 
cosmology, 
like alternate theories of gravity or anisotropic world models. 
Others restrict themselves to the standard cosmological models, but 
allow for some departures from the SM of particle physics. 
The interested reader is referred to e.g. \cite{Sarkar96,Malaney93} for 
thorough presentations of these alternatives. 
Let us just note that various remedies to the SBB crisis have been 
proposed, some of them maintaining the SBB well alive. On the other hand, 
a Big Bang model which allows for some inhomogeneities at the epoch of 
nucleosynthesis has raised much excitement in part of the nuclear 
astrophysics community (Sect.~6.1). 
%
 
\subsection{The chemical evolution of galaxies}
%
\begin{figure}
\center{\includegraphics[width=1.00\hsize,height=0.8\hsize]{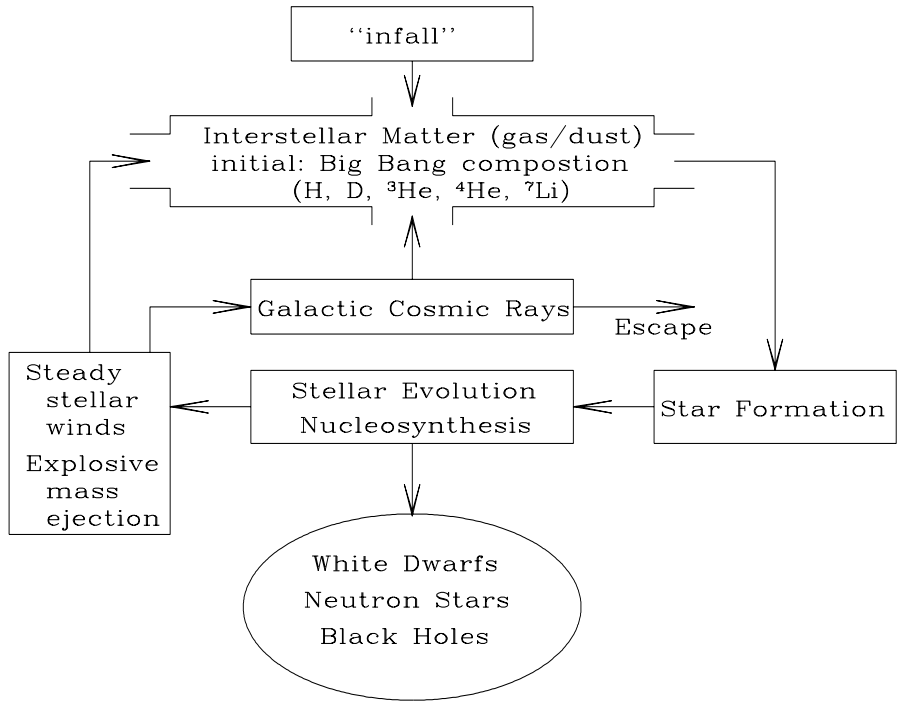}}
\vskip-2.5truecm                                   
\caption{A very schematic picture of the galactic ``blender'' (see text)
}
\end{figure}
%
The modelling of the evolution of the nuclear content of galaxies 
(classically termed the``chemical'' evolution of galaxies) is without any 
doubt one of the most formidable problems astrophysics has to face. 
This question has been tackled at various levels of sophistication, 
ranging from so-called ``chemo-dynamical models'' to simple `` one-zone'' 
models, the latter ones  limiting themselves to the description of the 
evolution of the abundances in the solar neighbourhood. 
Because of 
their complexity, the first types of models are cruder in their 
nucleosynthesis aspects than the latter ones, which do not address any 
galactic thermodynamics- or dynamics-related  issues. 
This immense problem
(\cite{Pagel97}, \cite{Shore97} - \cite{Prantzos97})
 cannot be reviewed here, and we limit
ourselves to a very crude overview of its chemical aspects.

The way a galaxy evolves chemically is represented in a very sketchy 
manner in  Fig.~8.  
Let us consider the ISM (made of gas and dust) right after galaxy 
formation.  
Its composition is assumed to be essentially the one emerging from
the Big Bang, the standard model of which predicts the presence of 
significant amounts of just H, D, \chem{3}{He}, \chem{4}{He} and 
%
\chem{7}{Li}.\footnote{We neglect here any possibility of 
{\it pre-galactic}
nucleosynthesis of thermonuclear nature (\cite{Norgaard76}) by still 
putative pre-galactic very massive stars (\cite{Ferrara98}),
 or of spallation type  by ``cosmological cosmic 
rays'' \cite{Montmerle77}. 
These very early modifications of the Big Bang yields have been advocated 
at several occasions, but are usually not taken into account in galactic 
chemical evolution models
}
%
Part of the ISM material is used to form stars which, through a large  
variety of nuclear reactions, transform the composition of their 
constituent material during their evolution.  
At one point or another during that evolution, some material may be 
returned to the ISM through various mechanisms (Sect.~3.1).  
In general, all the stellar material with altered composition is not 
returned to the ISM. 
Part is locked up in stellar residues (white dwarfs, neutron stars, or 
black holes), and normally is not involved in the subsequent chemical
evolution of the galaxies, at least when only single stars are 
considered.
Also recall that a tiny fraction of the matter ejected by stars can be 
accelerated to galactic cosmic-ray energies.  
Spallation reactions by
these high-energy particles interacting  with the ISM material can
be responsible for the Li, Be and B
contents of the galaxies, and especially of
the disc of our own Galaxy (Sect.~3.2).
At least in spiral galaxies like our own, some fraction of the galactic 
cosmic-ray nuclei might escape the galactic disc. 
Part of the supernova ejecta might also be ejected from the disc 
(though not sketched in Fig.~8). 
In contrast, some material, possibly of Big Bang composition, might 
fall onto the galactic disc (``infall'') from the galactic halo to 
dilute the stellar-processed material. 

\begin{figure}
\center{\includegraphics[width=0.8\hsize,height=0.5\hsize]{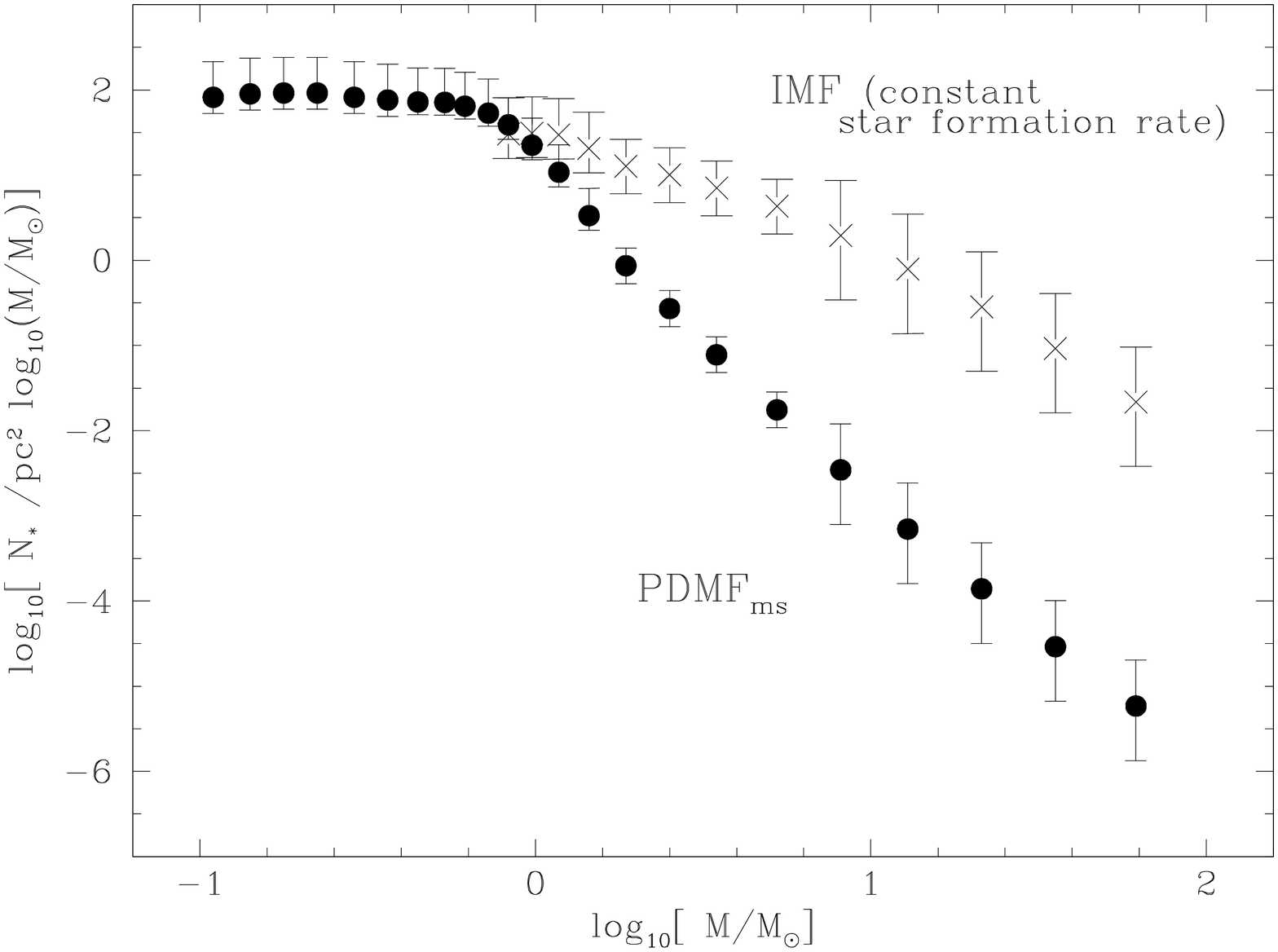}}
\caption{An Initial Mass Function (IMF) relevant to the disc of our
Galaxy in the solar neighbourhood (adapted from \cite{Miller79}). 
It is derived from the observed present-day mass function for Main 
Sequence stars (PDMF$_{\rm ms}$) under the assumption that the stellar 
birthrate is constant in time. 
The PDMF$_{\rm ms}$ counts stars that are just leaving the MS in the HRD.
The deviation of PDMF$_{\rm ms}$ from IMF reflects
the ratios of the MS lifetimes to the age of the Galaxy 
(assumed to be in the 9 - 15 Gy range)
}
\end{figure}
%
One basic ingredient of the models for the chemical evolution of the 
galaxies is the stellar creation function, that is, the number of stars 
born per unit area of the galactic disc (in spiral galaxies) per unit
mass range and unit time interval.  
That question has been discussed at length in \cite{Scalo86}.  
As supported to a large extent by phenomenological considerations,
and also for obvious reasons of modelling facilities, it is generally 
assumed that the star formation is separable into a function of time only 
(the stellar birthrate, which is found not to vary widely with time, at 
least in the neighbourhood of the Sun), and a function of mass only, 
referred to as the ``Initial Mass Function (IMF).''  
An IMF constructed in such a way from observational data is represented 
in Fig.~9.
The main result of such studies is that, in the solar neighbourhood at 
least and apparently also in more general situations, the IMF is more or 
less steeply decreasing with increasing stellar mass, at least in the 
$M \gsimeq $ M$_{\odot}$ range. 

Another main ingredient of the galactic chemical evolution models is the 
mass and composition of the matter ejected by a star with a given 
initial mass. 
%
\begin{figure}
\center{\includegraphics[width=0.8\hsize,height=0.5\hsize]{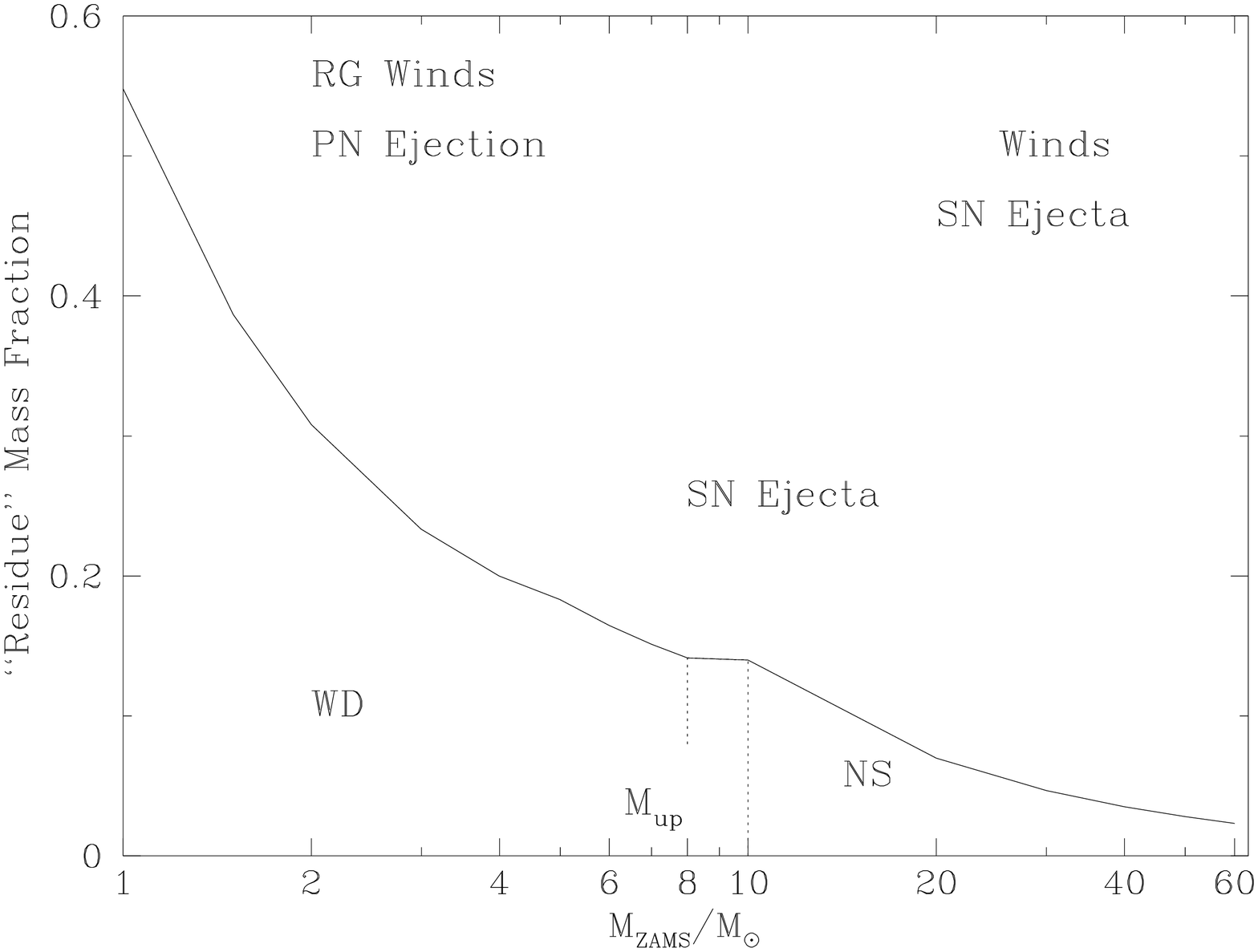}}
\caption{Schematic representation of the fraction of the initial
(``zero-age main sequence'')
mass $M_{\rm ZAMS}$ of 
a single star that remains bound in a white dwarf (WD) or
neutron star (NS) residue at the end of its evolution.  
The rest of the stellar material is ejected mostly by the indicated 
mechanisms.  
[RG, PN and SN stand for ``Red Giant,'' ``Planetary Nebula'' and
``Supernova,'' respectively.]
Stars with initial masses up to $M_{\rm up} \approx 8$ M$_{\odot}$ 
are assumed to end their lives as WD with typical masses  
provided by stellar evolution models \cite{Bloecker95}.
Stars with $M \gsimeq 10$ M$_{\odot}$ are assumed to leave a 
1.4 M$_{\odot}$ neutron star (NS).
In the remaining $8 \lsimeq M \lsimeq 10$ M$_{\odot}$ range, stars might 
explode leaving either a low-mass NS, or an O+Ne WD,  
or even possibly no residue.  
This overall picture is just meant to be a sketch, and is subject 
to various uncertainties
}
\end{figure}
This quantity is loosely referred to here as the stellar ``yields.''
Its evaluation requires the modelling of the evolution of stars with 
initial masses in a broad range of values (essentially between about
1 M$_{\odot}$ and 100 M$_{\odot}$), as well as of the 
concomitant nucleosynthesis. 
The dominant mechanism(s) for the restitution of matter to the ISM by a 
star of given mass has(have) also to be known. 
These processes are depicted schematically in Fig.~10 as a function
of the initial stellar mass, along with the fraction of this mass that is
ejected.

It has to be noticed that some chemical evolution models also consider, 
though in a rather rough way, the specific role that could be played by 
certain binary systems. 
This concerns in particular the galactic enrichment by the explosively 
processed material ejected by novae or by  Type-Ia SNe associated 
with the explosion of accreting WD.
%
 
\section{Nuclear data needs for astrophysics: Nuclear masses, decay modes}
%
As made clear in the previous sections, the Universe is pervaded with 
nuclear physics imprints at all scales. 
In order to decipher these messages from the macrocosm, it is unavoidable 
to  start with a proper description 
of the basic properties, in particular 
masses and decay modes, of a very large variety of nuclei in laboratory 
conditions.  
This information, though necessary for the development of nuclear 
astrophysics, is by far not sufficient, however. 
A remaining essential task is to scrutinize how these properties can be 
affected under astrophysical environments, which are highly versatile and 
are often characterized by high temperatures and/or densities that are 
out of reach of laboratory simulations. 
Over the years these questions have been at the focus of a tremendous 
amount of experimental and  theoretical work, as will be described in 
Sect.~7.
 
\subsection{The masses of ``cold'' nuclei}
%
%
\begin{figure}
\center{\includegraphics[width=0.9\textwidth,height=0.45\textwidth]{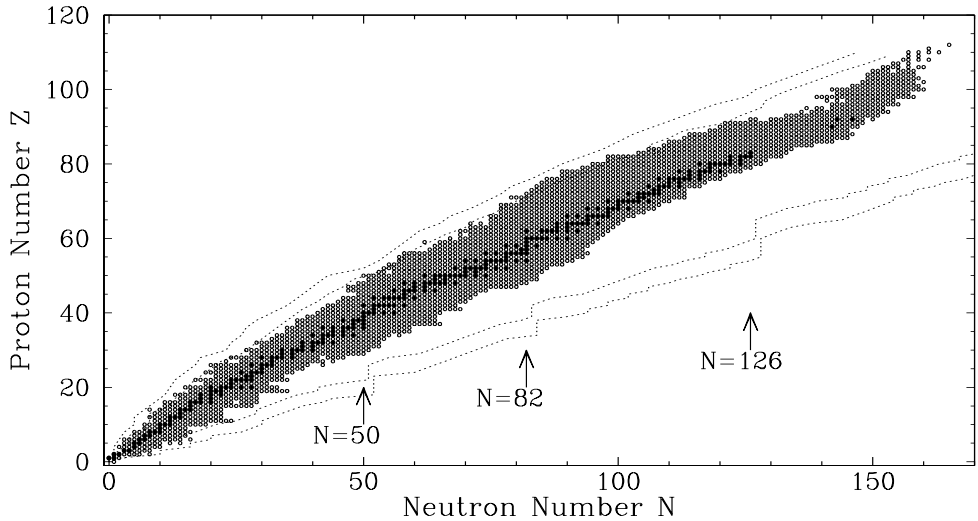}}
\caption{``Segr\`e chart'' of the naturally-occurring stable and long-lived nuclides
(laboratory half-lives $t_{1/2} \gsimeq 4.6$ Gy, the
approximate age of  the solar system) ({\it solid circles}). The radionuclide $^{234}$U
($t_{1/2} \approx 2.5 \times 10^5$ y) is also found naturally on earth as a result of the
$^{238}$U decay chain; it is omitted in the figure.  Artificially produced isotopes
with known half-lives, and in many cases  with measured masses as well, are represented
by open circles.  These data are from
\cite{Horiguchi96} supplemented by
\cite{Bohlen95}  for the lightest nuclides, and by \cite{Hessberger98} for the heaviest 
ones. 
The naturally-occurring nuclides form the ``line of $\beta$ stability'' 
(or ``Heisenberg's  valley''). 
The species located to the right (left) of this line are  
referred to as ``neutron-rich'' (``neutron-deficient''), and undergo
$\beta$-decays. Very heavy nuclides may additionally $\alpha$-decay and/or
fission. Far enough away from the line of stability, nuclei may 
even become
unstable against the emission of a neutron or a proton. 
Some such emissions
have already been observed in the laboratory. 
The ensembles of nuclei whose
neutron or proton separation energies tend to zero define the neutron or
proton ``drip lines'' ({\it dotted lines})
predicted from a mass model (Sect.~7.1.1).
Double drip lines are shown
to account for the nuclear odd-even effects.
 (Note that a
negative proton 
separation energy does not necessary imply spontaneous proton
emission, the reason being the Coulomb barrier effects.)}
\end{figure}
Nuclear masses (equivalently, binding or separation energies) enter all
chapters of nuclear astrophysics. Their knowledge is indispensable in 
order to evaluate the rate and the energetics of any nuclear transformation. 
 
Figure~11 displays the approximately 2500 nuclides that have been 
identified by now in the laboratory. Among them, 286 are naturally 
occurring, the remaining ones being artificially produced. 
As extended as it
is, this  data set does not quite meet the astrophysics requirements.
 This is
especially true when dealing with the r-process
nucleosynthesis (Sect.~8.1), which involves a large number of
nuclei unidentified 
in the laboratory. Theory is thus a mandatory complement to
the experimental efforts.
 
\subsection{Nuclei at high temperatures}
%
The special conditions prevailing in astrophysical environments, and 
particularly in stellar interiors,  bring their share of additional and 
sometimes major difficulties. 
Even the very basic concepts of nuclear ``binding'' or stability have to 
be handled with great care when dealing with certain astrophysical 
conditions. 
This comes about from the fact that nuclei exist not only in their 
ground state, but in excited states as well.
These states are populated through particle (especially electron) or 
photon interactions. 
In stellar interiors (even in explosive situations), thermodynamic
equilibrium holds in general, 
at least locally, to a high level of accuracy,
so that the relative 
populations of the nuclear excited states are very well approximated by 
a statistical Maxwell-Boltzmann distribution law (\cite{Cox68}). 
Following this law, the thermal population of a nuclear excited state 
starts being significant as soon as the temperature [$kT \approx 8.6
\times (T/10^8$K) keV] becomes commensurable with its excitation energy. 
At a given temperature the low-lying excited states of an odd-$Z$, 
odd-$N$ heavy nucleus thus have an especially high equilibrium population.

Specific problems arise when dealing with isomeric states. 
In certain temperature regimes, their populations may indeed depart 
from the equilibrium values, in particular as a result of the 
selection-rule hindrance of the electromagnetic transitions between 
ground and isomeric levels.
Interesting situations of this type concern, for example,
\chem{26}{Al} (\cite{Coc98}), \chem{176}{Lu} (\cite{Klay91}), and 
\chem{180}{Ta} (\cite{Lesko91}).
In all these cases, the thermalization requires a multi-step 
electromagnetic link involving higher excited states, which can be 
efficient at sufficiently high temperatures. For example, the ground 
state of \chem{26}{Al}  and its short-lived 228 keV
isomer ($t_{1/2} \sim 6.3$ s) are expected to be in thermal equilibrium 
only at temperatures in excess of about $5 \times 10^8$ K \cite{Coc98}.

The existence in a stellar plasma of nuclei in their ground as well as 
excited states has an important bearing on various decay modes or 
nuclear transmutations, and consequently on different nucleosynthesis
processes. 
In many cases, therefore, the determination of the ground state
mass is insufficient and the evaluation of ``nuclear partition 
functions,'' i.e. sums of the equilibrium populations of the states of 
a nucleus, has to be carried out. 
This is typically the case when  abundances have to be evaluated in 
conditions where reactions and their reverses equilibrate.
An extreme case of this situation is the ``nuclear statistical 
equilibrium (NSE)'' regime.

The fact that nuclear excited states enter various nuclear astrophysics
calculations obviously makes indispensable the knowledge of nuclear spins
and energies of these states. 
Such information is often missing experimentally, 
especially when dealing
with ``exotic" nuclei far from stability, or even with
 stable nuclei when high 
temperatures have to be considered. In such cases, relatively high-energy
levels may indeed be significantly populated.
 
\subsection{Nuclei at high densities}
%
At high densities, such as encountered in supernovae or neutron stars, 
the meaning of the nuclear binding has to be  understood in terms of the 
nuclear equation of state (EOS), which describes the energy density and 
pressure of a system of nucleons and/or nuclei as a function of matter 
density.  

At densities $\rho \lsimeq 10^{-2} \rho_0$, where $\rho_0 \approx 3 \times
10^{14}$ g/cm$^3$ is the nuclear saturation density, NSE generally holds in
realistic astrophysical conditions,  so that the EOS can be derived from the laws of
equilibrium statistical  mechanics.
A troublesome problem  here is related to the fact that the matter in 
question may involve very neutron-rich nuclei
whose partition functions, if not binding energies (which are essential 
quantities entering the statistical-mechanics equations), are unknown 
in the laboratory.

The nuclear EOS at densities $\rho$ ranging from 0.01 to 10 times 
$\rho_0$ (or even slightly higher) is one of the most important 
ingredients for both supernova and neutron star models. In particular, 
the properties of the EOS around $\rho_0$ dictate to a large extent if 
a massive star can explode as a supernova or not. 
To cover the physics involved in the EOS in this wide density domain 
(\cite{Hillebrandt91,Pethick95}) is far beyond the scope of the current
review. 
Later in Sect.~7.1.2, however, we present a very brief description of 
EOS needs for studies of supernovae and neutron stars.
   
\subsection{Nuclear decays and reactions via weak interaction}
%
Weak interaction processes  play a decisive role in a wide variety of 
astrophysical questions (\cite{Grotz90}).  
Here we discuss very shortly  a few 
problems of direct relevance to stellar evolution and nucleosynthesis,
omitting such an important topic as the properties of neutrinos  and their
meaning in 
cosmology.\footnote{This exclusion applies to double-$\beta$ decays, for 
instance, whose study is motivated by particle physics, although its 
methodology is of pure nuclear-physics nature
}
 
A very specific example of the importance of weak interaction concerns 
the starting transmutation H + H $\rightarrow$ \chem{2}{H} + e$^+$
  + $\nu_e$ 
of the p-p chain of  reactions, which is the essential H-burning
 mode in the
very large  galactic population of stars with masses $M \lsimeq 1$
M$_{\odot}$.  In more advanced stages of evolution, various sorts of weak
interaction  processes occur, some of which are unknown in the laboratory,
just as  the H + H reaction. 

\subsubsection{Various $\beta$-decay modes in astrophysics}\ \ \ 
%
\begin{figure} 
\center{\includegraphics[width=0.8\textwidth,height=0.55\textwidth]{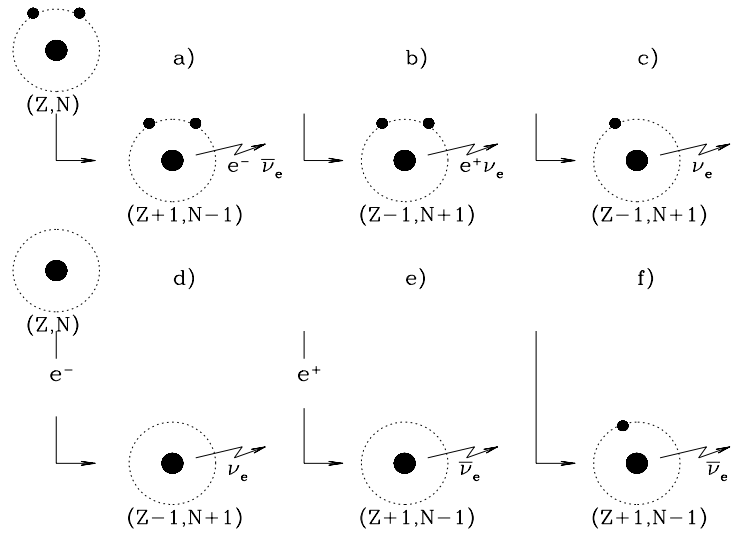}}
\vskip-1.0truecm
\caption{Various nuclear $\beta$-decay modes: ({\it a}) $\beta^-$ decay,
({\it b}) $\beta^+$ decay, ({\it c}) orbital-e$^-$ capture, ({\it d}) 
continuum-e$^-$ capture, ({\it e}) continuum-e$^+$ capture, and ({\it f})
bound-state $\beta$ decay. 
The processes a) - c) are the usual decay modes in the laboratory, while 
d) - f) can occur under stellar conditions
}
\end{figure}
%
The most familiar forms of weak interaction in the laboratory are the 
$\beta^-$-decay (e$^-$ emission), $\beta^+$-decay (e$^+$ emission), 
and orbital-e$^-$ capture processes depicted in Fig.~12~(a)-(c). 
The probabilities of these three processes may be quite different in 
laboratory and astrophysical conditions. One effect that can contribute 
to a deviation from the laboratory $\beta^-$-decay half-lives relates to 
the possible reduction of the electron phase-space as the result of the 
degeneracy of the Fermi-Dirac electron gas that
 is encountered in a variety of
stellar 
situations.\footnote{Under local thermodynamic equilibrium conditions that
prevail in most stellar interiors (\cite{Cox68}), the electron gas is
well described by the classical Maxwell-Boltzmann distribution law. 
In various
situations, however, use of the Fermi-Dirac distribution is made 
necessary. The electrons are then
 referred to
as ``degenerate,'' the degree of this degeneracy increasing with its
 ``Fermi
energy" (that is, with increasing density and decreasing temperature)}
%
Additional effects
concern the contribution 
to the decay process of  excited states of a nucleus,
or the ionization.  This loss of bound electrons takes place in
a large variety of stellar  conditions, and especially at high 
temperatures. 
Atoms accelerated to relativistic cosmic-ray energies are also stripped 
of their electrons. Ionization may influence the nuclear half-lives in 
several ways. 
It has first the obvious effect of reducing the probability of capture 
of bound electrons.
A less trivial consequence relates to the possible development of
the process of ``bound-state $\beta$-decay,'' which creates an electron 
in an atomic orbit previously vacated (in part or in total) by ionization 
[Fig.~12~(f); Sect.~4.4.4]. (This process accompanies
$\beta^-$ decays even in the absence of ionization, but its 
relative contribution is quite
insignificant in such conditions.) 

Other $\beta$-decay modes develop specifically in stellar plasmas. The
``continuum-e$^-$ capture'' process [Fig.~12~(d)] is quite common, 
and often
overcomes orbital-e$^-$ captures in highly-ionized stellar material. 
In these
conditions, atoms are indeed immersed in a sea of free  electrons. 
Continuum-e$^-$ captures display their most spectacular effects in 
situations where the electrons are degenerate. 
Highly stable nuclei in the laboratory may well become $\beta$-unstable 
if indeed the electron Fermi energy is large enough for allowing 
endothermic transitions to take place through the captures of free
electrons with energies exceeding the energy threshold for these 
transitions.
Clearly, the higher the electron Fermi energy, the more endothermic the
transitions may be. The evaluation of the free-e$^-$ capture rates may 
benefit from the laboratory knowledge of the rate of the inverse 
$\beta^+$-decays. 
In many instances, this information is insufficient, and additional 
theoretical predictions are required.  
Among the most spectacular endothermic free-e$^-$ captures that can 
be encountered in astrophysics, let us mention those on \chem{14}{N}, 
\chem{16}{O}, \chem{20}{Ne} or \chem{24}{Mg} in certain WDs  
(\cite{Bravo83} - \cite{Hashimoto93}), or the 
transformations of protons into neutrons in the highly-condensed 
collapsing core of a massive star on the verge of experiencing a 
supernova explosion (Sect.~8.3).
As a result, the neutron fraction may become as high as 90\% in these 
locations.
As the density increases further with the continuation of the collapse, 
the pressure exerted by this neutron gas may become sufficient for 
counter-balancing the gravitational forces, opening the possibility of 
forming a stable neutron star.

Positrons captures [Fig.~12~(e)] are also of importance in certain stellar
situations,  and especially in high-temperature 
(typically $T > 10^9$ K) and 
low-density locations.
In such conditions, a rather high concentration of positrons can be 
reached from an e$^-$ + e$^+  \leftrightarrow \gamma + \gamma$ 
equilibrium which favours the e$^-$e$^+$ pairs. 
The competition (and perhaps equilibrium) between positron captures on
neutrons and electron captures on protons is an important ingredient of 
the modelling of Type-II supernovae. 
  
\subsubsection{Neutrino reactions}\ \ \ 
%
Aside from the capture of continuum electrons on protons (and on 
iron-group nuclides), various weak-interaction processes involving
neutrinos also have an important bearing on Type-II supernovae.  
The probabilities of production of all sorts of (anti-)neutrinos at the 
centre of a nascent (hot) neutron star, certain reaction cross sections 
that determine their transport rate to the neutron star surface, and the 
interactions  of the emerging neutrinos with neutrons and protons near 
that surface are expected to be essential ingredients of the Type-II 
supernova models (\cite{Bethe90} - \cite{Janka93}).
Neutrinos emerging from the neutron star could also interact with 
pre-existing heavy nuclei as they  pass through the outer supernova 
layers. 
This interaction might lead to a limited nucleosynthesis referred to as 
the $\nu$-process 
(\cite{Woosley90,Nadyozhin91}).\footnote{Also note that the
cross sections for the captures of high-energy neutrinos by certain heavy 
nuclei, like \chem{71}{Ga} and \chem{205}{Tl}, are needed for designing 
neutrino detectors
} 

While most of the $\beta$-decay processes of astrophysical interest 
mentioned above can be dealt with in the classical ``V $-$ A'' theory 
of the weak interaction (\cite{Konopinski66}), the evaluation of 
neutrino interaction cross sections requires due consideration of both 
the charged- and neutral-currents of the unified electro-weak 
interaction (\cite{Tubbs75,Bruenn85}). 
Many interesting and complicated problems are also raised by  various aspects
of nucleon correlations and spin fluctuations in neutrino scattering at high 
densities (\cite{Raffelt96} -
\cite{Yamada98}).
  
\subsubsection{$\beta$-decays from nuclear excited states}\ \ \ 
%
As we have emphasized earlier, the excited levels and ground state 
of a nucleus are very often populated in thermal equilibrium, possible 
exceptions being certain isomeric states. 
These various levels may thus contribute to the decay of a 
nucleus, so that its effective $\beta$-decay half-life may strongly 
depart from the laboratory value. 

A classical example of the importance of the $\beta$-decays from nuclear
excited states concerns \chem{99}{Tc}. 
Technetium observed at the surface of certain AGB stars 
is considered to be \chem{99}{Tc}, which is a product of 
the s-process of neutron captures 
developing in the He-burning shell of these
stars (Sect.~8.1). 
Under such conditions (typical temperatures in excess of about $10^8$ K) 
the 141 and 181 keV excited states of \chem{99}{Tc} can contribute 
significantly to the effective decay rate. 
Figure~13 shows a dramatic drop from the laboratory (ground state)
$t_{1/2} \approx 2.1 \times 10^5$ y   to as
short as some 10 y at $T \gsimeq 3 \times 10^8$ K.
%
\begin{figure} 
\center{\includegraphics[width=0.8\textwidth,height=0.4\textwidth]{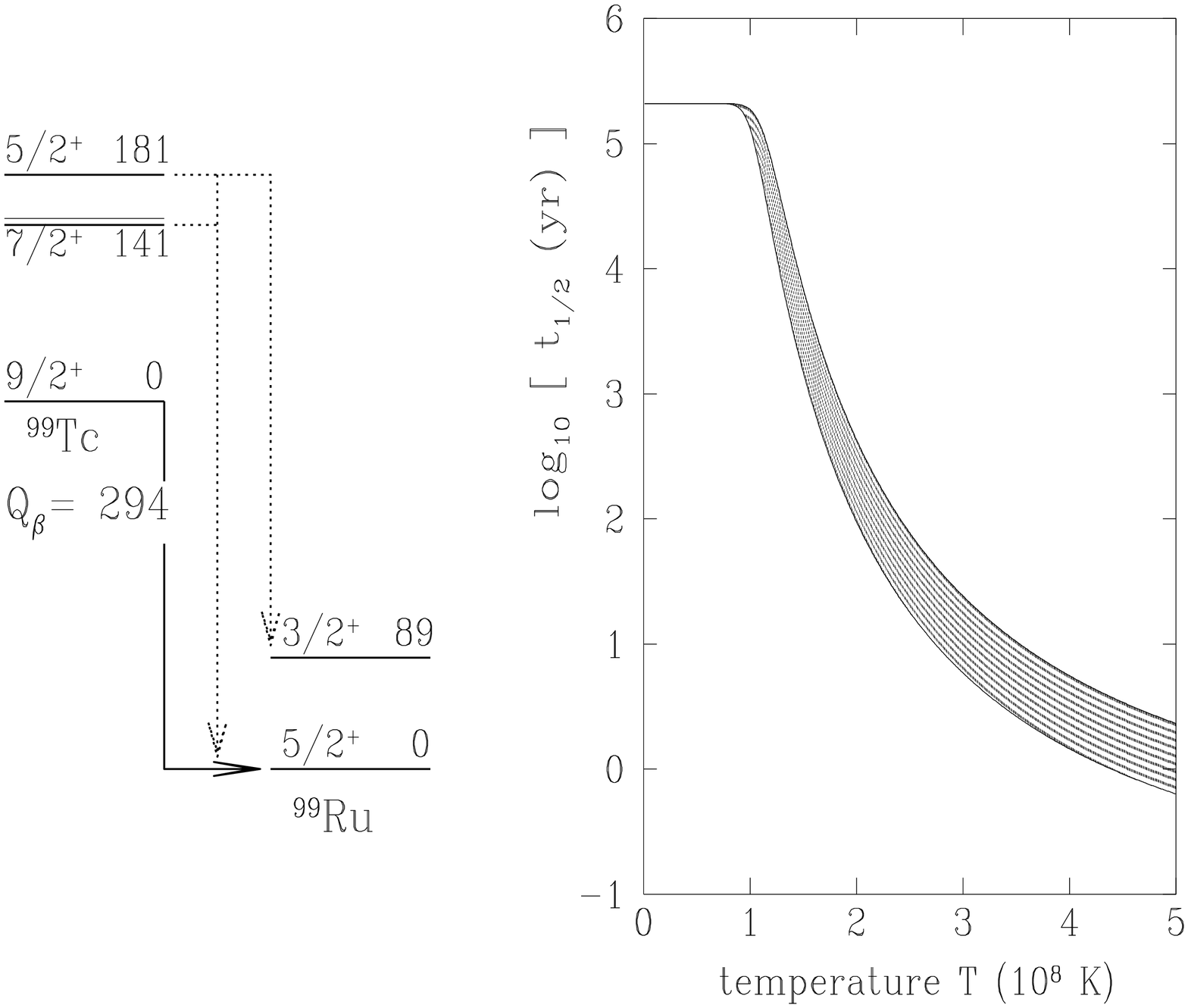}}
\vskip-0.3truecm
\caption{The low-lying levels of \chem{99}{Tc} and of the $\beta$-decay 
daughter \chem{99}{Ru} labelled by excitation energies (in keV) and 
spin-parity $J^\pi$ ({\it left panel}), and its estimated effective 
$\beta$-decay half-lives versus temperature ({\it right panel}). 
The \chem{99}{Tc} ground state decays to the \chem{99}{Ru} ground state 
through a slow ``second forbidden'' 
transition. 
The $\beta$-decay of the 143 keV isomeric state is observed to be slow 
as well. 
When thermally populated, the levels at 141 and 181 keV are expected to  
undergo the fast ``allowed'' (``Gamow-Teller'')
transitions indicated by the dotted arrows 
of the left panel. 
The effective half-lives shown on the right panel are adapted from 
\cite{Takahashi86}. 
They are obtained under the assumption of thermally equilibrated 
population of the \chem{99}{Tc} states, and from estimates of the
experimentally unknown excited-state $\beta$-transition matrix elements.
(Evaluated uncertainties translate into the displayed  shaded area.) 
In this specific example, the rates are almost density-independent
}
\end{figure}
%
This example illustrates quite vividly that the decay of 
thermally-populated 
excited states may alter laboratory half-lives most strongly in the 
following conditions: (1) the ground-state decay is slow, as a result 
of selection rules, and (2) (low-lying) excited states can decay through
less-forbidden transitions to the ground and/or excited states of the 
daughter nucleus. 
Needless to say, the temperatures have to be high enough for the relevant 
excited levels to be significantly populated.
  
\subsubsection{$\beta$-decays at high ionization}\ \ \ 
%
It has already been stressed that ionization may affect $\beta$-decay 
lifetimes in various ways. 
This is illustrated here by way of two examples. 
\vskip3truemm
\noindent {\it bound-state $\beta$-decay}\ \ \ 
A theoretical conjecture of the existence of this process goes back to 
more than 50 years ago \cite{Daudel47}, but its experimental confirmation 
had to await until quite recently (Sect.~7.2.2). 
In fact, it had already been realized in the early 1980s that 
bound-state $\beta$-decay can be responsible for the transformation of
some stable nuclides, like \chem{163}{Dy}, in hot enough 
stellar interiors 
\cite{Takahashi81,Takahashi83}. 
Subsequently the interest for this process in astrophysics has been 
growing, in particular with regard to specific aspects of the s-process,
and in relation to some cosmo-chronological studies 
(\cite{Takahashi87}; Sect.~8.2).
\vskip3truemm
\noindent{\it $\beta$-decays of radionuclides in cosmic-rays}:\ \ \ 
The cosmic-ray abundances of some radioactive nuclides can be profitably 
used for estimating the age of these high-energy particles 
(\cite{Simpson88}), or more precisely the time the cosmic rays have been 
confined within the disc, and possibly the magnetic halo of the 
Galaxy.\footnote{There is substantial evidence from the observation of
synchrotron emissions that a galaxy containing relativistic cosmic-ray 
electrons in its disc develops a magnetic halo.
The galactic cosmic rays could spend part of their confinement time 
within such halos
}  
This concerns in particular \chem{54}{Mn}, the neutral atoms of which  
are known to undergo orbital electron captures ($t_{1/2} = 
312$ d).
In high-energy cosmic rays, those orbital electrons are stripped off to 
leave bare \chem{54}{Mn}, which is expected to transform very slowly 
via $\beta^-$ and $\beta^+$-decays.  
The theoretical evaluation of the rates of these laboratory-unknown 
transitions encounters enormous difficulties \cite{Casse73}, and their
measurements are eagerly awaited. 
Recently, \chem{26}{Al} 
has been resolved from the stable \chem{27}{Al} in the cosmic radiation 
\cite{Simpson98}. 
Due to the suppression of 
its e$^-$-capture mode resulting from its complete 
ionization, the cosmic-ray  \chem{26}{Al} half-life is increased up to 
$8.7 \times 10^5$ y from the laboratory value of $7.2 \times 10^5$ y.
Another interesting case, in particular for $\gamma$-ray astronomy, 
concerns \chem{44}{Ti}. Its half-life, the laboratory value of which 
has become well known to be close to
60 y 
\cite{Ahmad98} - \cite{Fulop98}, 
may be increased in
young supernova remnants 
 because of its possibly substantial ionization
\cite{Mochizuki98}.
  
\subsection{Nuclear decays and reactions via electromagnetic interaction}
%
Nuclei immersed in a high-temperature stellar photon bath  
may be subjected to photodisintegrations 
of the ($\gamma$,n), ($\gamma$,p) 
or ($\gamma$,$\alpha$) types.
Because of the experimental (and theoretical) difficulties raised by 
the direct determination of photodisintegration rates, especially under 
the constraint that the photons obey a 
Plank distribution 
law,\footnote{This is
the case in stellar interiors, 
where local thermodynamic equilibrium holds
to a very good approximation (\cite{Cox68})}
 use is usually made
of the detailed balance theorem 
applied to the reverse radiative captures of
nucleons or $\alpha$-particles. This procedure makes clear that the
photodisintegration rates depend on temperature $T$ and on the reaction
$Q$-value as exp(-$Q/kT$). Photodisintegrations thus play a more and more
important role as the evolution of a star proceeds, i.e., as 
temperatures get
higher (Fig.~7). They start contributing to the energetics and the
nucleosynthesis  during the Ne-``burning'' stage of massive stars
(Sect.~5.2.4),  which is truly a Ne-photodisintegration 
phase dominated by 
\reac{20}{Ne}{\gamma}{\alpha}{16}{O}. 
The importance of photodisintegrations culminates at the 
supernova stage (Sect.~8.3). 
    
\subsection{Nuclear decays via strong interaction}
%
As fissions, and in particular neutron-induced 
fissions, may dictate the  nature of the heaviest nuclides that can be
synthesized in the r-process (Sect.~8.1), it is no surprise that 
they have been under active astrophysical scrutiny in relation with
the possibility of stellar production of super-heavy nuclei. Fissions,
especially of the $\beta$-delayed type (i.e. those from
levels fed by $\beta$-decays) have the
additional interest of possibly being responsible for the cycle-back to
lighter 
species of the heaviest r-process produced nuclides. This can also be
achieved by $\alpha$-decays.  Finally, the energy released by
fissions may play an important role  in some scenarios of ``quasi
r-processing,'' such as the one associated  with neutron stars
(\cite{Sumiyoshi98}; Sect.~8.1.4).

Nuclear $\alpha$-decays have been studied extensively in the laboratory. 
Save some exceptional cases \cite{Perrone71}, the available data appear 
to be sufficient for 
astrophysical purposes. The situation is quite different
regarding fission.  In particular, the fission barriers of the very
neutron-rich nuclides  
involved in the r-process cannot be measured in the
laboratory with  present techniques, and theory is the resort.

\section{Thermonuclear reactions in non-explosive events}
%
To the items of Sect. 4, one has certainly to add the knowledge of
the rates of nuclear reactions in energy domains of astrophysical 
relevance. 
This is the focus of a very intense laboratory activity. 
As we have previously noted, the rest of this review is concerned with 
thermonuclear reactions only.
For the details about spallation reactions, the readers are referred to 
the works quoted in Sect.~3.2.

The thermonuclear reactions of astrophysical interest concern mainly
the capture of nucleons or $\alpha$-particles. 
A limited number of fusion reactions involving heavy ions 
($^{12}$C, $^{16}$O) are also of great importance. 
As mentioned earlier, charged-particle induced reactions are 
essential for the energy budget of a star, as well as for the production 
of new nuclides in stellar and non-stellar (Big Bang) situations. 
In contrast, the role of neutron captures is largely restricted to 
nucleosynthesis, their energetic impact being negligeable. 
 
In non-explosive conditions, corresponding in particular to the quiescent 
phases of stellar evolution which take place at relatively low 
temperatures, most of the reactions of interest concern stable nuclides.  
Even so, the experimental determination of their cross sections and the  
evaluation of the corresponding stellar reaction rates face enormous  
problems, and represent a real challenge (\cite{Rolfs88}).
\vskip3truemm
\noindent {\it The energy region of ``almost no event''}\ \ \
The experimental difficulties relate directly to the fact that the 
energies  of astrophysical interest for charged-particle induced 
reactions are much lower than the Coulomb barrier. 
As a consequence, the cross sections can dive into the nanobarn to  
picobarn abysses. 
Thanks to their impressive skill and painstaking efforts, experimental 
physicists involved in nuclear astrophysics have been able to 
provide the  
smallest nuclear reaction cross sections ever measured in the laboratory. 
However, in very many cases, they have not succeeded yet in reaching
the region of ``almost no events'' of astrophysical relevance.
Theorists are thus requested to supply reliable extrapolations from the  
experimental data obtained at the lowest possible energies.
\vskip3truemm
\noindent{\it Electron screening corrections}\ \ \
As if the evaluation of stellar reaction rates were not complicated
enough, a whole new range of problems has opened up with the discovery 
through a series of remarkable experiments that the reaction cross 
sections measured at the lowest reachable energies are in fact 
``polluted'' by atomic or molecular effects induced by the experimental 
conditions (\cite{Rolfs93} - \cite{Angulo93}).
As a result, the situation appears even more intricate than previously 
imagined, necessitating a multi-step process in order to go from  
laboratory  data to stellar rates: before applying the usual electron  
screening corrections relevant to the stellar plasma conditions  
(\cite{Dzitko93}), it is required first to extract 
the {\it laboratory} electron screening effects from the experimental 
cross-section data in order to get the reaction  probabilities for
bare nuclei.  
In spite of heroic laboratory efforts and complementary theoretical 
modelling, much obviously remains to be done in order to get reliable 
estimates of the laboratory electron-screening factors. 
Uncertainties also remain in the evaluation of the stellar plasma  
screening, and may impede an accurate enough treatment  of some specific 
problems, as exemplified in the following section.
 
\subsection{Energy production in the Sun, and the solar neutrino problem}
%
The Sun serves as a very important test case for a variety of problems
related to stellar structure and evolution, as well as to fundamental  
physics. 
Surprisingly enough for a star that has all reasons to be considered as 
one of the dullest astrophysical objects, the Sun has been for years at  
the centre of various controversies. 
One of them is the solar neutrino problem, referring to the fact  
that the pioneering \chem{37}{Cl} neutrino-capture experiments carried  
out over the years in the Homestake gold mine observe a neutrino flux  
that is substantially smaller than the one predicted by the solar models. 
That puzzle has led to a flurry of theoretical  
activities, and to the development of new detectors (namely Kamiokande II/III,
Superkamiokande, and the SAGE and GALLEX gallium experiments). These activities have
transformed the original solar neutrino problem into {\it problems}. The relative
levels of `responsibility' of particle physics, nuclear  physics or astrophysics in
these discrepancies have been debated ever  since (\cite{Haxton95} -
\cite{Castellani97}\footnote{In \cite{Bahcall98}, the discussion is conducted in
particular in the light of the Superkamiokande experiment supporting the ideas of
`oscillations' between different neutrinos types \cite{Fukuda98}}).  
In what follows we address some questions that concern nuclear physics 
directly.

\setcounter{footnote}{0}

Much experimental and theoretical work has been devoted to the reactions 
of the p-p chains that are the main energy and neutrino producers in the 
Sun. (See \cite{NACRE98} for a compilation and evaluation of the existing
data (hereafter referred to as NACRE).)
In  spite of that, problems remain concerning the astrophysical  rates 
of some of the involved reactions.  
This is especially the case for \reac{7}{Be}{p}{\gamma}{8}{B}, which  
provides the main neutrino flux detectable by the chlorine detector, and 
is considered by some as one of the most important nuclear reactions for 
astrophysics. 
Improved low-energy data are also required for other reactions,
like $^3$He($^3$He,2p)$^4$He. They would additionally shed some light
on the question of the laboratory electron screening. 
Since such low-energy measurements are predominantly hampered by the 
cosmic-ray background, improved data could be obtained by underground 
measurements. 
This was the aim of the installation of a 50 kV accelerator at the
Gran  Sasso laboratory within the pilot Italian-German project LUNA 
(Laboratory Underground for Nuclear Astrophysics), and the motivation 
for the extension of this project to new facilities \cite{JunkerHir98}.

Another nuclear physics activity that has an important bearing on the 
solar neutrino problem concerns the calibration through (p,n) reaction 
studies (Sect.~7.2.1)
of the $\beta$-decay nuclear matrix elements involved in the 
rates of neutrino captures to excited states of \chem{71}{Ge},
 which  are responsible for the largest uncertainties 
in the analyses of the SAGE and GALLEX experiments 
\cite{Taddeucci87,Bahcall88}.

\subsection{Non-explosive stellar evolution and concomitant 
nucleosynthesis}
%
As made clear in Sect.~3, many astrophysical problems, like the 
interpretation of the surface composition of chemically peculiar stars or 
the study of the chemical evolution of the Galaxy, require the modelling 
of the evolution of stars with initial masses in the approximate 
1 to 100 M$_{\odot}$ range, as well as of the concomitant 
nucleosynthesis (\cite{Weaver93,Iben91}). 
We present here a brief summary of nuclear data typically in quest for 
studies of the controlled thermonuclear burning phases.
 
\subsubsection{Hydrogen burning}\ \ \ 
%
In addition to the p-p chains that operate in solar-type stars,
energy production by H burning  can also occur non-explosively through
 the cold CNO cycle. 
 Some hydrogen can also be consumed by the NeNa and MgAl chains
(\cite{Rolfs88}).  
These last two burning modes most likely play only a minor role in the
stellar energy budget, but are of significance in the production of the Na
to Al  isotopes, especially in massive stars.  
Most important, the MgAl chain might synthesize \chem{26}{Al}, which is a 
very interesting radio-active nuclide for $\gamma$-ray astronomy 
and cosmochemistry.

Much experimental and theoretical effort has been devoted to the
reactions involved in these burning modes, as summarized in NACRE
\cite{NACRE98}, which also provides typical uncertainties still 
affecting the relevant reactions. 
The yields from the CNO, NeNa and MgAl burning modes based on previously 
adopted rates and their uncertainties have been analyzed in 
\cite{Arnould95}, and its update
\cite{ArnouldNACRE98} based on the NACRE data.
These predictions demonstrate that better determinations of certain 
reaction rates would be desirable in order to set up meaningful 
comparisons between certain abundance predictions and observations.

\subsubsection{Helium burning and the s-process}\ \ \
%
The main reactions involved in the He-burning stage have been discussed
 in many places (\cite{Rolfs88}, NACRE \cite{NACRE98}).
Of very special and dramatic importance for the theories of stellar  
evolution and of nucleosynthesis is the famed 
\reac{12}{C}{\alpha}{\gamma}{16}{O} reaction (\cite{Weaver93}), 
which has been the subject of a flurry of experimental investigations, 
as well as of theoretical efforts (\cite{NICIV}, NACRE \cite{NACRE98};
Sect.~7.3.2). In spite of that, uncertainties
remain, 
and preclude certain nuclear  astrophysics predictions to be made at a
satisfactory level  (\cite{Weaver93}).  

Other $\alpha$-particle induced reactions are of special importance, 
and have been the subject of many dedicated experimental and theoretical 
works. 
This is particularly the case with \reac{13}{C}{\alpha}{n}{16}{O} and 
\reac{22}{Ne}{\alpha}{n}{25}{Mg} \cite{Drotleff93}. 
These reactions are considered as the main sources of neutrons for the 
s-process nucleosynthesis (Sect.~8.1.4). 
A further reduction of the remaining uncertainties of those reaction 
rates would be highly desirable, as exemplified in \cite{Meynet92}. 

The proper treatment of the s-process also requires the knowledge of a
host of neutron-capture cross sections at typical energies from about 
10 to 100 keV on targets in the whole $12 \leq A \leq 210$ mass range.  
Much dedicated experimental work has led to a substantial improvement in  
our knowledge of relevant (n,$\gamma$), as well as (n,p) and (n,$\alpha$) 
cross sections. 
However, some of them are not yet determined with the required accuracy.  
This concerns especially reactions on unstable targets close to the  
valley of nuclear stability (Sect.~7.4). 
In these cases, the knowledge of the neutron-capture cross sections has 
to be complemented by the astrophysical rates of the competing 
$\beta$-decays.
 It has also to be emphasized that the bound-state $\beta$-decay process 
may well come into play for ionized atoms, and 
significantly  
affect the production of some specific s-nuclides.
 
\subsubsection{Carbon burning}\ \ \
%
The C-burning phase raises the very interesting question of the fusion of
light heavy ions below the Coulomb barrier, and in particular of the
origin  of the very pronounced structures observed in the \chem{12}{C} + 
\chem{12}{C} fusion cross section at low energies.  
The devoted experimental and theoretical works do not provide fully  
satisfactory solutions 
(\cite{Rolfs88,ArnouldHoward76,Descouvemont89}).
 
\subsubsection{Neon, oxygen, and silicon burning}\ \ \
%
The Ne-burning phase is initiated by 
\reac{20}{Ne}{\gamma}{\alpha}{16}{O}, 
the first major energetically significant photodisintegration reaction 
experienced by a star in the course of its evolution. 
Its rate is evaluated by applying the detailed balance principle to the 
inverse  \reac{16}{O}{\alpha}{\gamma}{20}{Ne}
reaction. 
A complementary \chem{20}{Ne} destruction channel is 
\reac{20}{Ne}{\alpha}{\gamma}{24}{Mg}. 
Both $\alpha$-capture reactions have been studied experimentally and 
theoretically (\cite{Hammer98} - \cite{Descouvemont87}, 
NACRE \cite{NACRE98}).  

The \chem{16}{O}+\chem{16}{O} fusion reaction that governs the O-burning
phase does not exhibit intricacies comparable to those encountered with  
\chem{12}{C}+\chem{12}{C}.  
Yet, much has still to be learned about heavy-ion reaction mechanisms at 
low energy, and about the way to extrapolate the yields to 
the energy regions of astrophysical interest 
(\cite{ArnouldHoward76} for references).

Silicon burning comprises a very complex pattern of nuclear reactions,
and evolves into a nuclear statistical equilibrium (NSE) 
regime when the remaining Si has sufficiently low abundances
\cite{Woosley73}.   Much effort has been devoted over the years to the
measurement of 
$\alpha$- or p-capture rates of interest in that burning phase 
(\cite{Arnould92,Arnould94} for references). 
Such experiments are also important for evaluating the virtues of 
statistical model calculations (Sect.~7.5.4) that have to be used in 
order to calculate the host of unmeasured reaction rates involved in the 
Si-burning modelling.

\section{Thermonuclear reactions in explosive events}
%
In stellar (such as nova and supernova) or non-stellar (Big Bang) 
explosions, the energies of astrophysical interest are typically larger 
than in the non-explosive situations, and can be of the order of the 
Coulomb barrier. In such conditions, the relevant cross sections are 
also larger, and may range between the micro- and the millibarn.  
Unfortunately, there is a very high price to pay in order to enter that  
cross section range.
\vskip3truemm
\noindent {\it The realm of exotic nuclei}\ \ \  
The thermal neutron, proton and $\alpha$-particle bath present in 
an explosive astrophysical site is able to drive the nuclear flows to 
either very neutron-deficient or very neutron-rich nuclides. 
The precise description of these flows requires the knowledge of the 
rates of captures of neutrons,  protons or $\alpha$-particles by 
highly $\beta$-unstable nuclei. 
Except in some specific cases, those reactions have not lent themselves  
yet to a direct experimental scrutiny, so that their rates have to be  
evaluated theoretically. 
\vskip3truemm
\noindent {\it Targets in excited states}\ \ \  
The reaction rate evaluation gets even much more complicated 
as a huge variety of target states can be significantly populated in very 
hot explosive sites, and thus contribute substantially to the 
effective  reaction rates.
In such conditions, a large resort to theory is mandatory in order to 
predict these rates, even in the case of stable targets.

Another general distinctive feature of the nuclear processes in 
explosive situations relates to the fact that the burning time-scales 
are much shorter (of the order of seconds to hours) than in quiescent 
conditions. 
In fact, a given burning phase may in certain cases stop well before 
exhaustion of the reactants, in contrast to the situation prevailing in 
non-explosive stellar evolution. 
This is a manifestation of the ``freeze-out'' of the charged-particle 
induced reactions. 
This regime develops when the typical expansion time-scales of the 
explosively processed material become shorter than mean lifetimes of 
nuclei against charged-particle captures. 
These nuclear time-scales
 indeed increase dramatically with the decreasing 
temperatures in an expanding material as a result of the decrease of the 
Coulomb-barrier  transmission probability, which in turn owes to the 
decrease of the mean relative energies of the interacting particles. 

The corresponding explosive nucleosynthesis has in many 
instances been studied in the framework of simplistic parametrized 
astrophysical models which allow avoiding the many intricacies of the 
(often unknown) explosion characteristics. 
The approximations are sometimes going so far as to adopt constant 
temperatures 
and densities for a given time-scale, with the presumed values of 
these quantities, as well as of the 
initial composition of the material to 
be processed, being just ``inspired" by more detailed models (if they 
indeed exist !). 
In contrast, nuclear reactions are followed by detailed networks (see 
\cite{Arnett96} Chap.~9). 
This approach, hybrid in the degrees of sophistication, has certainly 
some virtues. 
In particular, it provides clues to the general characteristics of 
a burning mode, and possibly some identification of the nuclear input of 
importance. 
It has also obvious shortcomings, and there is a danger for the 
conclusions drawn from such an approach to be over-interpreted in its 
astrophysics or nuclear physics aspects.
 
\subsection{Big Bang nucleosynthesis}
%
As reviewed in great detail by \cite{Smith93} 
(also NACRE \cite{NACRE98}), 
the dedicated experimental and theoretical efforts 
for the determination of
the rates of the  reactions involved in the standard Big Bang 
nucleosynthesis model have succeeded in putting on a remarkably safe 
nuclear footing the conclusion that such a model is able to account 
for the pre-galactic abundances of the nuclides D, \chem{3}{He}, 
\chem{4}{He} and \chem{7}{Li} derived from various astrophysical 
observations and from models for the chemical evolution of galaxies.
A quite limited number of reactions, however, still suffer from some  
uncertainties. Even if they are relatively low, those uncertainties could 
have some impact on the yield predictions, and consequently on the 
acknowledged virtues of the standard Big Bang model.

Various non-standard Big Bang models, and in particular one invoking some  
inhomogeneities at the nucleosynthesis epoch, have raised much
excitement in a fraction of the nuclear astrophysics community
(\cite{Smith93,Orito97}).
These models are indeed calling for many reactions that are 
not involved in the standard model, and whose rates are largely 
unknown. 
Of course, the interest of pursuing specific nuclear studies in 
relation with the inhomogeneous Big Bang clearly depends on the outcome
of the investigation  concerning the validity of that model, which 
is far from being fully  ascertained yet (\cite{Sarkar96,Malaney93}).
 
\subsection{The hot modes of hydrogen burning}
%
Hydrogen can  burn explosively in various 
astrophysical events, like novae or
X-ray bursts. The corresponding 
``hot'' burning modes involve a
large variety of  unstable nuclei.  
These modes have specific nucleosynthesis
signatures, and raise many  difficult 
experimental and  theoretical astro- and
nuclear-physics  problems.  
 
\subsubsection{The hot p-p mode}\ \ \
%
This mode, as first recognized by \cite{Arnould75}, could in particular 
develop in nova explosions resulting from  the accretion on a white dwarf 
of material from a companion star in a  binary system. 
A variety of reactions of importance have been identified. 
One of the keys of this type of burning is the 
\reac{8}{B}{\gamma}{p}{7}{Be} photodisintegration.
This reaction impedes the transformation of \chem{7}{Be} into \chem{4}{He}
which characterizes the cold p-p chain, and may thus be responsible for 
some \chem{7}{Li} production (through the \chem{7}{Be} decay) in nova
 situations (\cite{Gehrz98}).

Several of the hot p-p reaction rates have been scrutinized both 
theoretically and with the help of indirect experimental techniques. 
Let us note in particular that the rate of transformation through
\reac{11}{C}{p}{\gamma}{12}{N} of the \chem{11}{C} which could be 
produced by \reac{7}{Be}{\alpha}{\gamma}{11}{C} has been measured in 
a Coulomb break-up experiment (Sect.~7.3.2),  the results of which differ 
quite substantially from the predictions of  a microscopic model
(Sect.~7.5.1). 
 Such a model has also been used to predict the rate of  
 \reac{8}{B}{p}{\gamma}{9}{C}, which might also have some importance 
in the nova scenario \cite{Boffin93}.
 
\subsubsection{The hot CNO and NeNa-MgAl chains}\ \ \
%
These hot H-burning modes develop when some produced
$\beta$-unstable nuclei decay more slowly than they capture protons, 
which is in contrast to the situation characterizing the corresponding 
cold burnings. 
The high temperatures encountered in explosive situations are again 
demanded. 
This request relates directly to the fact that the $\beta$-decays of 
the involved  light nuclei are essentially temperature-independent 
(in view of the paucity of  excited states that can be significantly 
populated at the temperatures of relevance for these processes), while 
the proton capture rates increase dramatically with increasing
temperatures (as a result of the larger Coulomb-barrier penetrability). 

More specifically, the cold CNO switches to the hot CNO mode when 
\reac{13}{N}{p}{\gamma}{14}{O} becomes faster than the \chem{13}{N} 
$\beta$-decay.
This occurs typically at temperatures in excess of $10^8$ K. 
The cold NeNa and MgAl chains evolve into a hot mode at
temperatures that are quite similar to the operating  conditions for the
hot CNO. 
In fact, novae could be favourable sites for the development of both
the  hot CNO and NeNa-MgAl chains.

As an immediate consequence of the definition of the hot H burning, 
many reaction rates on unstable nuclei come into play. 
Much theoretical and experimental effort has been devoted to a reliable  
determination of the rates of some of the reactions that have been 
identified as keys in the development of those processes. 
In general, 
those rates have not been measured directly (Sect.~7.3.1), and are
rather  evaluated indirectly (Sect.~7.3.2).
There are, however, some noticeable exceptions to this situation. 
The first one concerns \reac{13}{N}{p}{\gamma}{14}{O}. 
>From its direct measurement (which agrees with indirect determinations),  
it is concluded that the  
\reac{13}{N}{p}{\gamma}{14}{O} rate is now known well enough for practical
astrophysical purposes \cite{Arnould92a}. Direct experiments have also 
been conducted on \reac{22}{Na}{p}{\gamma}{23}{Mg} and  
\reac{26}{Al^g}{p}{\gamma}{27}{Si}, which are important 
destruction channels of the 
\chem{22}{Na} and 
\chem{26}{Al^g} radionuclides of astrophysical importance. 
These experimental efforts have succeeded in reducing drastically the 
nuclear physics uncertainties affecting the explosive (nova) 
\chem{22}{Na} and \chem{26}{Al} yields, as can be seen from a 
comparison between the present situation (\cite{Coc98,Coc95}) 
and the one prevailing in the mid-1980s.  
Note that large uncertainties still remain in the 
\reac{26}{Al^g}{p}{\gamma}{27}{Si} rate at lower temperatures 
characteristic of quiescent stellar evolution phases. 
Those uncertainties, however, do not have a significant impact on the 
corresponding \chem{26}{Al} yield predictions, as the \chem{26}{Al} 
$\beta$-decay is likely to be the main destruction channel in those 
conditions (as an illustration of this claim, see \cite{Meynet97} for a
calculation of the \chem{26}{Al} yields from non-exploding massive 
stars of the Wolf-Rayet type).  
On the other hand, in conditions where the \chem{26}{Al^m} isomeric  
state can have a significant thermal population, the  contribution of 
the proton capture by \chem{26}{Al^m} to the net stellar 
\chem{26}{Al} + p rate has to be taken into account. 
At present, this evaluation merely relies on a global statistical model 
calculation (NACRE \cite{NACRE98}; Sect~7.5.4). 
Its direct experimental determination would require the development of a
\chem{26}{Al^m} beam, which obviously represents an interesting 
 technological challenge.

\subsubsection{The rp- and $\alpha$p-processes}
%
The hot CNO and NeNa-MgAl modes can transform into the so-called rp- or
$\alpha$p-processes when \reac{15}{O}{\alpha}{\gamma}{19}{Ne} or
\reac{14}{O}{\alpha}{p}{17}{F} become more rapid than the corresponding
$\beta$-decays. 
Alternatively, \reac{18}{Ne}{\alpha}{p}{21}{Na} could play this role if 
its still-uncertain rate can indeed become faster than the \chem{18}{Ne} 
$\beta$-decay in appropriate astrophysical  conditions, which could be
encountered in certain Type-I supernovae, or in X-ray bursters resulting 
from the accretion of matter on a neutron star.  
In such conditions, it is expected that the nuclear flow can go all the  
way from the C-N-O region up to, or even slightly beyond, the iron peak  
through a chain of proton captures (and their inverse) and 
$\beta^+$-decays. 
This chain of transformations is termed the rp-process 
(\cite{Vanwormer94,Schatz98}). 
It could transform into an $\alpha$p-process at temperatures that are 
high enough for $\rm (\alpha\,,p)$ reactions to play a leading role by
 bypassing the proton-capture + $\beta$-decay chains. 
It is sometimes speculated that the rp-process can even reach the 
vicinity of the $A \approx 100$ mass range \cite{Schatz98}.

A host of reactions on unstable neutron-deficient nuclei, some of 
them being close to the proton drip line, are involved in the rp- or
$\alpha$p-process (a few already partake in the hot CNO 
and NeNa-MgAl chains).  
A direct measurement of a significant fraction of all those potentially
important reactions is difficult to conceive. 
To-date, and apart from the experiments on 
\reac{13}{N}{p}{\gamma}{14}{O} mentioned earlier, direct measurements 
have been conducted only for \reac{18}{F}{p}{\alpha}{15}{O} and
\reac{19}{Ne}{p}{\gamma}{20}{Na}, while preliminary results exist for 
\reac{18}{Ne}{\alpha}{p}{21}{Na} rate (Sect.~7.3.1). 
The \reac{18}{F}{p}{\alpha}{15}{O} reaction is an efficient path in the transformation
of  initially present \chem{12}{C} and \chem{16}{O} into \chem{15}{O}, while 
the \chem{19}{Ne} and possibly \chem{18}{Ne} destructions take
part in the breakout from  the C-N-O region. 
In fact, much of what is known on reaction rates involved in the  
rp-process has been derived from a variety of indirect techniques.

As a necessary complement to these experimental efforts, use is made of 
a global statistical model, which appears to be adequate 
except for reactions with low $Q$-values implying relatively low nuclear 
level densities.
Reaction networks appropriate to the modelling of the rp- or 
$\alpha$p-process also require the evaluation of $\beta^+$-decay rates 
(and possibly also of  continuum-e$^-$ capture rates if indeed high 
enough electron degeneracies are prevailing).

Obviously, much remains to be done in order to put the rp- and  
$\alpha$p-processes on a safer nuclear footing. 
The proper target of these efforts, and especially the reliable 
identification of reactions which have to be  painstakingly measured in 
the laboratory, is by far not a trivial matter. 
For example, no realistic astrophysical models have ever been used to 
analyze in a careful and detailed way the impact of reaction rate 
uncertainties on the observable properties of the highly complex 
objects, like X-ray bursts (\cite{Taam98}), where the rp- or 
$\alpha$p-process is expected to develop. 
Rather, the identification of nuclides and reactions of putative 
importance relies on grossly simplified and schematic models
(\cite{Vanwormer94,Schatz98}). 
In addition, the rates of some reactions  claimed thereby to be 
important are often evaluated in a quite approximate way. 
All this might lead to some misleading evaluations of the true 
importance of some nuclear data. 
As illustrative examples, let us pick the proton captures on 
\chem{27}{Si}, \chem{31}{S}, \chem{35}{Ar} and \chem{39}{Ca}. 
They are considered as key reactions for the rp-process by 
\cite{Vanwormer94}, while \cite{Iliadis98} find no compelling reason for
attempting to measure the corresponding rates using radioactive ion beams.
Thus, nuclear physicists may find it desirable to be fully aware of this 
type of situations before embarking on very difficult experiments 
\cite{Arnould95e}.

\subsection{The He to Si explosive burnings}
%
The explosive combustion of He, C, Ne, O and Si can also develop
in a variety of situations, like  the explosions in massive single stars, 
the accretion, or even merging, of white 
dwarfs in binary systems, as well as
possibly the accretion of matter at the surface of neutron stars. 
As in the case of explosive H burning, these hot burning modes involve 
a variety of reactions on unstable targets that do not play any 
significant role in the  corresponding cold combustions.
 
These various explosive burnings and the consequent 
``yields'' of
nuclides 
 have been studied in either parametrized or  
more realistic explosion models
(\cite{Arnett96,Nomoto97,WoosleyWeaver95}).

\subsection{The $\alpha$-process and the r-process}
%
The various non-explosive or 
explosive burning modes mentioned above cannot account for the 
synthesis of the neutron-rich $A > 60$ nuclides (called r-nuclides) at 
a level compatible with their bulk solar-system abundances.
 A specific mechanism, referred to as the r-process, is thus called
for. It relies on a chain of captures of neutrons whose concentrations 
are by far higher than in the s-process (Sect.~8.1). 
As this requirement is clearly impossible to meet 
in quiescent evolutionary 
phases, deep supernova layers in the vicinity of a forming neutron star 
residue have been quite naturally envisioned as a possible 
r-process site. 
However, for decades, it has not been possible to substantiate this 
connection on grounds of detailed supernova models.
 
In this respect, recent progress in the modelling of Type-II supernovae 
has raised a lot of excitement and hope in that the r-process might 
happen in the so-called ``hot bubble'' region created by neutrino 
heating at the periphery of a nascent neutron star (Sect.~8.3).
The hot bubble consists of a rapidly
expanding matter with high entropy and with a more or less   
significant neutron excess. At short times, or at temperatures of about 
$(10 \sim 7) \times 10^9$ K, the bubble composition is
determined by NSE favouring $\alpha$-particles and some neutrons.
As the temperature decreases further along with the expansion, the
$\alpha$-particles recombine to form heavier nuclides, starting
with the $\alpha\alpha$n$\rightarrow$$^9$Be($\alpha$,n)$^{12}$C 
reaction. It is followed by a complex sequence of $\alpha$-particle and
nucleon  captures [especially of the ($\alpha$,n), (p,n) and (n,$\gamma$)
type],  and of the  inverse transformations, 
synthesizing heavy nuclei even 
beyond Fe. This is the so-called 
``$\alpha$-process''\footnote{This must not be confused with the same 
term used by \cite{BBFH57}, which is now referred to as the ``Ne burning'' 
}
\cite{Woosley92, Witti94}.
If enough neutrons would be left at the freeze-out of the $\alpha$-process
at a temperature slightly in excess of $10^9$ K, the r-process could
start. 

In addition to the astrophysics questions, the $\alpha$- and r-processes 
raise major nuclear physics problems. They indeed involve a wealth of
very neutron-rich $A \gsimeq 12$  nuclei whose properties are 
little known
experimentally, 
a large fraction of them even remaining to be produced in the
laboratory. In particular, masses, $\beta$-decay and neutron
capture rates have in most instances to be estimated theoretically. 
The impact of the related uncertainties on abundance predictions has been
discussed in many places (a few of them are quoted in Sect.~7).
Neutron-induced fission, as well as $\beta$-delayed 
neutron emission or fission may also bring more uncertainties in the
calculated r-process yields, 
and even in the very energetics of the process.
These additional difficulties relate directly to our poor knowledge of
fission  barriers of very neutron-rich actinides.  
  
\subsection{The p-process}
%
It was realized very early in the development of the theory of
nucleosynthesis that the production of the stable heavy 
neutron-deficient nuclides requires a special mechanism, termed the 
p-process. 
It has by now become quite clear 
that this process can develop if pre-existing 
more neutron-rich (s- or r-) species can experience high enough 
temperatures ($T \gsimeq 2 \times 10^9$ K) for being photodisintegrated 
by more or less complicated sequences of ($\gamma$,n), ($\gamma$,p) and 
($\gamma$,$\alpha$) reactions. 
Such requirements are met in the O/Ne-rich layers of massive stars during 
their  pre-explosion or supernova phase (\cite{Arnould98}), as well as 
possibly in the  material accreted at the surface of an exploding white 
dwarf (\cite{Howard92}). The latter possibility remains, however, to be scrutinized
in the framework of detailed stellar models. 
   
The main nuclear physics uncertainties that affect the modelling of the
p-process concern the involved nucleon or $\alpha$-particle captures by 
more or less neutron-deficient nuclides, as well as the relevant 
photodisintegrations. 
Except for few cases (\cite{Fulop96} -
\cite{Somorjai98}), no experimental data are available, and resort is
classically made to statistical model predictions.

If it is not free from difficulties,
the astrophysical site of the p-process may be considered
as better known (at least in the limits of spherically-symmetric model 
stars)
than the (sometimes still putative) locations where the rp-, $\alpha$p- or
r-processes may develop. 
In addition, the p-process deals with nuclei that are closer to the line
of stability than those involved in 
the other mentioned mechanisms, and at least 
some of the basic 
properties (like
masses and $\beta$-decay half-lives) of a substantial fraction 
of them have
been measured. 
  
\section{Nuclear data acquisition for astrophysics}
%
We now turn to the on-going experimental 
and theoretical efforts devoted to
the acquisition of the nuclear data needed in a variety of
astrophysical problems. After a brief description of the 
advances concerning
nuclear binding and
$\beta$-decay properties, the current status of our knowledge of nuclear
reaction rates is discussed in some detail.
 
\subsection{Nuclear binding} 
%
Astrophysics requires approaching the question
of the nuclear binding energies either ``traditionally'' through  
experiments or mass models concerned with isolated nuclei, or through a
nuclear equation of state when a high-temperature and high-density 
plasma has to be dealt with.
 
In the last decade, much effort has been devoted to the measurement of
nuclear masses 
(\cite{Audi97}) by ever-improving and 
innovative techniques. 
In particular, the development of radioactive beam
facilities allows an increasing variety of proton- or 
neutron-rich nuclei to be
studied (\cite{RIB92} - \cite{Aberg97}), and
this tendency will certainly develop further. The accuracy of the
measurements is also significantly improving, and may now reach the 10 keV
level, even for nuclei relatively far from the line of stability. Roughly
speaking, 
these experiments involve high-precision  direct mass measurements
using high-resolution spectrometers, or indirect measurements based
on the study of the energetics of a nuclear transformation from 
which an unknown mass can be deduced from the knowledge of the other
participating  nuclei (\cite{Mittig97}). Despite the experimental
advances, 
many masses remain to be measured  in order to meet the astrophysics
needs, so that recourse has to be made to theory.

\subsubsection{Nuclear mass models}\ \ \ 
%
Much theoretical progress has been made through  the continued
sophistication of the Droplet Model, the most matured version of which
is the so-called ``Finite Range Droplet Model (FRDM)''
 \cite{Moeller95}, and
through the development of a global ``microscopic'' description of
the mass surface that approximates the  Hartree-Fock predictions. This
approach is referred to as the ``Extended Thomas-Fermi plus Strutinski
 Integral
(ETFSI)'' \cite{Aboussir95}. 
 
When compared with more sophisticated methods, those global models
are favoured for nucleosynthesis calculations that require unknown
nuclear masses.
For example, calculations of the Hartree-Fock-Bogoliubov (HFB) type
with the use of zero-range (``Skyrme'') or finite-range (``Gogny'')
effective forces have not been able to reproduce the measured masses
with the accuracies easily reached by those global models.
Because of their complexities, in addition, such computations so far
have been limited only to some selected sets of nuclei.
The situation is similar for the ``Relativistic Mean Field (RMF)''
 models,  
although systematic mass calculations appear less cumbersome.
(See \cite{Patyk97} for comparisons of those various methods and
their results.)
 
Despite these shortcomings, further calculations based on the HFB or
RMF methods 
are certainly of high value in
the continuous attempt 
to improve the FRDM and ETFSI predictions meant for 
astrophysical applications \cite{Pearson98}. 
Such improvements would certainly be of great value, given the sometimes large
discrepancies between the predictions  of the FRDM and ETFSI
models for the masses of nuclei very far from the line of nuclear
stability (\cite{Goriely93}). 
The question of the evolution of the magnitude
of  the  nuclear shell effects with neutron excess, as well as of the 
``magicity'' of certain neutron numbers itself \cite{Dobaczewski94}
has also raised some excitement
and debate (\cite{Kratz95,Pearson96}), and remains to be
settled by more nuclear physics investigations.

\subsubsection{Nuclear equation of state 
at high temperatures/densities}\ \ \ 
%
In supernovae or neutron stars, the thermodynamic 
conditions and neutron
richness of the material are
such that it appears difficult to extract useful nuclear
information from e.g. heavy-ion collision experiments. 
Resort has thus to be
made to theory (\cite{Hillebrandt91,Vautherin94}). 
 
At temperatures in excess of about $5 \times 10^9$ K, NSE is 
establishment to
a high level of accuracy, and the nuclear EOS can be appropriately 
calculated from statistical mechanics 
(``Saha equation"),
at least if the densities do not  exceed about one hundredth of 
the nuclear
matter density. The task is then reduced to the evaluation of 
binding energies
and nuclear partition functions, which 
is far from being
trivial as very neutron-rich nuclei may be involved.
 
At higher, subnuclear densities, the Coulomb (lattice) 
correlations have to be
properly taken into account. This has been attempted through a
temperature-dependent Hartree-Fock method 
\cite{Hillebrandt84}, or with the Thomas-Fermi (TF) approximation
(combined with the RMF theory)
\cite{Shen98}. The TF
approximations have often been used for simplicity in studies
of complex geometrical configurations that are expected  
just below the saturation density
(\cite{Ravenhall83,Laussaut87}).
 
The superfluidity that characterizes the crust and outer core of a 
neutron star has been investigated with the help of realistic 
nucleon-nucleon forces (\cite{Tamagaki93}).
Such studies may help understanding some puzzling phenomena, like the
sometimes spectacular changes in rotational period 
(``glitches") of certain
pulsars (\cite{Tamagaki93} - \cite{Mochizuki97}).
 
Beyond the saturation 
density, many microscopic calculations of the EOS have
been performed (\cite{Vautherin94}). 
As yet, many of them are concerned
with neutron or symmetric nuclear matter 
at zero temperature, and thus cannot
be applied directly to the  supernova or hot neutron star problems.
Similarly microscopic methods have been used 
to study the pion condensation
which is expected to occur in the inner core of a neutron star at a few 
times the nuclear density (\cite{Tamagaki93}).
The RMF model has been applied for studies of the EOS 
at even higher densities with the inclusion of various hyperon 
and lepton degrees of freedom (\cite{Glendenning97}).
 
Finally, a transition from the hadronic matter to a quark-gluon plasma is
predicted by some QCD lattice  calculations to occur at the extremely high
temperatures and/or densities  that could be 
reached in the innermost core of
a neutron star (\cite{Petersson91}).
Similar conditions might have prevailed in the early Universe, with some
nuclear astrophysics consequences if indeed the outcome has been the
persistence of inhomogeneities at the epoch of Big Bang nucleosynthesis.

\subsection{Nuclear decay properties} 
Of the various nuclear decay modes, $\beta$-decay processes enter
astrophysical problems most importantly.
A difficulty stems from the fact that, in many instances, 
$\beta$-decay strength functions have to be known in an energy 
range that
is out of reach of standard $\beta$-decay experiments.
Also, many 
$\beta$-decay processes may take place at temperatures 
that are inaccessible by
laboratory simulations.  
 
\subsubsection{Beta-decay half-lives and strength functions}\ \ \
%
As in the case of nuclear masses, much progress has been
made in the experimental determination of $\beta$-decay half-lives and
strength functions. 
In spite of this, a large variety of data required by
astrophysics are still lacking, 
and one has to rely heavily on theory. 
This concerns in particular the $\beta^-$-decays of very 
neutron-rich nuclei 
involved in the r-process,
or the captures of highly-degenerate free electrons by protons or 
iron-group nuclides, which play an important role in the late evolutionary
stages and final fate of a variety of stars (\cite{Aufderheide94}).
The latter
(laboratory-unknown) process involves $\beta^+$-strength functions, particularly of the
Gamow-Teller type, in a wide range of energies.
Such data can be gained from the analysis of
charge-exchange (n,p) reactions, but this approach has been applied so far in a limited
number of cases only (\cite{ElKateb94,Williams95}). 
 
As for $\beta$-decay properties of highly unstable nuclei,
the development
of radioactive beam facilities and of highly efficient detectors
(\cite{RIB92} - \cite{Aberg97}) has been quite beneficial. This is exemplified by 
the determination of the half-lives of the magic or near-magic nuclei \chem{130}{Cd},
\chem{79}{Cu} and of their  neighbours taking part in the r-process
\cite{Kratz86,Kratz91}, which has been made possible 
with the development at ISOLDE/CERN of the target technology and of a
highly efficient neutron-detection method.
The use of the LISE spectrometer at GANIL in combination with a very fast
in-flight separation technique  and of the neutron counter  developed at
ISOLDE has led to  the measurement of the $\beta$-decay 
properties of \chem{44}{S} and 
\chem{45-47}{Cl} \cite{Sorlin93}, which are of astrophysical interest.
 The $\beta$-decay
half-lives of neutron-rich Fe, Co, Ni and Cu isotopes of r-process
relevance have also been obtained following their production by 
neutron-induced fission at the Grenoble ILL high-flux
reactor \cite{Bernas92}. Measurements concerned with very
neutron-rich nuclei have also been performed at 
GSI Darmstadt following their
production in relativistic projectile fission (\cite{Bernas97}) or in
fragmentation reactions \cite{Ameil98}. On the proton-rich side,
experiments   conducted at several facilities 
(GANIL, GSI, MSU and RIKEN) have
helped clarifying the location of the proton drip line, 
and have provided an
ensemble of $\beta$-decay rates of relevance to the rp- 
or $\alpha$p-processes (\cite{Roeckl98}). 

On the theoretical side, the evaluation of $\beta$-decay rates raises
problems, in particular because this integrated nuclear quantity is highly
sensitive to the energy distribution of a small fraction
of the  sum rule
for the Gamow-Teller transition strength which is used up in the
$\beta$-decay energy window of a given unstable nuclide. In fact, a major part
of the sum rule is exhausted by the ``Gamow-Teller giant resonance.''  The early
prediction of this situation \cite{Ikeda63} has been confirmed by many (p,n) and
some ($^3$He,t) experiments  (\cite{Takahashi84,Takahashi92} for a retrospective).
 
The first attempt to consider the sum
rule in predicting the half-lives  of very unstable nuclides was of
statistical nature, leading to the so-called  
``Gross Theory''
(\cite{Takahashi73}). 
Various modifications to the original model have
been proposed (\cite{Tachibana98}).
From a microscopic point of view, 
large-scale shell-model calculations are required. Modern shell model diagonalization
techniques
(\cite{Martinez98}) using a huge, but still truncated, configuration space
make it possible to deal with finite-temperature continuum-e$^-$  captures by
pf-shell nuclei of interest in the late stages of the evolution of
 massive stars. 
A numerical approach of the
nuclear shell model, referred to as the ``shell model Monte Carlo" (\cite{Koonin97})
employs an even larger model space allowing the calculation of the $\beta$-decay
rates for nuclei with $N$ and $Z$ in the vicinity of 28 up to \chem{63}{Co}. However,
the method has a limited applicability to odd-$A$, and even more so to odd-odd nuclei.

A microscopic approach which makes feasible large-scale calculations of the
$\beta$-decay properties of heavy nuclei is based on the so-called ``quasi-particle
random-phase approximation (QRPA).''
 This method  has recently been adopted in conjunction with
mean field models meant for systematic calculations of 
nuclear masses, and in particular with the FRDM and ETFSI models 
(\cite{Moeller97,Borzov98}). 

In spite of these recent efforts, many difficulties obviously remain. 
They
take the form of discrepancies between the predictions of the 
existing models,
and deviations of all models with respect to one or another new
 measurement,
especially near neutron-shell closures, as made clear by the
 NUBASE
compilation \cite{Audi97}. 
Following recent experiments, a pessimistic view is 
 expressed 
that ``the large discrepancies found between the
measured and the theoretical values emphasize that most recent theoretical
work is not an improvement over calculations made almost a decade ago''
 \cite{Ameil98}.

The situation may look even more uncomfortable when dealing with the   
$\beta$-delayed neutron-emissions and fissions that have to 
be included in the detailed modelling of the r-process. The
probabilities of these processes are little known experimentally, 
with laboratory data
lacking also for the fission barriers of the neutron-rich actinides
of interest. 
In these conditions, the uncertainties in the $\beta$-decay models
combine to those in the barrier evaluations. 
There is some hope of improvement
in the reliability of the latter predictions 
with the use of the ETFSI
model. The results \cite{Pearson98} appear, at least, to
reproduce existing fission barrier data fairly well.
 
\subsubsection{Bound-state $\beta^-$ decays}\ \ \
%
Bound-state $\beta$-decay can drastically affect the half-lives in 
extreme cases where the $\beta$-decay $Q$-values are small enough to be 
substantially affected by the ionization (\cite{Takahashi87a}).
 Such an effect can even lead to the decay of terrestrially
stable nuclides.
This is notably the case for \chem{163}{Dy}.
A storage-ring
experiment at GSI has confirmed that the half-life of
fully-ionized \chem{163}{Dy}$^{+66}$ is 47 d \cite{Jung92},
in excellent agreement with the theoretical prediction of 50 d
\cite{Takahashi87a}. 
Bound-state $\beta$-decay can also dramatically modify
the terrestrial  half-lives of \chem{187}{Re} and of \chem{205}{Pb}.
 It has recently been 
confirmed  by another GSI storage-ring experiment that
the fully-ionized \chem{187}{Re^{+75}} has indeed a half-life of 34 y
\cite{Bosch96},
which is more than $10^9$ times shorter than the value for the 
neutral atoms,
and in fair agreement with the theoretical prediction of 12 y
\cite{Takahashi87a}. Several studies have enlightened the astrophysical
importance of those  
three nuclides and of their bound-state $\beta$-decay.
In particular, that mechanism is expected to have a considerable impact 
on the \chem{187}{Re} - \chem{187}{Os} cosmochronology (Sect.8.2.3).
On the other hand, bound-state $\beta$-decay may be used as a tool to 
determine unknown $\beta$-decay matrix elements influencing the design of 
a \chem{205}{Tl} neutrino detector 
\cite{Freedman88,Kienle88}, and perhaps even to set 
some meaningful limits on the electron neutrino mass 
\cite{Jung92,Cohen87}.
 
Finally, it has to be stressed that the evaluation of the bound-state
$\beta$-decay rate in stellar plasmas is far from being straightforward,
even if the relevant nuclear matrix elements are known. 
The problems relate in particular  to electron screening effects,
which have been estimated so far from a finite-temperature Thomas-Fermi 
model \cite{Takahashi83}.
  
\subsection{Charged-particle induced reactions: Experiments}
%
We have already stressed that much dedicated and
heroic effort has been devoted to the measurement of the rates of a 
wealth of thermonuclear reactions in order to 
put the astrophysical models on a safer footing.
In many instances, such an experimental activity has been the trigger of
new and exciting technological or physical ideas.
The difficulty of providing data in quest and the vast
diversity of the problems to be tackled have always made it 
unavoidable to 
use the most sophisticated experimental techniques of nuclear physics, or
even to develop novel approaches. 
In that adventure, practically all types of accelerators have been used, 
from the electrostatic ones delivering energies in the few keV 
range to high energy heavy-ion accelerators.
 
As also stressed earlier, the problems that experimentalists are 
facing when they try to measure cross sections for astrophysics (mostly 
proton- and $\alpha$-particle-induced reactions, a substantial fraction
of them being of the radiative capture type) are of different natures, 
depending in particular on the non-explosive or 
explosive character of the 
sites where the considered thermonuclear reactions are expected to take 
place. 
More specifically, non-explosive conditions necessitate the knowledge of 
extremely small cross sections inferred 
by the relevant very low energies. 
Except in some cases, existing techniques have been able to provide 
measurements only at higher energies, so that a theoretical guide is
required to extrapolate the data down to the energies of astrophysical 
interest.
On the other hand, explosive situations make the energy problem less 
acute, but very often require cross sections to be known on unstable 
species.
 
In order to obtain thermonuclear
reaction rates,
various experimental techniques have been used,
which can be classified into ``direct'' 
and ``indirect.'' 
 
\subsubsection{Direct cross-section measurements}\ \ \ 
%
Direct measurements concern reactions that really take place in 
astrophysical sites. 
Strictly speaking, they would also have to be conducted at the stellar 
energies (referred to in the following as the ``Gamow window''). 
This is very seldom the case, especially in non-explosive situations, 
in view of the very low energies involved.
 
Direct methods have been, and still are, widely utilized in the case 
of stable targets, all efforts being directed towards the development of
techniques permitting to reach smaller and
smaller cross sections and/or 
higher and higher accuracies (\cite{Rolfs88,Greife94}). 
Typically, use is made of a dedicated accelerator delivering for several 
weeks low-energy ion beams of high intensity (1 mA) on a target that is  
able to withstand the heavy beam load (hundreds of watts), and that is  
also of high chemical and isotopic purity. 
A few per mil atoms of  impurity can indeed be responsible for a noise 
exceeding the expected  signal.  
In the case of the commonly used inverse kinematics geometry, a
heavy-ion  accelerator is often used in conjunction with a windowless gas
target of  the static or supersonic jet type.
Detectors have generally been the same as those used in classical
nuclear  physics. 
This will probably remain true in the future. 
In particular, new generations of detector systems and pulse-processing 
electronics that have primarily been developed for nuclear structure 
studies  will certainly be most welcome in the attempt to measure 
sub-picobarn cross sections (Chap.~3.5 of \cite{Kaeppeler98}). 
However, in a few instances, new detector types (e.g.~a D$_2$O detector 
\cite{Rolfs88}) have been developed specifically for nuclear astrophysics 
experiments. 
 
It has to be emphasized that the particular conditions of astrophysical
interest require special considerations that are not encountered in
ordinary nuclear physics. 
This concerns namely the necessity of frequent checks of the purity and 
stochiometry of the targets, the beam intensity determination with 
calorimeters, or the cosmic-ray shielding. 
This last requirement becomes of major importance as the experiments get  
more and more performing as to be able to provide good quality data at  
lower and lower energies. As the cosmic-ray background may become a  
deterrent limiting factor, underground experiments have started to be
conducted with a 50 kV accelerator installed in the Gran Sasso Laboratory.
Remarkable results have already been obtained for the 
$^3$He($^3$He,2p)$^4$He cross sections down to centre-of-mass energies
of about 20 keV, which are right within the Gamow window corresponding 
to the central regions of the Sun \cite{Junker98}. 
Plans exist to extent this underground facility in order to conduct very 
low energy measurements of other reactions involved in the various cold 
H-burning modes \cite{JunkerHir98}. 
Another type of background of beam-induced origin may also hamper the 
measurement of very low energy cross sections. 
High-granularity detectors will certainly be of great help in limiting 
the inconvenience of this type of noise (\cite{Nolan94,Lutz96}). 

It is impossible to cite here the myriad of direct cross-section
 measurements performed for
astrophysical purposes. 
The NACRE compilation \cite{NACRE98} may give the reader a flavour of
the volume of dedicated experiments of this type.

In the case of unstable targets, two different direct approaches are  
envisioned, depending upon the lifetimes of the nuclides involved in the 
entrance channel 
(\cite{Rolfs88}).  
The radioactive {\it target} technique appears most profitable for  
radio-active nuclides with lifetimes in excess of about one hour. 
It has been applied in particular to \reac{22}{Na}{p}{\gamma}{23}{Mg} 
and $^{26}$Al$^{\rm g}$(p,$\gamma$)\chem{27}{Si}
(NACRE \cite{NACRE98}).
In contrast, the radioactive {\it beam} method
 is appropriate for shorter-lived 
species, and has without doubt to be seen as a new frontier in nuclear 
physics and astrophysics.
Two basic techniques can be used to produce the high-intensity,
 high-purity
radioactive beams that are required for the study of the low-energy 
resonances or non-resonant contributions of astrophysical interest: the 
ISOLDE post-accelerator scheme, and the projectile  fragmentation method 
(\cite{RIB92} - \cite{RIB97},
\cite{Aberg97}). Some examples of the pioneering application of those
techniques are  presented below. 
\vskip3truemm
\noindent {\it ISOLDE post-acceleration type experiments}\ \ \
A major breakthrough in experimental nuclear astrophysics has been the
first direct and successful measurement in inverse kinematics of the 
resonant $^{13}$N(p,$\gamma)^{14}$O rate using the ISOLDE 
post-acceleration scheme at the Belgian Radioactive Ion Beam facility of 
Louvain-la-Neuve. 
(See \cite{Galster98} for details about this facility and its 
various developments, and \cite{Leleux98}
for the experimental problems -- beam and target 
preparation, detection methods -- raised by the measurement of proton 
captures induced by radioactive beams.)

The $^{13}$N(p,$\gamma)^{14}$O measurement has been made 
possible thanks to a
pure and intense  ($\approx 10^8$ pps) \chem{13}{N} ion beam 
impinging on a polyethylene (CH$_2$)$_n$
target, the capture $\gamma$-ray being observed 
with an array of efficient Ge detectors \cite{Decrock91}. 
The data obtained in such a way are in good agreement with indirect  
measurements using the Coulomb break-up technique, as well 
as with predictions based on a microscopic model (see below).  
These resonance data have been complemented with the evaluation of the 
interfering non-resonant  direct-capture contribution  to the 
\reac{13}{N}{p}{\gamma}{14}{O} rate  based on the investigation of the
\reac{13}{N}{d}{n}{14}{O} excitation  function \cite{Decrock93a}.

More recently, direct experiments have also been conducted on 
\reac{18}{F}{p}{\alpha}{15}{O} \cite{Graulich97} and 
\reac{19}{Ne}{p}{\gamma}{20}{Na} \cite{Vancraeynest98} with the use of 
\chem{18}{F} and \chem{19}{Ne} beams. 
Preliminary measurements exist for 
\reac{18}{Ne}{\alpha}{p}{20}{Na} \cite{Bradfield98}. 
All these reactions are of interest for the development of the rp- or 
$\alpha$p-process. 
\vskip3truemm
\noindent {\it fragmentation technique}\ \ \
An interesting alternative and complement to the above
method is to directly use 
medium-energy or  relativistic radioactive ion beams from projectile 
fragmentation. This technique has been developed at Berkeley, and
then used at GANIL, RIKEN and GSI. In particular, it has been applied at
RIKEN to the direct measurement of 
$^{8}$Li($\alpha$,n)$^{11}$B \cite{Gu95},
which has been predicted to be of interest 
in an inhomogeneous Big Bang model.
 
\subsubsection{Indirect cross-section measurements}\ \ \ 
%
The indirect methods are a very important complement,
or even an inevitable alternative, to the direct measurements concerning
reactions on stable as well as unstable targets. 
They will certainly continue to play a leading role in the future 
(\cite{Champagne92}).
This situation relates in particular to the extreme smallness of the 
cross sections of astrophysical interest, or to the incapability of 
setting up radioactive beams of the required purity and intensity.
 
Different indirect approaches have been developed and applied to a more
or less large extent, like (1) the use of transfer reactions, (2) the 
study of the inverse reactions,
or (3) measurements relating to the decay of radioactive 
beams.
\vskip3truemm
\noindent {\it transfer reactions}\ \ \ 
In cases where resonances near or below the reaction threshold can 
contribute significantly to the reaction rate, extrapolations of the
 rates from high energies to the Gamow window may fail.
In such conditions, the Breit--Wigner parameters (energy, angular 
momentum, partial and total widths, and decay modes) of the involved 
resonances must be determined independently. 
In nuclear structure studies, this information is typically obtained via 
transfer reactions. 
This approach is also widely in use for nuclear astrophysics purposes 
(\cite{Iliadis98,Champagne92,Bardayan97};
\cite{NICIV,Hirschegg98}). 

The successful application of this technique requires particle beams 
with good energy resolution (provided by  e.g.~tandem accelerators) 
coupled with high resolution spectrometers  (Q3D, split-pole). 
Transfer reactions are also well suited for investigations using 
radioactive beams or targets. 
For example,  it is planned to study the important 
\reac{15}{O}{\alpha}{\gamma}{19}{Ne} reaction through experiments on 
d(\chem{18}{Ne},\chem{19}{Ne^*})p using a \chem{18}{Ne} beam 
\cite{Bradfield98}. 
\vskip3truemm
\noindent {\it Inverse reactions: the example of the Coulomb 
break--up}\ \ \
Direct experiments on radiative captures require the observation
 of very low $\gamma$-ray yields in the presence of intense backgrounds. 
This problem can become unsurmountable in experiments with radioactive 
beams. 
In such difficult cases, experiments on inverse reactions, which relate 
to the forward transformations by the principle of detailed balance, may 
be considered as an alternative. 
For example, instead of measuring the A(b,$\gamma$)X reaction cross 
section, experiments can be conducted on the inverse X($\gamma$,b)A 
reaction. 
The $\gamma$--flux is provided by the virtual photons of the Coulomb
 field, which are seen by nucleus X when passing at a suitable distance 
 from a heavy target.  
This Coulomb break-up technique has the advantage that the cross sections 
to be measured are larger than those of 
the direct process because of the high density of virtual photons. 
Of course, it requires the availability of heavy-ion beams of sufficient 
energy (several tens to several hundreds of MeV/u) 
(\cite{Bauer98,Motobayashi98}). 
 
That method has been successfully applied, among others, to the important
\reac{12}{C}{\alpha}{\gamma}{16}{O} reaction \cite{Tatischeff95}. 
With radioactive beams produced by the fragmentation of energetic  
heavy-ions, the Coulomb break-ups of \chem{14}{O}, \chem{12}{N} and 
\chem{8}{B} have been used to study the reactions 
$^{13}$N(p,$\gamma)^{14}$O \cite{Motobayashi91,Kiener93},
$^{11}$C(p,$\gamma)^{12}$N \cite{Lefebvre95}, and
$^{7}$Be(p,$\gamma)^{8}$B \cite{Motobayashi98,Suemmerer98}.
\vskip3truemm
\noindent {\it Decay of radioactive beams}\ \ \
In certain cases, radioactive decays may offer an interesting 
alternative to transfer reactions for exploring nuclear levels of 
astrophysical importance.  
For example, the $\beta$-decay of \chem{20}{Mg} has been studied in order 
to improve the knowledge of the \chem{20}{Na} level structure above the 
\chem{19}{Ne} + p threshold, and concomitantly of the 
\reac{19}{Ne}{p}{\gamma}{20}{Na} break-out reaction from the hot CNO 
cycle \cite{Piechaczek93}.  
 
Similarly, $^{12}$C($\alpha,\gamma$)$^{16}$O  has been 
investigated through the
$\beta$-delayed $\alpha$-emission from \chem{16}{N} 
\cite{Azuma94} -
\cite{Gai98}, and through the $\beta$-delayed proton emission from
\chem{17}{N} \cite{King98}.  
These experiments provide information on the E1
and E2 contributions to  the rate, respectively.
 
\subsection{Neutron capture reactions: Experiments}
%
Neutron captures or reverse ($\gamma,n$) photodisintegrations are 
essential reactions in the s-, r- and p-processes of  
heavy element production. 
The experiments on (n,$\gamma$) reactions for astrophysical purposes 
first require the availability of neutrons with energies in the 
keV range. 
The capture cross sections are then measured through the detection of 
prompt capture $\gamma$-rays in combination  with a time-of-flight (TOF) 
technique for neutron energy determination, or through the use of 
activation methods. 
These different experimental approaches are only very briefly sketched 
here (Chap.~4.2 of \cite{Kaeppeler98} for details).

Various methods can be used for the production of keV neutrons, and 
can be seen as complementary. They are (i) nuclear reactions in 
combination with low-energy particle accelerators, where the 
limitations in available neutron fluxes can be compensated to some 
extent by relatively short neutron flight paths, (ii) the bombardment 
of heavy-metal targets by beams of typically 50 MeV electrons from a 
linear accelerator, producing high intensities of energetic neutrons 
which are slowed down in a moderator, or (iii) spallation reactions,
which can produce the highest available fluxes of keV neutrons.

The TOF techniques can be applied to the measurement of neutron captures 
by most of the stable nuclei, but require a pulsed neutron source
[e.g. \reac{7}{Li}{p}{n}{7}{Be} with a pulsed proton beam]  to  determine the neutron
energy via the TOF between target and detector.  Though that technique is always
applicable, it suffers from certain  limitations owing to the decreasing flux along the
neutron flight path.  The corresponding loss in sensitivity is, to some extent,
compensated by  the development of detectors covering a solid angle of almost 4$\pi$. 
An example of such a setup is provided by the 4$\pi$BaF$_{2}$ detector at 
the Karlsruhe Van de Graaff. Such detectors allow for an accuracy of
about 1\% for the astrophysical rates, compared with  the
typical 3 to 8\% uncertainties inherent  to other techniques.  
This accuracy is especially welcome for those nuclides that can be  
produced exclusively by the s-process. 
Those ``s-only'' isotopes indeed serve as the normalization that is 
necessary to evaluate the relative s- and r-process contributions to the 
(solar-system) abundances of all the other stable heavy nuclides. 
They also help defining the stellar conditions of operation of the 
s-process (Sect.~8). At this point, one has to
remember, however, that the remarkable 
experimental accuracy mentioned above clearly
concerns the capture of neutrons by nuclei 
in their ground states,
and is spoiled by the contribution to the reaction 
of target excited states, which can be  evaluated by theory only.

When neutron capture leads to an unstable nucleus, the activation  
technique can serve as an alternative method for measuring stellar  
(n,$\gamma$) cross sections. 
Experiments can be conducted  with neutrons e.g. from 
$^{7}$Li(p,n)$^{7}$Be,  the spectrum of which 
very closely resembles the thermalized (Maxwellian) stellar 
spectra in typical
s-process conditions.  
This technique does not require a
pulsed beam for  neutron production, 
and the samples can be placed directly
onto the  neutron target, so that a neutron flux 
about $10^6$ times higher
than  in the direct detection method can be obtained.  This makes the
activation technique clearly superior where a good  
sensitivity is required. 
It allows, for example, measurements of extremely small (n,$\gamma$) 
cross sections
in the $\mu$barn range, which can be of importance in various 
astrophysical situations e.g.~the s-process in low-metallicity stars  
(Sect.~8.1.5), as well as the  determination of cross sections of 
radioactive isotopes, where extremely  small samples have to be used in 
order to keep the radiation hazard and  the sample-induced backgrounds 
manageable.
The limitation of the activation technique to cases in
which neutron capture produces an unstable isotope could in principle
 be overcome by
analyzing the irradiated samples via accelerator mass spectroscopy. 
However, this technique requires extremely pure samples, which are 
presently not available.

By 1992, experimental neutron-capture cross sections were available for 
more than 90\% of the (approximately 240) stable $A \gsimeq 56$ nuclides
involved in the  s-process.
In 15\% of those cases, the obtained accuracy was better than $\pm$4\%
\cite{Beer92}. 
Since that time, new high-precision data have become available
(\cite{Kaeppeler98a} - \cite{WisshakNd98}). These
data have been complemented with the measurements of relatively  
small (n,$\gamma$) cross sections on various light nuclides 
(\cite{Nagai94,Nagai98}), as well as of 
some (n,p) and (n,$\alpha$) rates  
(\cite{Koehler97} - \cite{Wagemans98}). 
Further improvement is called for, particularly concerning 
neutron-capture  
cross sections on the unstable nuclides that can be involved in the  
s-process branchings (Sect.~8.1).  
Such experimental data are now available in a few cases only 
(\cite{Kaeppeler98})
 because the samples are difficult to 
prepare, and as a result of the sample activity itself. 
They concern the long-lived radionuclides \chem{93}{Zr}, \chem{99}{Tc}, 
\chem{107}{Pd} and \chem{129}{I}. 
For shorter-lived species, the activation technique has been found to be 
especially well suited, as demonstrated by experiments carried out with 
\chem{155}{Eu} ($t_{1/2} = 5$ y). 
 
Of course, the knowledge of the rate of the neutron captures of relevance 
to the p- and r-processes is by far less satisfactory, the vast majority 
of them involving unstable targets. 
As a consequence, the rates entering  the astrophysical modelling come 
almost exclusively from calculations. 
Although the measurement of a sizable fraction of the relevant cross 
sections cannot be imagined in any foreseeable future, experimental 
efforts devoted to a cleverly selected sample of those reactions would be 
of the highest importance, particularly in order to help improving the 
reliability of global reaction rate models.

\subsection{Thermonuclear reaction rates: Models}
%
In a variety of astrophysical situations, and especially during the
hydrostatic burning stages of stars, charged-particle induced 
reactions proceed at such low energies that a direct cross-section 
measurement is often not possible with existing techniques. 
Hence extrapolations down to the stellar energies of the cross sections 
measured at the lowest possible energies in the laboratory are the 
usual procedures to apply. 
To be trustworthy, such extrapolations should have as strong a 
theoretical 
foundation as possible. Theory is even more mandatory when 
excited nuclei are
involved in the  entrance channel, or when unstable very neutron-rich or
neutron-deficient  nuclides (many of them being even impossible to produce
with present-day  experimental techniques) have to be considered. 
Such situations are often encountered in the modelling of explosive 
astrophysical scenarios.
 
Various models have been developed in order to complement the experimental
information. Broadly speaking, 
they can be divided into ``non-statistical''
and  ``statistical'' models, with the former one having
different variants (\cite{Descouvemont98}).
Each of these models has its own advantages, drawbacks, and domains of
applicability. Unfortunately, the important question of the evaluation of 
their reliability is often very difficult to answer. Non-statistical and 
statistical models are  appropriate for systems involving, respectively, 
low and high  densities of participating nuclear levels. 
In more practical terms, the applicability of the statistical models is 
limited to reactions involving targets with mass numbers $A \gsimeq 20$ 
if they lie close to the line of nuclear stability. 
This mass limit has to be increased more and more  when moving farther 
and 
farther away from stability. 
Indeed, the nucleon or $\alpha$-particle separation energies decrease to 
such an extent that the number of available nuclear states becomes 
insufficient to validate a statistical description of the reaction 
mechanism. 
As far as the non-statistical models are concerned, one has to 
distinguish (i) those involving adjustable parameters, such as the 
$R$- or $K$-matrix methods, or (ii) the ``ab initio'' descriptions, 
like the potential model, the Distorted Wave Born Approximation (DWBA), 
or the microscopic models. 
The first family of models is applicable only when enough cross-section 
data are available above the Gamow window for a reliable extrapolation 
to lower energies.
In contrast, the second one is useful even in absence of such 
information, and requires merely an experimentally 
based nucleus-nucleus or 
nucleon-nucleon
interaction. 

\subsubsection{Microscopic models}\ \ \ 
%
In recent years the ``microscopic cluster model,'' based on a 
first-principle approach, has become an established tool to perform 
low-energy cross-section extrapolations for both radiative captures and 
transfer reactions (\cite{Descouvemont93}).
In this model, the nucleons are grouped into clusters. 
Keeping the internal cluster degrees of freedom fixed, the totally 
antisymmetrized relative wave functions between the various clusters are 
determined by solving the Schr\"odinger equation for a many-body 
Hamiltonian with an effective nucleon-nucleon interaction. 
When compared with most others, this approach has the major advantage of
providing a consistent, unified and successful description of the bound, 
resonant, and scattering states of a nuclear system. Various improvements 
of the model have been made (\cite{Descouvemont98}).   
 
In spite of its virtues, the microscopic cluster model cannot,
 in general,
reproduce available experimental data as accurately as often desired for 
the astrophysically required extrapolations.  
A higher level of reliability of these extrapolations is usually obtained  
by adjusting the parameters of the nucleon-nucleon interaction.  
It should be stressed, however, that these manipulations represent only  
minor corrections, and do not influence the description of the 
physics of the reaction process in any major way. 
In particular, the model retains its predictive power, and remains
superior to other frequently used extrapolation procedures. 
Its major drawback lies in its rather difficult handling, and in its 
time-consuming computations.
 
The microscopic model has been applied to many important 
reactions involving light systems, and in particular to the various 
 p-p chain
reactions 
(\cite{Langanke96}).
The available experimental data can generally be well reporduced.
 The microscopic cluster model or its variant
(the microscopic potential  model) has also 
made an important contribution to
the understanding  of the key $^{12}$C($\alpha,\gamma$)$^{16}$O 
reaction rate 
(\cite{Descouvemont93a}).
 
\subsubsection{The potential and DWBA models}\ \ \ 
%
The potential model has been known for a long time to be a useful tool
in the description of radiative capture reactions.
It assumes that the physically important degrees of freedom are the  
relative motion between the (structure-less) fragment nuclei in the  
entrance and exit channels, and that the fragments themselves are just  
accounted for approximately by the introduction of spectroscopic factors  
and strength factors in the optical potential. 
The associated drawbacks are that the nucleus-nucleus potentials adopted
for calculating the initial and final wave functions from the 
Schr\"odinger equation cannot be unambiguously defined, and that the
spectroscopic  factors cannot be derived from first principles. 
They have instead to be obtained from more or less rough ``educated 
guesses.''

The potential model has been applied, for example, to
the p-p chain
\reac{3}{He}{\alpha}{\gamma}{7}{Be} reaction \cite{Descouvemont98}. 
Another interesting application is to a systematic calculation of the 
non-statistical ``direct capture (DC)''
 contribution to the ensemble of (n,$\gamma$) 
reactions on neutron-rich targets that may be involved in the r-process 
(\cite{GorielyT98}).
The direct captures may contribute to the total (statistical + DC)
(n,$\gamma$) cross sections significantly, and in some cases 
overwhelmingly, particularly in very neutron rich nuclei.

On the other hand, the Distorted Wave Born Approximation (DWBA) has 
become the standard  model to describe nuclear transfer reactions.
Recently, it has been applied to astrophysics by constructing an 
effective nucleon-nucleon interaction folded by the  nuclear densities 
of the collision partners, with 
the overall strength of the nucleus-nucleus 
potentials being adjusted in order to reproduce experimental data 
(\cite{Oberhummer91}). 
   
\subsubsection{Parameter fits}\ \ \ 
%
Reaction rates dominated by the contributions from a few resonant or 
bound states are often extrapolated in terms of $R$- or $K$-matrix 
fits, which rely on quite similar strategies.  
The appeal of these methods rests on the fact that analytical expressions 
which allow for a rather simple parametrization of the data can be
 derived 
from underlying formal reaction theories. 
However, the link between the parameters of the $R$-matrix model and the 
experimental data (resonance energies and widths) is only quite indirect. 
The $K$-matrix formalism solves this problem, but suffers from other 
drawbacks (\cite{Barker94}).

The $R$- and $K$-matrix models have been applied to a variety of
reactions, and in particular to the analysis 
of the \reac{12}{C}{\alpha}{\gamma}{16}{O} reaction rate 
(\cite{Azuma95,Hale97}).
 
\subsubsection{The statistical models}\ \ \
%
Many astrophysical scenarios involve a wealth of reactions on 
intermediate-mass or heavy nuclei. 
This concerns the non-explosive or explosive burning of C, Ne, O and Si, 
as well as the s-, r- and p-process nucleosynthesis. 
Fortunately, a large fraction of the reactions of interest proceed through
compound systems that exhibit high enough level densities for statistical 
methods to provide a reliable description of the reaction mechanism.  
In this respect, the Hauser-Feshbach (HF)
 model has been widely used with
considerable success.

A version of the model that has recently been applied to astrophysical 
calculations is described in \cite{GorielyNC98}. 
It features in particular a new global $\alpha$-nucleus optical potential 
\cite{Grama98}, which is one of the least-known physical ingredients 
of the HF model, as well as an approximation of the soft 
dipole mode that 
may be of special importance when considering very 
neutron-rich nuclei. When
dealing with such nuclei, 
it has to be remembered that both direct and 
statistical components may contribute to 
the (n,$\gamma$)  capture. 
Eventually, the evaluation of an (n,$\gamma$) rate necessitates a
proper combination of the DC and HF (interfering) contributions. 
No model to-date performs this superposition in 
a reliable way for a very large sample of nuclei, 
as the one of relevance to
the r-process. 
 
\section{Selected topics}
%
We pick here three topics that have in previous sections evaded clear 
definitions and/or the systematized discussions they deserve for their
specific or general nuclear astrophysics interest: the s- and r-process
synthesis of heavy elements, cosmochronology, and the 
current understanding of
the Type-II supernova
mechanism.

%
\subsection{Heavy-element nucleosynthesis by the s- and r-processes 
of neutron captures}
%
We have in several occasions referred to the s- and r-processes, which 
are invoked for the synthesis of the vast majority of the 
naturally-occurring nuclides heavier than the ``Fe-group.''
For the sake of clarity, as well as for the benefit of non-expert 
readers, we now go back to their definitions, and give a brief summary
of  the nuclear and astrophysical problems they incur.
 
\subsubsection{Defining the s-process}\ \ \
%
The ``slow'' neutron-capture process relies on the assumption that 
pre-existing (``seed'') nuclei are exposed to a flux of neutrons that
is weak enough for allowing a $\beta$-unstable nucleus produced by 
a (n,$\gamma$) reaction to decay 
promptly, except perhaps for very long-lived isotopes which may
instead capture a neutron.
This definition \cite{BBFH57,Seeger65} does not make any reference to 
the very origin of the required neutrons, or to a specific astrophysical 
site. 
The nuclear flow associated with the s-process (the ``s-process path'') 
has then 
to develop in the close vicinity of the line of $\beta$-stability.
The endpoint of this path is \chem{209}{Bi}, at which 
the \chem{209}{Bi}(n,$\gamma$)\chem{210}{Bi}($\beta^-$)
\chem{210}{Po}($\alpha$)\chem{206}{Pb} chain leads to some cycling
of the material in this mass region.

In a steady-flow regime, which is a good approximation at least locally
in mass number $A$, 
the relative abundances of the synthesized nuclides are
inversely  proportional to their stellar neutron-capture rates. 
Since a nucleus with a 
 closed neutron shell has a quite low neutron-capture
probability because of the relatively low 
 neutron-separation energy of the
$N+1$ isotope, the flow staggers there, 
 resulting in an abundance peak at a
magic number such as $N = 82$ or 126. This translates in Fig.~6 as
solar-system abundance maxima at
$A = 138$ or 208, respectively. If (n,$\gamma$) rates dictate the relative
abundances along the path, the 
intervening $\beta$-decay rates determine the
speed of the nuclear flow, and consequently the s-process time-scale. 
 
It is conceivable that some complications may arise at specific locations 
of the  s-process path because of the possible competition between
$\beta^-$-decays and neutron captures. Even continuum-e$^-$ captures may
come into play. 
As illustrated in Fig.~14, such a situation translates into
 ``branching points'' in the s-process path, which constitute an important
ingredient  of the s-process (\cite{Ward76,Kaeppeler89}).   The handling
of these branches necessitates the knowledge of the stellar $\beta$-decay
rates, as well as of the probabilities of neutron 
captures by unstable nuclei
close to the line of stability.

\begin{figure} 
\center{\includegraphics[width=1.0\textwidth,height=0.5\textwidth]{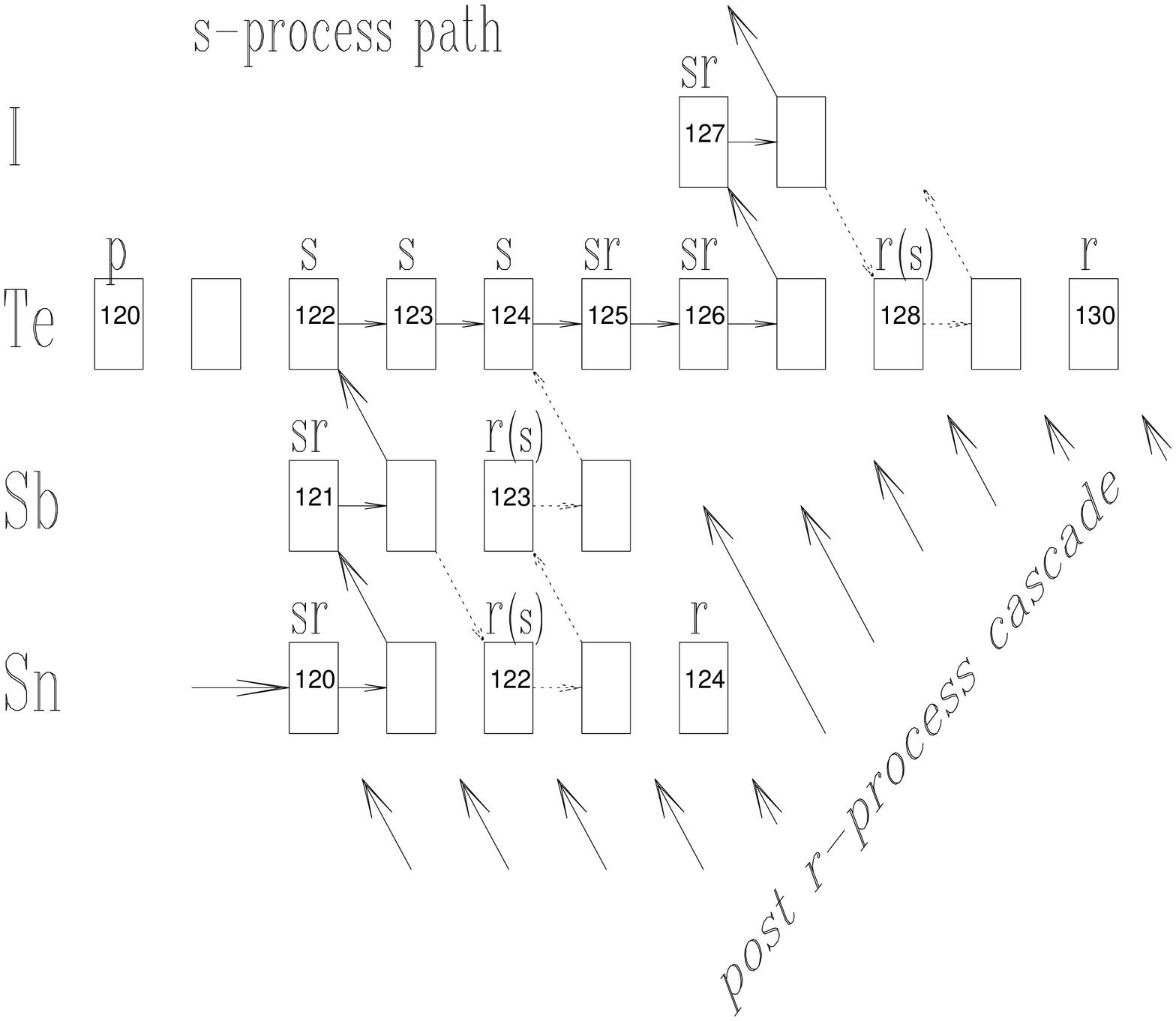}}
\caption{A typical s-process path in the $120 \leq A \leq 128$ region,
with some possible branches resulting from the 
competition between neutron 
captures ({\it horizontal}), $\beta^-$ decays ({\it upward} arrows) and 
continuum-e$^-$ captures ({\it downward} arrows) (taken from 
\cite{Takahashi95}). 
The stable isotopes are indicated by their mass numbers
in squares, 
 while the open squares identify the $\beta$-unstable isotopes. 
The $\beta^-$-decay  cascade from the r-process path (Fig.~15) towards
the line of stability after the r-process neutron irradiation is also 
schematized.  
The stable neutron-rich nuclides \chem{124}{Sn} and \chem{130}{Te} are 
the end-points of  the post-r-process cascade, and stay away 
from the s-process path. 
Such nuclides are referred to as ``r-only'' products (labels ``r'').
 In contrast, \chem{122}{Te}, \chem{123}{Te} and \chem{124}{Te} are 
located on the s-process flow, but are ``shielded'' from the r-process. 
They are thus ``s-only'' nuclides (labels ``s'').  
Other nuclides are located on both the s-process path and the post
r-process cascade, this situation  being in some cases the direct result 
of s-process path branchings. 
The prediction of the relative s- and r-process contributions to the
solar abundances of those nuclides of a mixed origin necessitates, in 
principle, the build-up of detailed  models for the 
neutron capture processes.
The notation ``sr'' is used when the s- and r-processes are expected to 
contribute almost equally to the abundance, while the label ``r(s)'' 
indicates that  the r-process contribution dominates.
The neutron-deficient nuclide \chem{120}{Te} is produced by neither of 
the neutron capture processes, and is classified as a ``p-only'' nucleus 
(label ``p'') 
}
\end{figure}
%
In general, 
the competition between neutron captures (whose rates depend on
temperature and neutron density) and
$\beta$-decays (which are mainly temperature-dependent) is incorporated in
semi-empirical analyses of the  s-process abundances to impose some
constraints, 
particularly on the  neutron density and the temperature for the
s-process (\cite{Kaeppeler98,Kaeppeler89}). The meaningfulness of these
branching point constraints of course relates directly to 
the reliability of
the (n,$\gamma$) and $\beta$-decay rate input. 
Additional nuclear data about
the rates of the neutron-producing reactions [mainly
\reac{13}{C}{\alpha}{n}{16}{O} and
 \reac{22}{Ne}{\alpha}{n}{25}{Mg}] and of
some (n,p) and (n,$\alpha$) reactions are 
also required when the s-processing
is followed in the framework of ``realistic" stellar models.

\subsubsection{Defining the r-process}\ \ \
%
In contrast to the s-process the ``rapid'' neutron-capture process, or
r-process, is based on the 
 hypothesis that the neutron density is so high that
neutron captures are  always faster than $\beta$-decays. 
The pre-existing isotopes of 
each element are converted by successive neutron
captures into  very neutron-rich ones, whose neutron separation energies
$S_{\rm n}$ are low enough for allowing the inverse ($\gamma$,n) reactions
to occur  efficiently and impede further progression.  
In the simplest picture
of the r-process \cite{BBFH57,Seeger65}, an (n,$\gamma) \leftrightarrow
(\gamma$,n) equilibrium is reached in each isotopic chain starting at iron
%
($Z = 26$).\footnote{This simplification is validated by the fact that the
r-process has traditionally been assigned to the Fe-rich inner core of a
massive star supernova}
%
In such conditions, the isotopic abundance distribution for a given
$Z$ is independent of (n,$\gamma$) or ($\gamma$,n) cross sections, and is
governed instead by the laws of statistical mechanics. 
More specifically, the
isotopic abundances obey a Maxwellian distribution, 
and thus depend  mainly on
$S_{\rm n}$, temperature
$T$ and neutron density $n_{\rm n}$ 
(and weakly on nuclear partition functions).
  Once
distributed in such a way, the isotopes 
then ``wait'' for $\beta^-$-decays to
occur (the ``waiting point approximation''), and to transport the material
from one ($Z$) isotopic chain to the next ($Z+1$). 
Beta-decays thus govern the
speed of the nuclear flow (the ``r-process path'').
 If $T$ and $n_{\rm n}$ were
kept constant in time, the  nuclear flow
  would follow an iso-$S_{\rm n}$ line
in the ($N$,$Z$) plane. The location of this line depends on $T$ and
$n_{\rm n}$. More specifically, it corresponds to $S_{\rm n}$-values
increasing with $T$ and decreasing with increasing  $n_{\rm n}$ 
\cite{Seeger65}: higher $n_{\rm n}$-values indeed tend to push material
farther away from stability, while higher temperatures have the reverse
effect, as they speed up the ($\gamma$,n) photodisintegrations. It is thus
conceivable that the r-process path could approach the neutron 
drip line if
high enough $n_{\rm n}$ can be obtained in an astrophysical
 site at
low enough temperatures. Some complication arises if enough neutrons are
available and if the 
neutron irradiation time is long enough for the r-process
path to reach the trans-actinide region, where 
 neutron-induced (or $\beta$-delayed) fissions could
interrupt the flow to higher $Z$-values, 
and could be responsible for a cycling-back
of the material to lower-mass nuclei.
 
In the simplest picture of the r-process the neutron flux and the
temperature are assumed to go abruptly to zero, 
as are the (n,$\gamma$) and
($\gamma$,n) rates, after a given irradiation time. As a result, the very
neutron-rich unstable  nuclides on the r-process path 
start to cascade through
$\beta^-$-decays back to the line of stability. This cascade may be
complicated by $\beta$-delayed fission  or neutron emissions
 (in the simplest
models, the re-captures of the emitted neutrons are neglected).

The basic nuclear data needed for the modelling of 
the r-process in the simple
framework described above are masses and $\beta$-decay rates of very
neutron-rich nuclei. Their fission barriers are also required.
 Various classes
of more sophisticated r-process models have also been 
constructed. Some still
make use of schematic parametrized astrophysical conditions, but avoid
simplifications like the waiting point approximation 
(\cite{Goriely96}), the sudden freeze-out of the neutron captures
 and inverse
photodisintegrations 
 or the neglect of the re-capture of delayed neutrons
(\cite{Meyer93,Howard93}). 
These approaches require the solution of very large
nuclear-reaction networks for the derivation of the abundances. 
The build-up of
these networks necessitates the knowledge of neutron capture rates
 and of their
inverse photodisintegration rates on top of the nuclear data already 
needed by the
simple model. 
 Even some neutrino-interaction rates may have to be considered
\cite{Meyer95}. Other complementary approaches have explored in greater
detail more realistic astrophysical conditions for the development of the
r-process, and the possibility for the $\alpha$-process to be the
progenitor of the r-process. Figure~15 illustrates possible paths
of an $\alpha$- and subsequent r-process. 
%
\begin{figure} 
\center{\includegraphics[width=0.8\textwidth,height=0.68\textwidth]{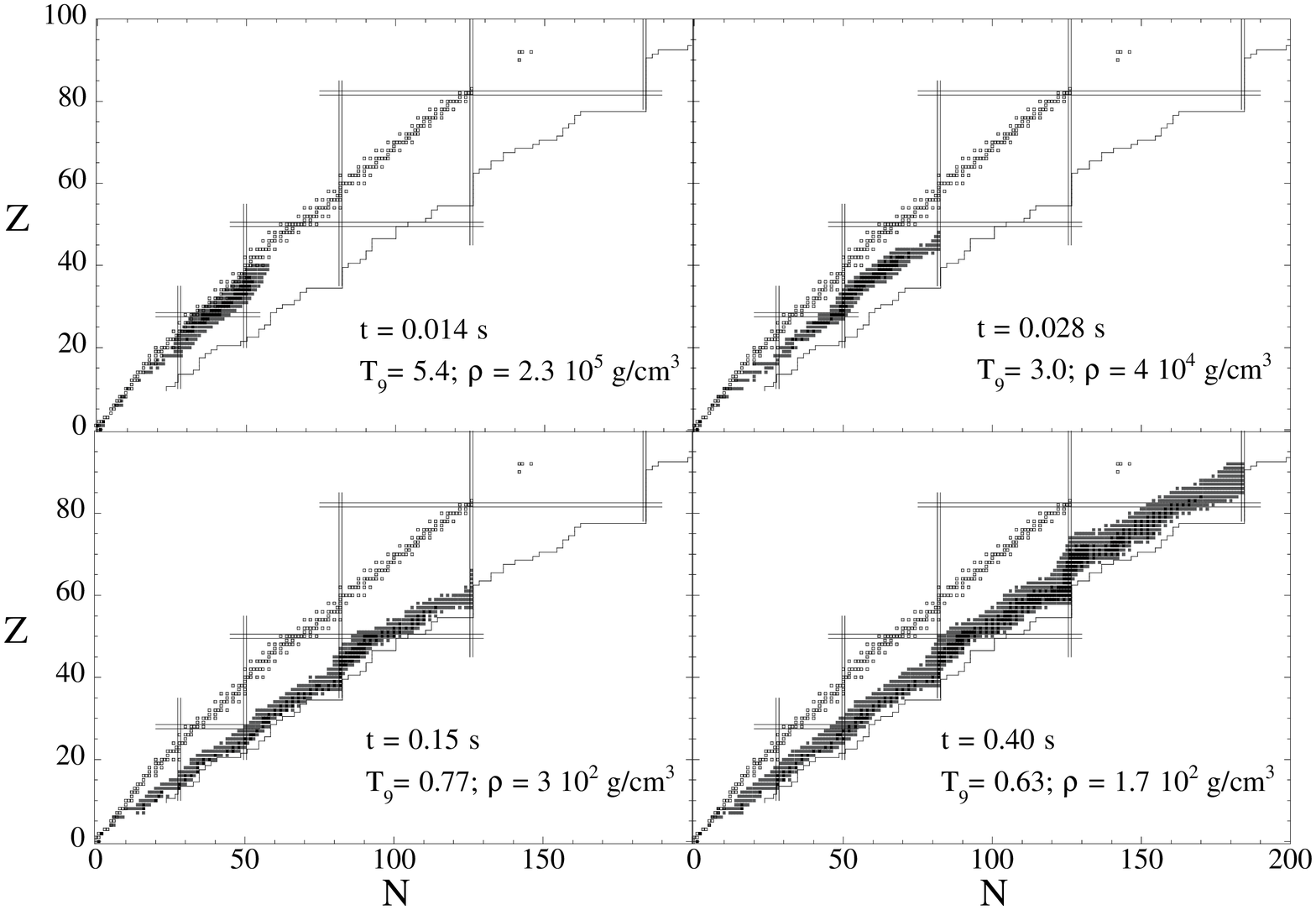}}
\caption{An example of the development of
$\alpha$- and r-process paths in sequence of time ($t$), represented 
by {\it strips} 
comprising synthesized nuclides 
between the valley of stability ({\it boxes}) and the
 neutron drip line ({\it stairs}), with
{\it double-lines} running along the neutron 
and proton magic numbers
(from \cite{Gorielypriv}). 
The evolution of a hot-bubble material is followed with the use of
a generic model for its dynamics \cite{Takahashi97} with a slight
modification, and of a detailed nuclear network.
The model parameters are adjusted so as to induce the r-process
(see Sect.~8.3.2).
The initial values of the temperature (displayed 
in units of $10^9$ K, $T_9$) and
 density ($\rho$) are $10 \times 10^9$ K and $1.6 \times 10^6$
 g/cm$^3$, respectively.
In the beginning, the $\alpha$-process synthesizes nuclides near the  
line of stability, and even beyond the ``Fe-peak.''
With the temperature decrease, it ``freezes out.'' 
Neutron captures then 
bring the path into the more neutron-rich region, and to more massive
nuclides as (relatively slow) $\beta$-decays intervene
} 
\end{figure}
%

\subsubsection{The s- and r-process contributions to the
solar-system composition}\ \ \
%
>From the above definitions, one can guess qualitatively which 
of the s- or/and r process(es) was (were) responsible for the production 
of a given heavy nuclide in the solar system.
The basic principles underlying such an identification are depicted in 
Fig.~14.
A more quantitative modelling of the s- 
(and to a lesser extent r-) process  
makes it possible to split the solar abundance curve beyond the Fe peak 
region into an s- and an r- (as well as a p-) component. The results are shown in
Fig.~16.  It has to be noted that this decomposition is affected by
uncertainties  of various  natures (\cite{Goriely98}).
Some of them are of cosmochemical and nuclear origins, and others
are of more purely astronomical venue. 
In this respect, one has to keep in mind
that the ``solar s- and  r-abundances'' of Fig.~16 rely on semi-empirical
analyses 
 (particularly of the s-process) that are far from being rooted in 
realistic astrophysical models.
%
\begin{figure} 
\center{\includegraphics[width=0.9\textwidth,height=0.45\textwidth]{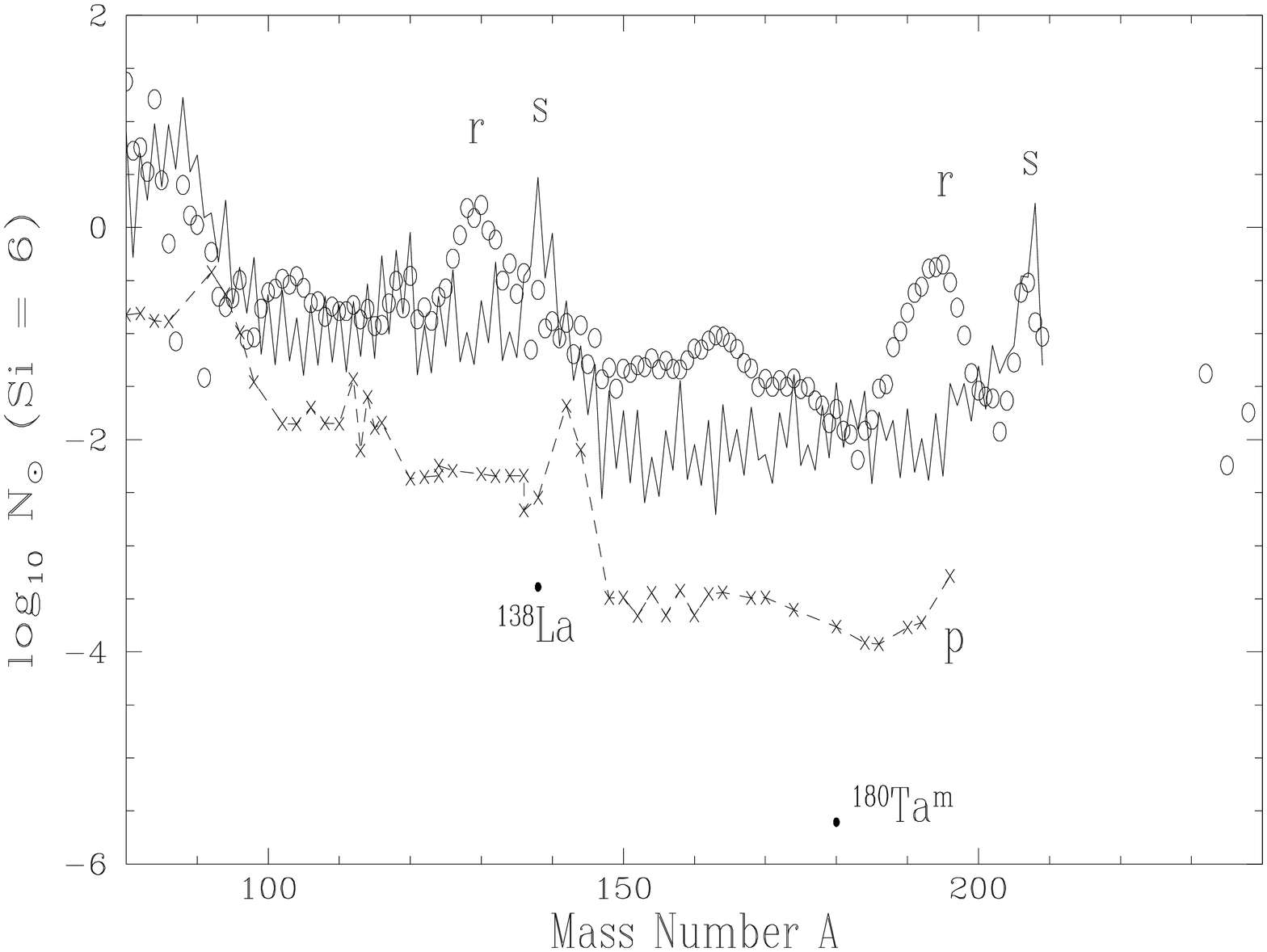}}
\caption{Decomposition of the solar abundances of heavy nuclides into
an s-process ({\it solid line}), an r-process ({\it open circles}) and a
p-process ({\it crosses}) contributions 
\cite{Takahashi95} (based on \cite{Kaeppeler89}).
The procedure of the decomposition is as follows: First, the
abundances of the ``s-only'' nuclei are fitted with the help of
a semi-empirical s-process model. 
This fit requires a suitable distribution of
neutron doses.  
The model predictions are then used to evaluate the s-process
contributions to the other nuclides. 
The subtraction of these s-process contributions from the observed solar 
abundances leaves for each  isotope a residual abundance that represents 
the r-process (if neutron-rich) or  p-process (if neutron-deficient) 
contribution. 
Normally, the ``r-only'' nuclides do not have to be subjected to this 
procedure. 
Such analyses come to the conclusion that about half of the heavy
nuclei in the solar material come from the s-process, and the other half 
from the r-process, whereas the p-process is responsible for the
production of the remaining low-abundance nuclides.
The rarest naturally-occurring nuclide, $^{180}$Ta$^m$ (indeed an  
isomeric state) is probably produced by 
the s-  and p-processes in relative quantities that have yet to be
determined precisely. 
It has been speculated that $^{138}$La may be synthesized by the 
$\nu$-process or by spallation reactions.
}
\end{figure}

\subsubsection{Astrophysical sites for the s- and r-processes}\ \ \
%
The results displayed in Fig.~16 do not help much in the 
attempt to identify
precisely the astrophysical (stellar)  s-, r- and p-process sites,
particularly because the solar-system bulk material is a mixture
of a large number of nucleosynthesis events that have taken place in
the Galaxy before the isolation of the solar nebula. At best, 
this mixture can provide a useful guide in the problem at hand. 
The observation
of the contamination of the surface of certain stars by heavy 
elements (mainly
s-nuclides) produced in their interiors may be of more help, 
as the data refer
to a single event. This is also the case with some isotopically-anomalous
meteoritic grains of suspected circumstellar origin, in which the precise
isotopic composition is known for a variety of heavy 
elements.\footnote{In contrast to stellar spectroscopy, 
the grain analysis can
provide very precise isotopic compositions, even for heavy elements 
(\cite{Hoppe97}). The drawback of these studies is that the
precise characteristics of the stars 
from which the grains may originate are
unknown}
%
The application of the semi-empirical s-process model outlined
above leads to the conclusion that the 
 s-component displayed in Fig.~16 can
be fitted quite satisfactorily if the neutron density is of about $10^8$
cm$^{-3}$ and if the temperature lies in the 
$(1 \sim 3) \times 10^8$ K for
some 10 to 100 y. In order to account for the r-component, strikingly
different conditions are required. 
More specifically, neutron densities well
over $10^{20}$ cm$^{-3}$ have to be maintained 
at temperatures $\gsimeq 10^9$ K up to 
a second or so. The best fits to the 
observations (mainly of s-nuclides) at the surfaces 
of individual chemically-peculiar
stars generally necessitates conditions that differ to some
extent from those derived from the consideration of the solar-system
abundances. One of the most tantalizing problems in stellar physics and
nuclear astrophysics is to obtain the development of the heavy element
nucleosynthesis processes able to account for the 
observations as a {\it
natural} consequence of 
stellar evolution on the  basis of realistic models of
various stars 
(with different initial  masses and metallicities), and of the
best possible nuclear-physics input. This requirement is clearly most
difficult to meet for the r-process, 
but also raises serious problems for the
s-process, while the situation for the p-process is seemingly less
severe.

In what follows we sketch the current status of the continued
search of the s- and r-process sites, which is far from being satisfactory 
(\cite{Meyer94}).
\vskip15pt
\noindent
{\it s-Process sites}\ \ \
As demonstrated by many observations, the surfaces of a variety of low-mass ($M \lsimeq
3$ M$_\odot$) asymptotic-giant-branch (AGB) stars are enriched  with certain
``s-elements,''\footnote{An element that  has an overwhelming either s- or
r-component in its 
 solar-system  abundance is customarily referred  to as an
s- or r-element.  For example, Ba is an s-element, whereas Eu is an
r-element. This naming may not always properly  reflect the true origin of
an element outside the solar system, as exemplified for Ba in Sect.~8.1.5}
%
implying that they have been synthesized in the interiors of their 
own and dredged-up. These observations also make it
plausible that AGB stars
eject part of their synthesized s-nuclides into the ISM 
through their winds,
and thus contribute to the galactic, and in particular 
solar-system, s-nuclide
enrichment. 
As dust particles are known from astronomical observations to form
in their ejecta, AGB stars could also be the source of certain anomalous
meteoritic grains containing various heavy elements with an
 s-process isotopic
pattern.  

The s-process in AGB stars is thought to occur in their He-burning shell
surrounding a nuclearly inert C-O core, either during recurrent and short
convective episodes (``thermal pulses''), or in between these pulses. A
fraction of the produced s-nuclides (along with other He-burning products)
could then be
brought by convection to the surface shortly after each pulse. A
similar scenario might develop in intermediate-mass (3 - 8 M$_\odot$) AGB
stars, even if this possibility does not receive a strong observational
support 
(\cite{Lattanzio97,Gallino97}).

It is generally considered 
that the necessary neutrons for the development of
the AGB s-process are mainly provided by
 \reac{13}{C}{\alpha}{n}{16}{O}, which
can operate at temperatures around $(1 \sim 1.5)\times 10^8$ K. 
In contrast,  the
\reac{22}{Ne}{\alpha}{n}{25}{Mg} could at best only slightly contribute
because the AGB He shell does not appear to reach the necessary
temperatures ($T \gsimeq 3 \times 10^8$ K) for \chem{22}{Ne} to burn. This
situation is at the origin of the most acute problem raised by 
the AGB star
s-process. If \chem{22}{Ne} finds a natural origin in the classical
\chem{14}{N}($\alpha,\gamma$)\chem{18}{F}($\beta$)\chem{18}{O}($\alpha,
\gamma$)\chem{22}{Ne}
burning of the \chem{14}{N} produced in the preceding CNO cycle, the
\chem{13}{C} originating from this cycle
 is by far not abundant enough for
providing the amount of neutrons necessary for a
 full s-process to develop. An
additional \chem{13}{C} source is thus mandatory. 
It has been proposed that,
under certain conditions, protons and \chem{12}{C} could be
 brought together
at high enough temperatures for \chem{13}{C} to be produced by
\reac{12}{C}{p}{\gamma}{13}{N}($\beta^+$)\chem{13}{C}. Then, standard
stellar models, which do not provide any such source, 
have to be modified in some
{\it ad hoc} manner (\cite{Lattanzio97}).
 
The dredge-up of the produced s-nuclides to the AGB star surfaces 
(as demanded
by the observations) is also far from being well understood. 
This subject is in
fact a matter of debate, different recent models reaching in
some cases quite different conclusions concerning the characteristics, and
even the very existence, of this transport episode 
(\cite{Lattanzio97,Straniero97,Mowlavi98}). 

The AGB star s-process has been widely discussed in attempts to explain
the spectroscopic (\cite{Busso92}) and meteoritic 
(\cite{Gallino97}) data referred to above, as well as to account for the
$A \gsimeq 100$ solar-system distribution of s-nuclides
(\cite{Kaeppeler90}). In these studies, essential quantities,
 like  the amount
of available \chem{13}{C}, 
and thus the neutron density, or the efficiency of
the assumed dredge-up are treated as adjustable parameters.

Massive stars, and more specifically 
their He-burning cores, are also predicted
to be s-nuclide producers through the operation of the
\reac{22}{Ne}{\alpha}{n}{25}{Mg}. 
This neutron source can indeed be active in
these locations as their central regions are hotter than the
 He shell of AGB
stars. Many calculations performed in the framework of realistic stellar
models demonstrate that this site is responsible for a
 substantial production
of the $A \lsimeq 100$ s-nuclides, and can in particular account for the
solar-system abundances of these species 
(\cite{Rayet98}).
 This success does not have to hide some difficulties,
however. 
Uncertainties in the efficiency of this s-process relate directly to
remaining nuclear physics uncertainties in the
\chem{22}{Ne} burning rate \cite{Meynet92} or in (n,$\gamma$) rates,
 as well
as to stellar model uncertainties, even if they may appear less 
severe than in
the AGB star case. In this respect, it may be of interest to note 
that the Ba
enhancement derived \cite{Hoeflich88} from the early spectra 
of the supernova
SN1987A could not be explained by a core He-burning s-process 
model
\cite{Prantzos88}. This difficulty may lie in the uncertainty in the
spectroscopic analyses \cite{Mazalli92}, 
or in the adopted astrophysical model.
\vskip15pt
\noindent
{\it r-Process sites}\ \ \
The search for astrophysical sites for the r-process has been quite
unsuccessful despite the fact that many  possibilities have been
suggested. Of course, this search does not receive any 
really useful guidance
from the direct observation of an object where r-nuclides are produced
{\it in situ}. 
On the theoretical side, all attempts have just revealed more or less
severe shortcomings (\cite{Meyer94}). The
supernova scenarios are no exception, even if their Type-II supernova 
hot-bubble version has at first 
 raised a lot of optimistic excitement, and
remains an adequate site for the $\alpha$-process (Sect.~8.3).
  
The quite old speculation that the r-process might develop in the  
mergence of two neutron stars, or of a neutron star and a black hole
(\cite{Schramm82}) has recently become fashionable again in relation
with the demonstrated capabilities of performing multi-dimensional
hydrodynamical simulations of the merging phenomena, and with 
the potentiality
of these systems to be detectable gravitational wave, neutrino
 or $\gamma$-ray
burst emitters (\cite{Ruffert98}). That these events are certainly
less frequent than Type-II supernovae  would be compensated by
 the presumably
larger masses of the ejecta, and thus possibly of the r-process
 yields, than 
in the hot-bubble case.

Several speculative aspects of the conditions under which an r-process
might accompany a neutron star merger
(\cite{Sumiyoshi98},
\cite{Latimer77} - \cite{Ruffert97}) remain to be worked
out in detail. Some models might show some similarity with the hot bubble
scenario.  Others emphasize the role of a neutron star heating by the
$\beta$-decays or fissions of extremely neutron-rich ``nuclei" beyond the
neutron-drip 
line that may populate the very cold high-density neutron star
crusts. In such a situation, a reliable nuclear EOS is required to set the
initial conditions for the nucleosynthesis, and the weak interaction rates
of nuclei in highly exotic configurations are needed. 

All in all, much remains to be explored before confirming or disregarding
neutron star mergers as possible sites of the r-process 
(or, in fact, of any
other form of nucleosynthesis).  We add here that the outcome of 
the hydrodynamical simulations is sensitive to information
 that is still
statistically missing, like the initial masses 
and spins of the
mergers (\cite{Ruffert96}). Further complications might 
arise from the
necessity of performing the calculations in a general relativity framework
(\cite{Shibata97,Wilson96}), as opposed to a Newtonian scheme largely
adopted so far. Preliminary attempts lead to the conclusion 
that interacting
neutron stars, instead of merging, might well transform into
 individual black
holes \cite{Wilson96}, in which case the hope to make an r-process would 
of course vanish,
again.

\subsubsection{Heavy elements in low-metallicity stars}\ \ \
%
The search for the astrophysical 
sites for the s- and r-processes may be
helped by the rapidly-accumulating  observational data on the surface
abundances of heavy elements in metal-poor stars (\cite{McWilliam97}).
Such analyses 
attempt to correlate the observed abundances
of ``typical s- or r-elements'' with the stellar metallicities,
which may or may not be theoretically
 explained in terms of a possible impact
of  the initial compositions on the heavy-element yields in a given
scenario. 
Of course, the sometimes large scatters in abundances observed at
one metallicity endanger too naive interpretations of the observations
(\cite{Baraffe93}).

>From the observation that the Galaxy
may have been enriched with r-elements earlier than with s-elements
(\cite{McWilliam97}),
it is classically inferred that the bulk galactic content
of r-nuclides comes from more massive, shorter-lived, stars than the bulk
s-nuclides. 
A metallicity change in a star of a given mass may also modify the
number of neutrons made available for the production of trans-iron nuclei
through  variations in the amounts of both the produced neutrons and the
neutrons captured by light 
(even as light as oxygen) poisons. This may affect
the global efficiency of a 
neutron-capture process and the relative yields of
heavy elements 
(\cite{Rayet98}).

As a word of caution, we recall that the classical
terminology of ``s-element" or ``r-element" 
refers in fact to the solar-system
composition, and may be quite misleading when 
one deals with low-metallicity
objects. An interesting illustration of this danger concerns 
recent attempts to analyze the isotopic composition of Ba in metal-poor
stars. 
The Ba observed in the classical
metal-poor subgiant HD140283 was on one hand claimed \cite{Magain95}
 to be of s-process venue in concordance with the conventional
classification of Ba as an s-element.
On the other hand, just the opposite conclusion was reached 
\cite{Gacquer98} in that  its isotopic
composition was consistent with an r-process origin. The development of
spectroscopic techniques enabling the measurement of the isotopic 
composition of a variety of heavy elements
is obviously of major interest for the theory of
nucleosynthesis.

\subsection{Cosmochronometry}
%
The dating of the Universe and of its various constituents is another
tantalizing task in modern science, referred to as ``cosmochronology.''
This field is in fact concerned with different ages, 
each of which corresponding to an epoch-making event in the past
(\cite{Flam90}).
They are in particular the age of the Universe  $T_{\rm U}$, of the 
globular clusters $T_{\rm GC}$, 
of the Galaxy [as (a typical?) one of many 
galaxies] $T_{\rm G}$, of the galactic disc $T_{\rm disc}$, and of the 
non-primordial nuclides in the 
disc $T_{\rm nuc}$, with $T_{\rm U} \gsimeq 
T_{\rm GC} \approx$ ($\gsimeq$?) $T_{\rm G} \gsimeq T_{\rm disc} \approx
 T_{\rm nuc}$. As a consequence, cosmochronology involves not only
cosmological models and observations, 
but also various other astronomical and
astrophysical studies, and even invokes some nuclear physics information. 

The cosmological models can help determining  $T_{\rm U}$, as well as 
$T_{\rm GC}$ and $T_{\rm disc}$, at least to some extent 
(\cite{Tayler86} - \cite{Arnould90} for  brief 
accounts). The HRD has also been used quite extensively in order to
evaluate $T_{\rm GC}$. In short, the method relies on the confrontation
between stellar 
 evolution predictions and the position of the turnoff point
(TOP) of the MS,
 complemented with the largest possible ensemble of observed 
 properties of the HRD of globular clusters (including in particular the
magnitude 
 difference between the TOP and the location of the HB) 
(~\cite{VandenBerg96,Jimenez98a}). The
HRD technique
 has also been applied to the determination of $T_{\rm disc}$.  
So-called ``luminosity functions,'' which provide the total
number of stars per absolute magnitude interval as a function of absolute
magnitudes, 
have also been used to evaluate  $T_{\rm GC}$, as well as  $T_{\rm
disc}$. 
The luminosity function of white-dwarf stars has also been proposed
as $T_{\rm disc}$ evaluators. 
These stars, the end products of the evolution
of the very abundant  single or binary low- and intermediate-mass stars, 
are supposedly the oldest stars in the galactic disc. 
In very short, the method relies on a confrontation between the observed 
sharp fall-off in the number of WDs with a luminosity lower than
a given value and the predicted times taken by WDs to cool 
enough for being so weakly luminous.

Each of those methods has advantages and
weaknesses of its own  (\cite{Arnould90}). The
age estimates they provide are sketched in a synopsis form in Fig.~17,
which 
also displays the limits derived 
from the nucleo-cosmochronology discussed
below. 
%
\begin{figure}
\center{\includegraphics[width=0.9\hsize,height=0.54\hsize]{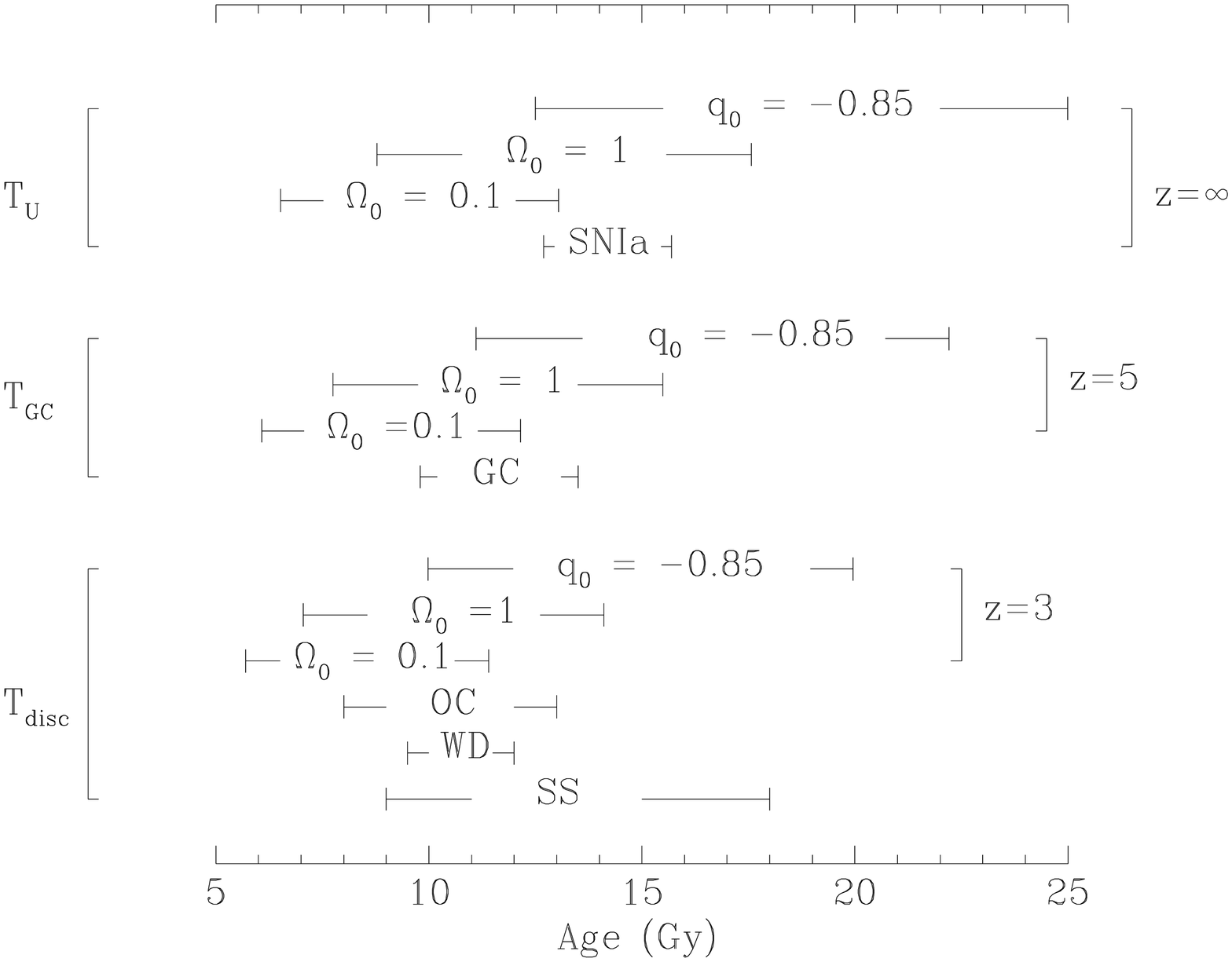}}
\caption{Ages of the Universe ($T_{\rm U}$), of galactic globular 
clusters 
($T_{\rm GC}$) and of the galactic disc ($T_{\rm disc}$). 
Estimates from cosmological models are specified by $\Omega_0$ 
values for the 
Standard Model, and by $q_0 = \Omega_0/2 - \lambda_0 = 3\Omega_0/2 - 1$
 in 
the $\Lambda \not= 0$, $k = 0$ Model.
In both cases, the displayed ranges correspond to
 $100 \geq H_0$(km/s/Mpc)$ 
\geq 50$. 
The adopted $z$ (redshift) values are in no way meant to be precise.
(Note, for instance, that a galaxy with $z = 5.34$ has recently been
reported \cite{Dey98}.) 
The label SNIa defines the age limits derived from values of those  
cosmological model parameters that are determined 
 with the use
of Type-Ia supernovae as standardized candles
(\cite{Riess98,Tripp97}).
The  ages derived from HRD analyses of globular clusters 
\cite{Chaboyer98,Jimenez98} or open clusters 
\cite{Grenon90,Phelps97} are 
in the 
ranges labelled GC and OC, respectively. 
The predictions based on white dwarf luminosity functions are noted WD
\cite{Hernanz94,Oswalt96}, 
while the nucleo-cosmochronological evaluations of 
the age $T_{\rm nuc}$ of the nuclides in the solar system are marked with 
the label SS
}
\end{figure}

\subsubsection{Nucleo-cosmochronology: generalities}\ \ \
%
The dating method that most directly relates to nuclear astrophysics is 
referred to as ``nucleo-cosmochronology.'' 
It primarily aims at determining
the age $T_{\rm nuc}$ of the nuclides in the
 galactic disc through the use of
the observed bulk (meteoritic)  abundances of radionuclides with lifetimes
commensurable with presumed $T_{\rm disc}$ values. 
Consequently, it is hoped
to provide at least a lower limit to $T_{\rm disc}$. The most studied
chronometries, and some of their main  characteristics, 
are summarized elsewhere 
\cite{Arnould90a}. The discovery of isotopic anomalies attributed to the
{\it in situ} decay in some meteoritic material of radionuclides with
half-lives in the  approximate
 $10^5 \lsimeq t_{1/2} \lsimeq 10^8$ y range has
broadened the original scope of nucleo-cosmochronology. 
It likely provides some
information on discrete nucleosynthesis  events that 
presumably contaminated
the solar system at times between about $10^5$ and $10^8$ y prior to the
isolation of the solar material from the general 
galactic material, as well as
constraints on the chronology of nebular and planetary events in the early
solar system.  This chronometry using short-lived radionuclides
(\cite{Podosek97,Arnould97}) is not reviewed here.

If indeed the bulk solar-system composition witnesses the operation of the
galactic blender, a reliable evaluation of $T_{\rm nuc}$ requires: 
(i) the build-up of models for the evolution of nuclides in the Galaxy, 
primarily in the solar neighbourhood, that account for
as many astronomical 
data as possible;
 (ii) the construction of nucleosynthesis models that are 
able to provide the
 isotopic or elemental yields for the nuclides involved 
in the chronometry; as well as (iii) high quality data for the meteoritic 
abundances of the 
relevant nuclides.  All these requirements clearly make the
chronometric task especially demanding.

\subsubsection{The trans-actinide clocks}
%
The most familiar long-lived chronometers are the 
\chem{232}{Th}-\chem{238}{U} and \chem{235}{U}-\chem{238}{U} pairs
\cite{Fowler60} developed on grounds of the present meteoritic content of
these nuclides. 
Their use as reliable nuclear clocks raises major problems,
however, as it depends heavily, among others, on the
availability of precise production ratios. 
Such predictions are out of reach
at the present time. One is indeed dealing 
with nuclides that can be produced
by the r-process only, which suffers from 
very many astrophysics and nuclear
physics problems, as we have emphasized in many 
occasions. In addition, these nuclides 
are the only naturally-occurring ones beyond
\chem{209}{Bi}, so that any extrapolation relying 
on semi-empirical analyses
and fits of the solar r-process abundance curve is in danger of being
especially unreliable. The difficulty is further 
reinforced by the fact that
most of the r-process precursors of U and Th are 
nuclei that are unknown in
the laboratory, and will remain so for a long time to come.  Theoretical
predictions of properties of relevance, like masses, 
$\beta$-decay strength functions and fission barriers, are extremely
difficult, particularly as essentially no calibrating points exist.
This problem would linger even if a realistic r-process model were given.
Last but not least, most of the huge
amount of work devoted in 
the past to the trans-actinide chronometry (\cite{Cowan91})
has adopted simple functionals for the time dependence of the r-process
nucleosynthesis rate  (a.k.a. ``Mickey Mouse Models'' coined by Pagel
\cite{Pagel90}) with little consideration of the chemical evolution in the
solar neighbourhood.

The so-called Th-chronometry \cite{Pagel89} attempts to use the relative
abundances of Th and Eu 
(which is presumed to be dominantly produced by the
r-process) observed at the surface of stars with various
metallicities.\footnote{Originally, 
an attempt was made to use the observed
Th/Nd ratios \cite{Butcher87}, albeit 
the disadvantage of Nd being possibly
produced also by the s-process}
Under the assumption, which may sound reasonable but has not at all to be
taken for granted, 
that any r-processes in the past have produced Th and Eu
with a constant ratio, the age determination is reduced to the problem of
mapping the metallicity on time through a chemical evolution model.
High-quality observational Th/Eu abundance data in stars of various
metallicities are accumulating \cite{daSilva90} - \cite{Sneden96}.
Though some attempts have already been made
\cite{Cowan97,Pfeiffer98}, much remains to be done in
the difficult task of deriving $T_{\rm nuc}$ from some of these
observations.

The Th-chronometry could be put on safer grounds if the Th/U ratios  
would be known in a variety of stars with a high enough accuracy 
(\cite{Pfeiffer98} regarding
the current status of 
the U abundance determination 
 in a metal-poor star). These nuclides are indeed likely to be
produced simultaneously, so that one may hope to be able to predict their
production ratios more accurately than Th/Eu. Even in such relatively
favourable circumstances, one would still face the severe question of
whether Th and U were produced  
in exactly the same ratio in presumably a few
r-process events (a single one?) that have contaminated the material from
which metal-poor stars formed. Even if this ratio would turn out to be the same
indeed, its precise value remains to be calculated (see 
\cite{Arnould90a} for an illustration 
of the dramatic impact of a variation
in the predicted Th/U ratio on predicted ages).

\subsubsection{The \chem{187}{Re} - \chem{187}{Os} chronometry}\ \ \
%
First introduced by Clayton \cite{Clayton64}, the chronometry using the
\chem{187}{Re} - \chem{187}{Os} pair is able to avoid the difficulties 
related to the r-process modelling. True, \chem{187}{Re} is an r-nuclide.
However, \chem{187}{Os} is not produced  directly by the r-process, but 
indirectly via the $\beta^-$-decay of \chem{187}{Re} ($t_{1/2} 
\approx 43$ Gy) over the galactic lifetime. This makes it in principle
possible to derive a lower bound for $T_{\rm nuc}$
from the mother-daughter abundance ratio, provided that the ``cosmogenic"
\chem{187}{Os} component is deduced from 
the solar abundance by subtracting
its s-process contribution. This chronometry is thus in the first instance
reduced to a question concerning the s-process. 
Other good news come from the
recent progress made in the measurement of the abundances of the concerned
nuclides in meteorites (\cite{Faestermann98} for references). 
This input
is indeed essential for the establishment of a reliable chronometry.
 
Although the s-process is better understood than the r-process, this
chronometry is facing specific problems. They may be summarized as follows
(\cite{Takahashi98a}): 1) the evaluation of the
\chem{187}{Os} s-process 
component from the ratio of its production to that
of the s-only nuclide \chem{186}{Os} is not a trivial matter, even in the
simple local  steady-flow approximation (constancy of the product of the
abundances by the stellar neutron capture rates over a restricted
$A$-range). The difficulty relates to the fact that
 the \chem{187}{Os} 9.75
keV excited state can contribute significantly to the
 stellar neutron-capture
rate because of its thermal population in s-process conditions ($T \gsimeq
10^8$ K) \cite{Woosley79,Winters86}. The ground-state capture rate
measured in the laboratory has thus to be modified by a theoretical
correction. In addition, the possible branchings of the s-process path in
the $184 \leq A \leq 188$ 
region may be responsible of a departure from the
steady-flow predictions for 
the \chem{187}{Os}/\chem{186}{Os} production ratio
\cite{Arnould84,Kaeppeler91}); and 2) at the high temperatures, and thus
high 
ionization states, \chem{187}{Re} may experience in stellar interiors,
its 
$\beta$-decay rate may be considerably, and sometimes enormously, enhanced
over the laboratory value by the bound-state $\beta$-decay of its ground
state to the 9.75 keV excited state of 
\chem{187}{Os} \cite{Takahashi83,Yokoi83}.
Such an enhancement has recently been beautifully confirmed by the
measurement of the decay of fully-ionized 
\chem{187}{Re} at the GSI storage
ring \cite{Bosch96,Kienle98}. The inverse transformation of \chem{187}{Os}
via free-electron captures is certainly responsible for further
corrections to the stellar \chem{187}{Re}/\chem{187}{Os} abundance
ratio \cite{Yokoi83,Arnould72}. Further complications arise
because these two nuclides can be concomitantly destroyed by
neutron captures \cite{Yokoi83}.
 
All the above effects have been studied in the framework of realistic  
evolution models for $1 \lsimeq M \lsimeq 50$ M$_\odot$ stars
\cite{Takahashi98a,Takahashi98}. The range of ages labelled SS in 
Fig.~17 gives the plausible upper   
limit on $T_{\rm nuc}$ derived from \chem{187}{Re} - \chem{187}{Os} 
within a galactic chemical evolution model that is constrained by
observational data in the solar neighbourhood \cite{Takahashi98}.
This work, which is an up-date of \cite{Yokoi83} with regards to  
meteoritic abundances, nuclear input data, stellar evolution models and  
observational constraints, leads to a lower limit of  about 11.5 Gy for
$T_{\rm nuc}$.  However, as even lower values cannot 
conclusively be excluded
within the remaining uncertainties in the chemical 
evolution model parameters,
the lower limit adopted for the SS range in Fig.~17 is 
given by the so-called
``model independent approach"  \cite{Schramm70} -\cite{Schramm90}.
These results may imply that the
 \chem{187}{Re} - \chem{187}{Os} chronometry
has not yet much helped narrowing the age range. There is still ample room
for improvements, however, and it is reasonably hoped that the Re -
Os chronometry will be able to set some meaningful 
limits on $T_{\rm nuc}$ in a
near future, and independently of other methods.

\subsection{Type-II supernovae}
%
A star more massive than about 10 M$_\odot$ is 
expected to end its life as a Type-II supernova explosion with a neutron
star left behind as its ``residue.'' We have discussed bit by bit the
nuclear physics involved during the evolution of such stars. Let us gather
here all these pieces of information to try understanding 
Type-II supernovae. These objects are indeed the most
remarkable astrophysical events in terms of the diversity of the nuclear
physics questions which play a decisive role in a sequential manner before
reaching the explosion stage. For simplicity, we consider only spherically
symmetric models (\cite{Bethe90} - \cite{Janka93}), except for some
limited considerations about a recent study of convective motions by
multi-dimensional hydrodynamical simulations (\cite{Janka96}).
%
%
\begin{figure}
\center{\includegraphics[width=1.00\hsize,height=0.72\hsize]{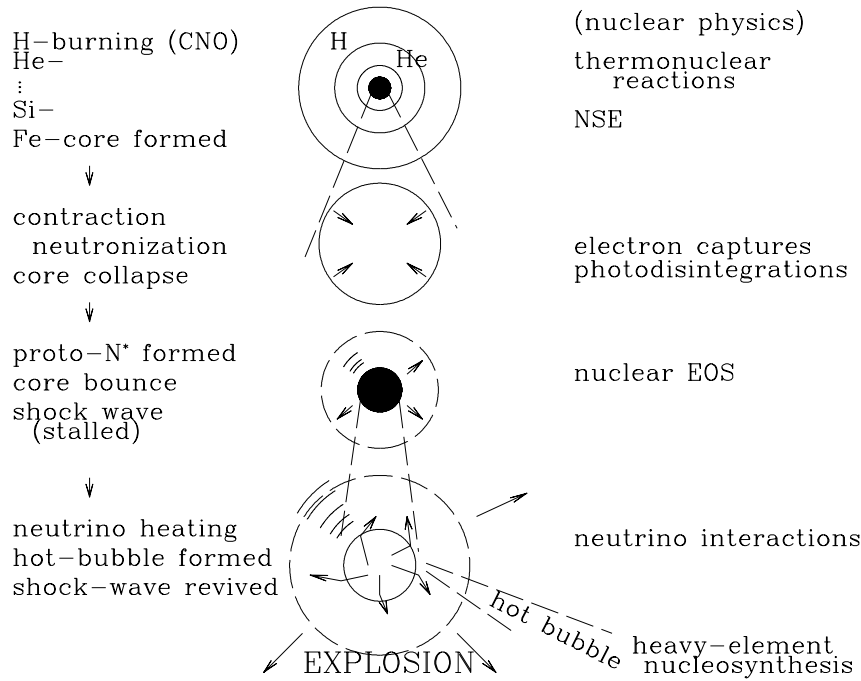}}
\vskip-1.2truecm
\caption{Schematic description of 
the evolution of a massive star towards a
supernova explosion}
\end{figure}

\subsubsection{Evolution of massive stars leading to neutrino-driven 
supernovae}\ \ \
%
Figure 18 depicts schematically the evolution of a massive star towards a
Type-II supernova. The time sequence (running from top to bottom) is
labelled with some nuclear physics phenomena that are decisive at each 
major epoch:  1) After having 
experienced various burning phases governed by
thermonuclear reactions, 
the star exhibits an ``onion-skin''  structure with an 
``iron'' core in its central region, this ``Fe peak'' composition being
dictated by the laws of statistical mechanics (NSE). 
The nuclear physics raised in these phases is already discussed in several
of the previous sections, and is not repeated here;
2) As the nuclear fuel is exhausted in the core 
(any possible nuclear transformation of its 
iron content is endothermic), it starts to contract 
(as usual when not enough nuclear
energy is available for covering the star energy losses). The
resulting increase of density and temperature near the 
centre makes possible
free electron captures on protons and some heavy nuclei, as well as the
photodisintegration of heavy nuclei. Both of these nuclear 
processes reduce
the pressure, and thus accelerate the contraction. 
The nuclear physics of
interest at this stage is somehow under control,
 except perhaps the evaluation
of free-e$^-$ capture rates on heavy nuclei;   3) As its mass nears the
``Chandrasekhar mass" of about 1.2 to 1.5 M$_{\odot}$, the core undergoes
a gravitational collapse to form a  proto-neutron 
star.\footnote{The Chandrasekhar mass is the limiting mass of a
configuration whose mechanical equilibrium is obtained by the compensation
of the gravitational forces by the pressure exerted by fully degenerate
electrons}
After a sufficient increase in density, the pressure from
the (non-relativistic) nucleons dominates the (relativistic) electron
pressure, so that the EOS becomes ``stiffer" (as mentioned 
earlier, there are tremendous difficulties 
in constructing ``the'' EOS and in
evaluating its stiffness in particular.) 
The mechanical equilibrium is restored
in the innermost parts of the core, which stop collapsing, leading to the
formation of a hot nascent neutron star.  An outward moving
shock develops at the interface of this configuration and the outermost
``infall'' material. Numerical simulations show,
 however, that the shock stalls
before it really reaches the outer edge of the initial Fe core, meaning no
``prompt" (i.e. on hydrodynamical time-scales) destruction of the star; 
4) Various kinds of (anti-)neutrinos are produced 
very near the  centre of the
nascent hot neutron star through various mechanisms which have been well
studied. Because of the high matter densities, those neutrinos cannot
escape freely (like in most previous evolutionary stages),
 but are gradually
transported towards the periphery of the nascent neutron star. 
The simulation
of this neutrino transport is a difficult task, particularly if one
attempts to describe the process by solving the Boltzmann transport
 equations.
An additional huge nuclear-physics complication arises 
in connection with the
required evaluation of neutrino-nucleus interaction cross sections, 
which necessitates the
careful treatment of various many-body effects (\cite{Raffelt96} -
\cite{Yamada98}).
 Those streaming neutrinos interact with the material near
the surface of the nascent neutron star that 
consists by then of neutrons and
protons. 
Very important, energy is deposited mainly by the
absorptions of
$\nu_e$ by neutrons 
and of $\bar{\nu_e}$ by protons, leading a portion of that
material to be expelled in the form of a ``neutrino(-energized) wind." A
rapidly expanding high-temperature and low-density region results, the
``hot bubble.'' 
Several hydrodynamical simulations support the idea
that the neutrino energy 
deposition of as little as  1\% of the approximate
$10^{53}$ erg of the gravitational 
potential energy of the forming neutron star is
sufficient for reviving the stalled shock wave, and make it drive an
explosion. 
The nuclear uncertainties associated with this stage largely reflect
those affecting the earlier phase, and in particular those concerning the
neutrino energy spectra. 

\subsubsection{Nucleosynthesis in the hot bubble: Can the r-process
 occur ?}\ \ \
%
The expelled material, or ``neutrino wind,'' is made of neutrons and 
protons. By the time the expansion has cooled the material below
about (10 to 7)$\times 10^9$ K, the weak 
interaction processes have largely
ceased, leading to a somewhat fixed 
neutron/proton ratio $n/p$ in excess of
unity\footnote{The ``electron concentration" 
$Y_{\rm e} = 1/(n/p+1)$ is often
used instead of $n/p$}
and to a somewhat fixed entropy $s$. 
The entropy is essentially determined by
the contributions from photons, electrons and positrons, and is very high 
because of the high temperatures 
and low densities. 
The expansion of this material favours the recombination of the
nucleons into heavier and 
heavier species starting with $\alpha$-particles and
ending with nuclei heavier than 
iron when the temperatures have reached values
slightly in excess of about $10^9$ K, at which point the charged-particle
induced reactions essentially freeze-out. 
A nuclear flow associated with these
transformations, termed the $\alpha$-process, is depicted in Fig.~15 
for a
generic hot-bubble model.

There has been
much hope that the captures of the neutrons left at the time of
the freeze-out would switch the
$\alpha$-process into an r-process \cite{Woosley92}. For this to happen,
and for the neutron-capture flow to possibly reach the
trans-actinide region, about $100 \sim 150$ 
neutrons have to be available per seed nuclei produced by the
$\alpha$-process. This translates into the requirement of
either very high $s$, low
$Y_{\rm e}$, short expansion time 
scale $\tau_{\rm exp}$, or of any combination
thereof (\cite{Witti94,Hoffman97,Meyer97}). 
For example, the model used to derive Fig.~15 assumes a set of 
 parameter values leading to a very short $\tau_{\rm exp}$, even if
the assumed $s = 200k$ per baryon ($k$ is the Boltzmann constant) and $Y_{\rm e} = 0.4$
are not  unthinkable.
 
Save
a claim of success \cite{Woosley94}, it has become a general consensus
that the currently available hydrodynamical 
simulations  of the hot bubble are
unable to provide suitable conditions 
for the occurrence of the r-process 
(\cite{Takahashi97,Hoffman97,Qian96}). The situation gets
even more unfavourable when considering the possible destruction of
$\alpha$-particles by neutrinos via the neutral current
of weak interaction \cite{Meyer95}. 
It remains, however, that a twist of 
$s, Y_{\rm e}$ or $\tau_{\rm exp}$ from
the model values can lead to an r-process that very well matches
the solar r-process abundance curve \cite{Takahashi94}. In this respect,
further scrutiny of the neutrino wind conditions (especially the neutrino
energy spectra) at the onset of nucleosynthesis may still be of value.

\subsubsection{Signatures of a large-scale mixing of nucleosynthesis
 products}\ \ \
%
The revived shock wave ejects into the ISM the  
nucleosynthesis products available at the pre-supernova stage,
save the possible modification of some of their abundances 
by the ``explosive
nucleosynthesis" associated with the brief period of shock 
re-heating of the
expelled layers.  Here we pick an example illustrating how 
observed nuclear
imprints in supernova remnants could help deepening (or at 
least, challenging
to deepen) our understanding of the very mechanism of the explosion.  

Several observations in SN1987A (early detection of X- and $\gamma$-rays, 
and strongly Doppler-shifted broad infrared emission lines,
 particularly of Fe 
and Ni) suggest that some portion of the nuclides have been transported
to the outer hydrogen envelope from as deep as close to the
surface of the nascent neutron star
much more quickly than expected
(\cite{Janka96}). The unpredicted short mixing
time-scale was attested 
in particular by the observation rather soon after the
explosion of the $\gamma$-ray lines emitted following the decay of the
\chem{56}{Ni} and \chem{57}{Co}  synthesized deep  in the interior.
 Along with other symptoms, such as the clumps and anisotropies in the
remnant, the above observations have prompted multi-dimensional 
hydrodynamical
simulations  which indeed predict large-scale convective motions
 in the
post-shock region
\cite{Janka96}.  This makes it technically difficult, if not
 impossible, to
predict the fate of the nucleosynthesis products in the material
 expelled from
a Type-II supernova.  We add here that the same simulations lead to
 nearly spherical
hot-bubbles after the cease of these convective motions.  

\section{Summary}
%
%
\begin{figure}
\center{\includegraphics[width=1.00\hsize,height=0.8\hsize]{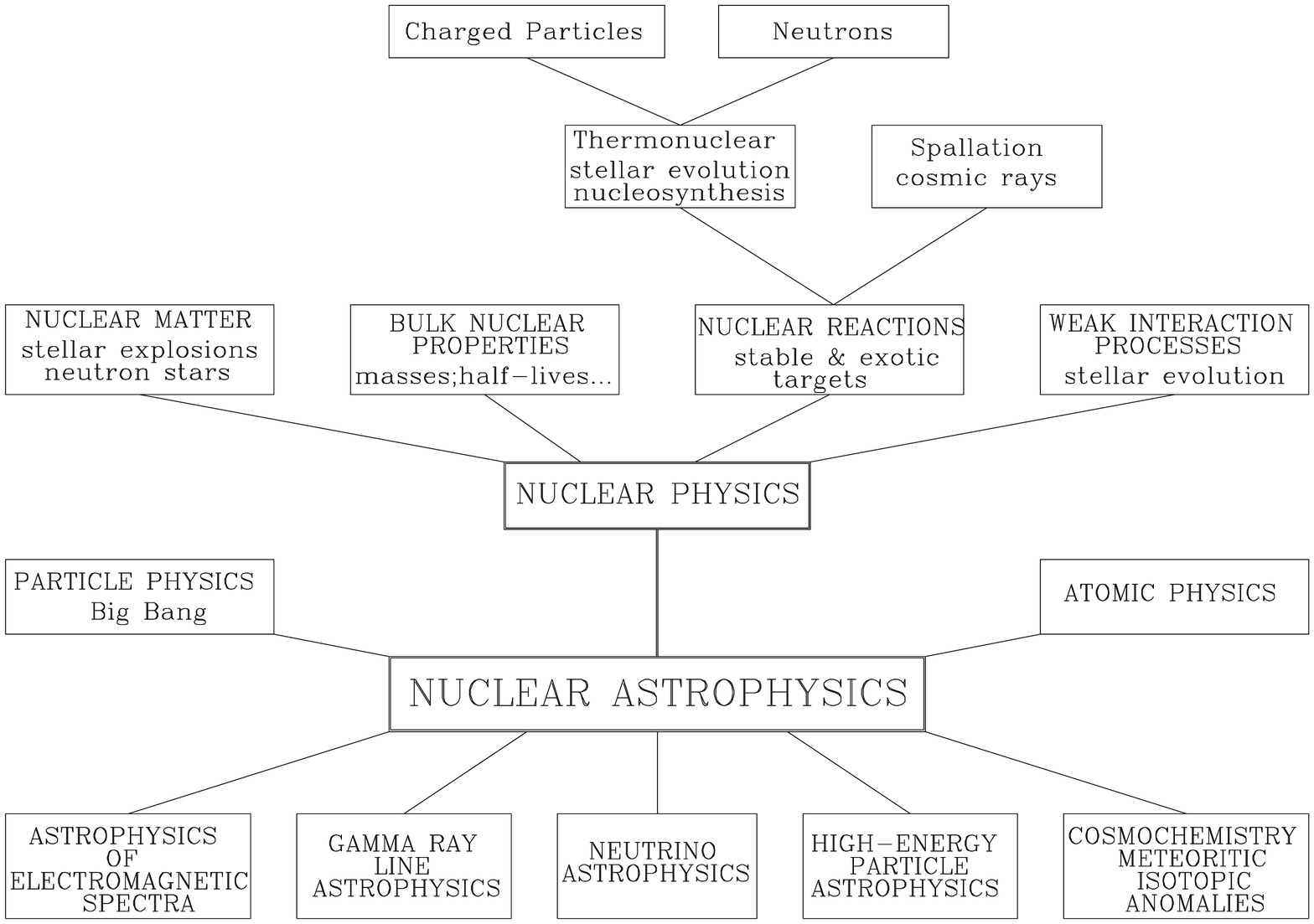}}
\vskip-1.9truecm
\caption{Diagrammatic presentation of possible connections of nuclear
astrophysics with other major fields of physics, and with various main
astrophysics subfields \cite{Arnould94N}}
\end{figure}
%
Nuclear astrophysics is without any doubt vastly interdisciplinary. 
As schematized in Fig.~19, it is in fact intimately connected with a 
huge variety of different and complementary 
research fields that are essential
in our understanding of a large diversity of classes of problems. 
As such, the nuclear astrophysics quest identifies 
itself to the everlasting
search for an El Dorado, and opens up at the same 
time a Pandora Box of
scientific questions. This makes the field especially
 exciting, but at the
same time remarkably demanding. This review has obviously
 not been able to go
beyond a quite limited overview of some of the many connections
drawn in Fig.~19.

One of our goals has been to demonstrate that 
nuclear physics and astrophysics
bring their share to the common adventure of  understanding the
ever-growing body of observations concerning the 
structure and composition of the
Universe and of its various constituents, 
ranging from sub-millimetre grains in
meteorites to galaxies at different redshifts. At all these scales,
observational data obtained with 
the help of an impressive panoply of advanced
techniques identify the strong imprints of 
the properties of atomic nuclei, 
and of their interactions. In such conditions, careful and dedicated
experimental and theoretical studies of a 
large variety of nuclear processes
are indispensable tools for the
 modelling of ultra-macroscopic systems such as
stars.

A major common challenge to modern nuclear 
physics and to nuclear astrophysics
is the exploration of {\it terra incognita}  located on the 
sides of the chart
of the nuclides well away from the valley of  nuclear stability
(and/or at moderately-high nuclear excitation). When
Marie and Pierre Curie discovered a century ago the mysterious
phenomenon of radioactivity, they certainly 
could not have imagined that they had just
opened the way to a better understanding 
of the Universe.  
Despite the impressive experimental as well as theoretical progress having
been  made since that time,
much obviously remains to be done in order to meet the
unusual request from 
 astrophysics regarding the exploration of the nuclear
exoticism. Even along the valley of nuclear stability, nuclear astrophysics 
has a lot of
scientific ebullience in store. There, astrophysics forces to 
enter the world
of ``almost no event,'' and to look for a needle in a haystack. 
Indeed, the
charged-particle induced reaction cross sections of astrophysical
 interest are
without doubt
 among the smallest one may ever dream measuring in the laboratory.
This has led 
nuclear experimentalists to start going deep underground to do
astrophysics, in  the search for a ``background noise deterrent,''
as vividly witnessed by the advent  of neutrino astrophysics.

 Where nuclear astrophysics will stand after a century from now is 
hard to imagine.
For sure, however,  a closer collaboration and an
improved mutual understanding between the nuclear- and astro-physics
communities would carry success in promoting further excitements 
in the field of nuclear astrophysics in the coming millennium.
\vskip1.0truecm
\noindent{\bf References}
\vskip0.5truecm


\end{document}